\documentclass[fleqn,usenatbib]{mnras}

\usepackage{newtxtext,newtxmath}

\usepackage[T1]{fontenc}

\DeclareRobustCommand{\VAN}[3]{#2}
\let\VANthebibliography\thebibliography
\def\thebibliography{\DeclareRobustCommand{\VAN}[3]{##3}\VANthebibliography}

\usepackage{graphicx}	
\usepackage{amsmath}	
\usepackage{amssymb}	
\usepackage{xcolor}
\usepackage{multirow}
\definecolor{exsitu}{RGB}{100,149,237}
\definecolor{migrated}{RGB}{255,127,80}
\definecolor{insitu}{RGB}{247,83,148}

\title[Centers of TNG50 Galaxies]{The Origin of Stars in the Inner 500 Parsecs in TNG50 Galaxies}

\author[Boecker et al.]
{Alina Boecker$^{1,2}$\thanks{E-mail: aboecker@iac.de},
Nadine Neumayer$^{1}$,
Annalisa Pillepich$^{1}$,
Neige Frankel$^{1,3,4}$,
Rahul Ramesh$^{5}$, 
\newauthor
Ryan Leaman$^{6}$,
Lars Hernquist$^{7}$
\\
$^{1}$Max-Planck Institut f\"{u}r Astronomie, K\"{o}nigstuhl 17, D-69117 Heidelberg, Germany \\
$^{2}$Instituto de Astrof\'{i}sica de Canarias, C/ V\'{i}a L\'{a}ctea s/n, E-38205 La Laguna, Spain \\
$^{3}$Canadian Institute for Theoretical Astrophysics, University of Toronto, 60 St. George Street, Toronto, ON M5S 3H8, Canada \\
$^{4}$Department of Astronomy and Astrophysics, University of Toronto, 50 St. George Street, Toronto, ON M5S 3H4, Canada \\
$^{5}$Universit\"{a}t Heidelberg, Zentrum f\"{u}r Astronomie, Institut f\"{u}r theoretische Astrophysik, Albert-Ueberle-Str. 2, D-69120 Heidelberg, Germany \\
$^{6}$Department of Astrophysics, University of Vienna, T\"{u}rkenschanzstrasse 17, 1180 Wien, Austria \\
$^{7}$Harvard–Smithsonian Center for Astrophysics, 60 Garden Street, Cambridge, MA 02138, USA
}

\date{Accepted 2022 December 16. Received 2022 November 25; in original form 2022 April 7}

\pubyear{2022}

\begin{document}
\label{firstpage}
\pagerange{\pageref{firstpage}--\pageref{lastpage}}
\maketitle

\begin{abstract}
\noindent
We investigate the origin of stars in the innermost 500\,pc of galaxies spanning stellar masses of $5\times10^{8-12}\,\mathrm{M}_{\odot}$ at $\mathrm{z=0}$ using the cosmological magnetohydrodynamical TNG50 simulation. Three different origins of stars comprise galactic centers: 1) in-situ (born in the center), 2) migrated (born elsewhere in the galaxy and ultimately moved to the center), 3) ex-situ (accreted from other galaxies). In-situ and migrated stars dominate the central stellar mass budget on average with 73\% and 23\% respectively. The ex-situ fraction rises above 1\% for galaxies $\gtrsim10^{11}\,\mathrm{M}_{\odot}$. Yet, only 9\% of all galaxies exhibit no ex-situ stars in their centers and the scatter of ex-situ mass is significant ($4-6\,\mathrm{dex}$). Migrated stars predominantly originate closely from the center ($1-2\,\mathrm{kpc}$), but if they travelled together in clumps distances reach $\sim10\,\mathrm{kpc}$. Central and satellite galaxies possess similar amounts and origins of central stars. Star forming galaxies ($\gtrsim10^{10}\,\mathrm{M}_{\odot}$) have on average more ex-situ mass in their centers than quenched ones. We predict readily observable stellar population and dynamical properties: 1) migrated stars are distinctly young ($\sim2\,\mathrm{Gyr}$) and rotationally supported, especially for Milky Way mass galaxies, 2) in-situ stars are most metal-rich and older than migrated stars, 3) ex-situ stars are on random motion dominated orbits and typically the oldest, most metal-poor and $\alpha$-enhanced population. We demonstrate that the interaction history with other galaxies leads to diverse pathways of building up galaxy centers in a $\Lambda$CDM universe. Our work highlights the necessity for cosmological context in formation scenarios of central galactic components and the potential to use galaxy centers as tracers of overall galaxy assembly.

\end{abstract}




\section{Introduction}\label{sec: intro}

The center of a galaxy depicts its brightest and densest region. Thus observations of galaxy centers provide us with the highest data quality, which should enable us to make the most precise predictions about their formation. On the other hand, being also the deepest point of the potential well, the center witnessed the galaxy's overall stellar assembly from the earliest cosmic times onward, as understood from the inside-out formation scenario of galaxies within a $\Lambda$CDM (Lambda-Cold-Dark-Matter) Universe. Therefore, many transformative processes of galaxy evolution influence a galaxy's center until the present day, which need to be taken into account to uniquely interpret even the highest quality observations. \par
As a consequence, a variety of central stellar structures are found in galaxies. Decreasing in size from the order of one kpc to sub parsec scales, these range from bars and (pseudo)bulges \citep[see e.g.][for a summary]{bulges,reviewbulge}, which can include other structures such as nuclear rings and disks, to nuclear star clusters \citep[NSCs; see e.g.][for a summary]{reviewnsc} and supermassive black holes \citep[SMBHs; see e.g.][for a summary]{reviewbh}. Some galaxies may exhibit more than one of these components or none at all. Many of these components possess scaling relations of their structural parameters, such as the \citet[][]{sersic} index and effective radius of bulges \citep[e.g.][]{gadotti09a,fisherdrory} and the luminosity/mass-size relation of NSCs \citep[e.g.][]{boker04,cote06,georgiev14}, as well as scaling relations with each other, such as the bulge-SMBH-mass \citep[e.g.][]{neumayer04,sani11,lasker16} and NSC-SMBH-mass relations \citep[e.g.][]{ferrarese06,georgiev16}, which also scale with the stellar mass of their underlying host galaxy \citep[e.g.][]{scott13,reines15,sanchez19}. Some of these scaling relations can differ for early-type and late-type galaxies, or depend on the bulge type or the presence of a bar \citep[e.g.][]{gadotti09b,georgiev16,davis19,sahu19}. \par
As diverse as the structural properties of central components are, so are the formation scenarios trying to explain them. Broadly speaking, all of these formation scenarios can be divided into internal and external processes. For example, bulges are thought to form from merger events \citep[e.g.][]{hopkins09,hopkins10}, from rapid early-on star formation \citep[e.g.][]{guedes13,okamoto13}, from secular evolution \citep[e.g.][]{reviewbulge,athanassoula05} or from the migration of clumps formed in the disk at high redshift \citep[e.g.][]{elmegreen09,dekel09}; bars form through disk instabilities either in isolation \citep[e.g.][]{bottema03,athanassoula13} or in a cosmological context \citep[e.g.][]{romano08,kraljic12,peschken19}; nuclear star clusters are thought to form through either star formation \citep[e.g.][]{maciejewski04,aharon15} or through the migration and successive merging of globular clusters in the center \citep[e.g.][]{hartmann11,agarwal11}; SMBHs can grow by accreting gas and by merging with other SMBHs \citep[e.g.][]{croton06,malbon07,fanidakis2011,lapiner21}. In many cases, the formation of any one component will also influence the others. For example, once a bar is formed it can re-arrange the orbits of stars causing radial migration, or it can efficiently funnel gas to the center, which can trigger star formation in the center and also feed the SMBH. In turn, the AGN (active galactic nucleus) feedback caused by the SMBH will then influence the gas supply and hence truncate the formation of stars. Thus, it is important to also understand the interplay between the presence and formation of several central components.\par
Observationally, we can only indirectly deduce constraints on any of these formation scenarios from the stellar population and dynamical properties of a galaxy's central structure(s). For external galaxies, such necessary measurements are only possible with integral field units (IFUs) that provide spatially resolved stellar population and kinematical maps \citep[e.g.][]{gadotti20,bittner20}. While major progress has been made in producing these maps with increasing quality, it is still difficult to disentangle stars from centrally overlapping galaxy components due to the line-of-sight integration - let alone identify stars of different origins within \emph{a} given central component. This is possibly further complicated by the fact that stars with properties characteristic of one formation scenario might be subdominant in luminosity or mass compared to the bulk stellar population. \par
Even in the Milky Way, it has only become evident fairly recently that all major central components contain metal-poor subpopulations of stars that also exhibit different kinematics. For the Galactic bulge \citep[see e.g.][for a summary]{barbuy18} there is a smooth transition from rotation to dispersion dominated kinematics for stars decreasing from (super-)solar metallicity all the way to the lowest metallicities ($[\mathrm{Fe/H}]<-2.0\,\mathrm{dex}$) \citep{ness13,zoccali17,arentsen20}. To a lesser extent this decrease is also seen for the nuclear stellar disk \citep{schultheis21} with additional evidence of recent star formation activity ($<1\,\mathrm{Gyr}$) on top of the overall old bulk population ($>8\,\mathrm{Gyr}$) \citep{paco20,paco21}. The nuclear star cluster, which hosts the most metal-rich stars in the Milky Way, also has a subpopulation of sub-solar metallicity stars, which show an asymmetric spatial distribution and a higher degree of rotation \citep{fk20,do20}. \par
Generally, signs of young, metal-rich and kinematically cold stars in these central structures such as bulges and NSCs, are associated with being formed in-situ from gas infall, while old, metal-poor and dispersion dominated systems are thought to originate from merger processes. However, stars formed in-situ at the beginning of a galaxy's lifetime are also metal-poor and might as well become dispersion dominated over time through various processes, such as resonances created by the bar. Therefore, even though observed properties of stars in the centers of galaxies act as a fossil record of their origin, we need simulations to disentangle which (combinations of) formation scenarios are able to predict those observations. \par
Cosmological, hydrodynamical galaxy simulations \citep[see e.g.][for a summary]{reviewsim,vogelsberger20} are ideal to study the complex formation pathways of galaxy centers as they encompass the most complete conglomeration of galaxy formation processes in a $\Lambda$CDM framework, thus capturing internal and external formation processes alike. The most recent simulations are able to produce a realistic, diverse population of galaxies \citep[see e.g.][and references therein for Illustris/TNG specifically]{vogelsberger14b,publictng}. Typically, large simulation boxes are used to study global galaxy properties across an array of different galaxies (e.g. \citealp[Illustris:][]{genel14,vogelsberger14a,vogelsberger14b}; \citealp[EAGLE:][]{schaye15,crain15};  \citealp[Horizon-AGN:][]{dubois14,dubois16}; \citealp[Magneticum:][]{hirschmann14,teklu15,bocquet16}; \citealp[IllustrisTNG:][]{weinberger17,pillepich18a}; \citealp[SIMBA:][]{dave19}), while zoom-in (re-)simulations focus on internal galaxy structures and dynamics (e.g. \citealp[ERIS:][]{guedes11}; \citealp[NIHAO:][]{wang15}; \citealp[Latte:][]{wetzel16}; \citealp[Auriga:][]{grand17}; \citealp[FIRE-2:][]{hopkins18}; \citealp[NIHAO-UHD:][]{buck20}). To understand the mass build-up of galaxy centers we need the advantages of both: a big enough box to probe many different assembly histories and thus galaxy demographics, and a zoom-in like resolution to focus on the center of galaxies and capture internal dynamical processes. \par
We therefore focus our analysis on the origin of stars in the central few hundred parsecs of galaxies in TNG50 \citep{pillepich19,nelson19} from the IllustrisTNG simulations. The 51.7$^3$\,cMpc$^3$ volume captures two $10^{14}\,\mathrm{M}_{\odot}$ halos and hundreds of Milky Way like galaxies, whereas the spatial resolution provides hundreds to tens of thousands stellar particles inside the central 500\,pc for a four dex range in galaxy stellar mass. Importantly, TNG50 starts to capture the diversity of central components, such as low and high S\'ersic index bulges in Milky Way like galaxies \citep{gargiulo21}, and performs well in a statistical comparison of simulated and observed bar properties \citep{rosasguevara21,frankel22}; both which were previously not possible with zoom-in simulations. Hence, TNG50 offers the unique opportunity to study the contribution of stars with different (internal or external) origins to the formation of the galaxy center across diverse galaxy formation pathways and demographics, while predicting the observable imprint that the different formation scenarios impose on the stars in a galaxy's center. \par
The goal of this study is to appeal to different scientific communities that focus on various central stellar structures of the Milky Way and external galaxies to provide an understanding where the most central stars of galaxies originate across a wide range of galaxy masses inside the TNG modelling framework. Specifically, we also study, for the first time, stars that have migrated towards the center to address formation scenarios of central structures that include the necessity for these processes such as NSC formation. Even though NSCs are not explicitly resolved in TNG50, we hope to offer new incentives for simulations \citep{antonini12,perets14,guillard16} and (semi-)analytical models \citep{antonini13,antonini15,leaman21} that are tailored towards NSC formation channels. Lastly, we aim to demonstrate that there are possibilities to use the bright centers of galaxies as a tracer of the galaxy's overall assembly history with readily available observables from current surveys such as SDSS \citep[e.g.][]{gallazzi20}. \par
This paper is organized as follows. In Section \ref{sec: analysis} we briefly describe the TNG50 simulation and the definition of properties of galaxies and stars (i.e. stellar particles) that we will analyze at $\mathrm{z}=0$. We also provide a detailed description and verification of selecting stars belonging to a galaxy's center and our galaxy sample selection. In Section \ref{sec: results2} we present the three different possible origins for stars residing in a galaxy's center and discuss their birth locations. In Section \ref{sec: results} we show the results of the different contributions of central stars of different origins across different galaxy population demographics and their observable stellar population and dynamical properties at $\mathrm{z}=0$. In Section \ref{sec: discuss}, we discuss our findings and implications from TNG50 on the central mass assembly of galaxies in a cosmological context. We also provide outlooks in the context of the formation of central galaxy components as well as the assembly of the overall host galaxy tailored towards measurements of extragalactic observations. Finally, we conclude our study in Section \ref{sec: summary}.

\section{Tools and Methods}\label{sec: analysis}

We briefly introduce the TNG50 simulation below as well as the properties of TNG50 galaxies and their stars (Section \ref{sec: properties_short}). We then describe in Section \ref{sec: center} how we define stellar particles that belong to a galaxy's center. \par

\subsection{The TNG50 simulation}

In this work we primarily study galaxies in TNG50 \citep{pillepich19,nelson19}, which is the highest resolution installment of the IllustrisTNG (Illustris \textit{The Next Generation}) \citep{pillepich18b,springel18,nelson18,naiman18,marinacci18} suite of cosmological, magnetohydrodynamical simulations\footnote{IllustrisTNG also encompasses two larger volume runs, namely TNG100 and TNG300 with subsequently coarser resolution.}. It provides unprecedented zoom-in like resolution within a representative cosmological volume with a box of 51.7\,cMpc on each side. \par
The simulation was performed with the \textsc{Arepo} code \citep{springel10,pakmor11,pakmor13,pakmor16}, which employs a finite-volume method on a moving-mesh to solve the equations of magnetohydrodynamics coupled with a tree-particle-mesh method for self-gravity. TNG50(-1) has a mass resolution of $4.6\times 10^5 \, \mathrm{M}_{\odot}$ for dark matter and $8.4\times 10^4 \, \mathrm{M}_{\odot}$ for baryonic particles. The softening length is 288\,cpc for collisionless particles for $\mathrm{z}\leq1$ and 576\,cpc for $\mathrm{z}>1$, whereas the softening length of the gas particles is adaptive depending on the local cell size of the moving mesh with a floor value of 74\,cpc. TNG50 is accompanied by three additional simulation runs (-2,-3,-4) that decrease the spatial resolution each time by half. The initial conditions are set according to cosmological parameters measured by \citet{planck15}. \par
Additionally, the TNG simulations implement a list of physical subgrid models, which describe galaxy formation and evolution, such as stellar formation and feedback, chemical enrichment, galactic winds, supermassive black hole growth and feedback. Details can be found in \citet{weinberger17,pillepich18a}.\par
Importantly, the TNG framework successfully reproduces key observational results such as the galaxy stellar mass function up until $\mathrm{z}<4$ \citep{pillepich18b}, bi-modality in galaxy color distribution \citep{nelson18}, the fraction of quiescent galaxies \citep{donnari19,donnari21b}, scaling relations, such as the galaxy mass-size relation \citep{genel18}, the gas-phase mass-metallicity relation \citep{torrey17} and certain element abundances \citep{naiman18}, as well as the clustering of galaxies \citep{springel18} and magnetic fields of massive halos \citep{marinacci18}. Specifically, the resolution of TNG50 allows for the study of internal dynamics and structures of galaxies \citep{pillepich19} as well as the influence of stellar and black-hole driven outflows on galaxy evolution \citep{nelson19}. \par
Results from the TNG simulation are output in 100 snapshots ranging from $\mathrm{z}=20$ until today with an approximate time step of 150\,Myr since $\mathrm{z}=4$. For each snapshot dark matter halos are identified by the friends-of-friends (FoF) algorithm \citep{fof} with a linking length of 0.2, with baryonic particles being attached to the same FoF group based on their nearest dark matter particle. Substructures within these halos, i.e. subhalos, are found through the \textsc{Subfind} algorithm \citep{subfind1}, which is run on both dark matter and baryonic particles. To track the mass assembly of subhalos/galaxies through cosmic time, merger trees are constructed based on the \textsc{Sublink} algorithm \citep{rodriguesgomez15}. The merger trees were constructed twice, once based on dark matter and once based on baryonic matter alone. \par
The entire simulations' particle information for the 100 snapshots, the halo and subhalo catalogues, merger trees as well as many more additional supplementary data catalogues are made publicly available on the TNG website\footnote{\url{https://www.tng-project.org}} \citep[see also][for the public data release]{publictng}.

\subsection{General note on calculations}

Unless otherwise stated we employ the following definitions in our subsequent calculations and plots. To center the coordinate system on a galaxy of interest we choose the position of the particle (of any type) with the minimum gravitational potential energy as the galaxy's center, as given by \texttt{SubhaloPos} in the subhalo catalogue. For the systemic velocity of a galaxy we use the median velocity of the 5\% most bound stellar particles. For face-on or edge-on projections, galaxies are oriented such that the z-axis is aligned with the total angular momentum of stellar particles within twice the stellar half mass radius. To track back galaxies in time we exclusively use the merger trees based on following baryonic particles (`\textsc{Sublink}\_gal'). Plots that display summary statistics of galaxy populations use a running median with a bin size of 0.25-0.3 dex, which is adapted, if necessary, to ensure a minimum number of ten galaxies per bin. Furthermore, all displayed quantities are in physical units and all provided \texttt{SubfindIDs} refer to galaxies at $\mathrm{z}=0$. \par
Throughout this study the terms in-situ, migrated and ex-situ always refer to stars within the central 500\,pc of galaxies unless otherwise stated.

\subsection{Galaxy characteristics and properties of their stars}\label{sec: properties_short}

\begin{table*}
    \centering
    \begin{tabular}{l|l|l|l}
         Property & Short description & Detailed description & Results \\ 
         \hline
         \hline
         \textbf{Overall galaxy} & & & \\ 
         \hline
         \textit{Mass} & total stellar or dynamical (i.e. stars+gas+dark) mass & \multirow{7}{*}{Appendix \ref{sec: galprops}} & \multirow{7}{*}{Section \ref{sec: demo}} \\
         \textit{Environment} & central or satellite & &\\
         \textit{Star formation activity} & star forming or quenched & &\\
         \textit{Morphology} & (kinematically) disk or bulge dominated & &\\
         \textit{Bar-like feature} & present or not, based on Fourier decomposition & &\\
         \textit{AGN feedback} & above or below average AGN feedback based on mass of the SMBH & &\\
         \textit{Physical Size} & compact or extended with respect to the mass-size relation & &\\ 
         \hline
         \hline
         \textbf{Individual stellar particle} & & &  \\
         \hline
         \textit{Age} [Gyr] & the lookback time when the star was born & \multirow{4}{*}{Appendix \ref{sec: starprops}} & \multirow{4}{*}{Section \ref{sec: pop}} \\
         \textit{Metallicity} $[\log_{10}Z/Z_{\odot}]$ & the total amount of metals & &\\
         \textit{[Mg/Fe]} [dex] & the abundance of magnesium as a proxy for $\alpha$-elements & &\\
         \textit{Circularity} $\epsilon$ & indicates the type of orbit the star is on & &\\
    \end{tabular}
    \caption{Properties of galaxies and their individual stars (stellar particles) in TNG50 at $\mathrm{z}=0$ investigated in this study. A detailed description on the exact calculation of the properties as well as the results with respect to the centers of galaxies are found in the indicated sections.}
    \label{tab: proptab}
\end{table*}

Throughout this study we are interested in two sets of demographics: 1) How does the central mass assembly of galaxies change as a function of a galaxy's overall bulk properties?, 2) How do the intrinsic properties of stars in the center of galaxies differ for different origins? \par
To address the first question we do not only study the central 500\,pc of galaxies as a function of the galaxy's total stellar (dynamical) mass, but we also divide our galaxy sample into different types of galaxies characterized at $\mathrm{z}=0$. To address the second question we study individual properties of stars (i.e. stellar particles) in the center of galaxies at $\mathrm{z}=0$. These investigated characteristics are briefly summarized in Table \ref{tab: proptab}, whereas a detailed description on their calculations can be found in Appendix \ref{sec: properties_long}. \par

\subsection{Defining stars belonging to a galaxy's center}\label{sec: center}

The most straightforward way to define a galaxy's center at $\mathrm{z}=0$ is to select all stellar particles within a 3D spherical aperture with a given radius $r_{\mathrm{cut}}$ around its center. This simple selection will give us knowledge about stellar particles that have an \emph{instantaneous} radius smaller than the selected aperture. However, as we are interested in the mass assembly of the center of galaxies, we want to make sure that selected particles roughly stay inside the spherical aperture over their orbital time at $\mathrm{z}=0$. This ensures that we track particles that changed their orbit, should they have migrated to the center, and not particles that are just on more eccentric orbits. \par 
To estimate, whether the particles are on such orbits confined to the center at $\mathrm{z}=0$, we calculate the specific energy $E_{\mathrm{cut}}$ a particle on a circular orbit with guiding radius $r_{\mathrm{cut}}$ would have, i.e.:
\begin{equation}\label{eq: ecut}
    E_{\mathrm{cut}}=\frac{v_{\mathrm{circ}}(r_{\mathrm{cut}})^2}{2}+\Phi(r_{\mathrm{cut}}).
\end{equation}
The circular velocity $v_{\mathrm{circ}}$ is calculated from the spherically enclosed mass (stellar, gas and dark matter particles) $v_{\mathrm{circ}}(r)^2=\frac{GM(<\,r)}{r}$, whereas the gravitational potential energy $\Phi$ is given by the simulation and interpolated to $r_{\mathrm{cut}}$. Stellar particles with total energies $E=\frac{|\mathbf{v}|^2}{2}+\Phi(\mathbf{x})$ less than $E_{\mathrm{cut}}$ should roughly be confined on orbits that are within the spherical volume with radius $r_{\mathrm{cut}}$, whereas particles with higher energies are able to move to larger radii and hence spend less time in the center. \par
We additionally enforce that the specific angular momentum in the z-direction $L_\mathrm{z}$ of stellar particles in the center lies between $L_{\mathrm{cut}}=\pm v_{\mathrm{circ}}r_{\mathrm{cut}}$, as we noticed that some lower mass galaxies with stellar masses $\lesssim 10^{10}\, \mathrm{M}_{\odot}$ had very large $L_\mathrm{z}$ and hence large radii with $E<E_{\mathrm{cut}}$, which probably stems from the fact that they are undergoing tidal stripping at present time. If particles with large orbital radii ($>2\,\mathrm{kpc}$) still persisted after this cut, we disregard them as well. These additional steps do not significantly affect the amount of central particles selected for galaxies with stellar masses $\gtrsim 10^{10}\, \mathrm{M}_{\odot}$. \par
In general, the selection based on Equation \ref{eq: ecut} is a simplification as it assumes a spherical mass distribution, but it gives a good enough estimate of which particles are truly confined to the center without actually integrating their orbits; see Appendix \ref{appendix: validate1} for validation of this with two galaxies contained in the subbox with higher time cadence. We visualize the difference between a simple selection in radius and the one in energy using Equation \ref{eq: ecut} in Figure \ref{fig: ecutvsradcut}. \par

\subsubsection{Choice of the central region}\label{sec: center_choice}

The last step in selecting stellar particles belonging to a galaxy's center is to set a value for $r_{\mathrm{cut}}$, with which we can in turn calculate $E_{\mathrm{cut}}$. \par
We choose a fixed value of 500\,pc\footnote{Due to the selection of stellar particles belonging to the 500\,pc center based on their energies, some particles will have instantaneous radii larger than 500\,pc at $\mathrm{z}=0$, but typically not larger than 1\,kpc.} for $r_{\mathrm{cut}}$ across all galaxies to avoid running too close into the numerical softening length (see Section \ref{sec: moresims} for further elaboration on this). We explicitly do not choose to adopt a mass-dependent size, as already with a 10\% scaling of the mass-size relation of TNG50 galaxies, we are at the softening length of $10^{10}\,\mathrm{M}_{\odot}$ in stellar mass galaxies, while for the highest mass galaxies we approach 5\,kpc, which we do not deem to be central anymore. We also refer the reader to Section \ref{sec: moresims} and Appendix \ref{appendix: validate5} for a more detailed discussion and investigation about numerical resolution effects and the choices of $r_{\mathrm{cut}}$. \par

\begin{figure}
    \centering
    \includegraphics[width=\columnwidth]{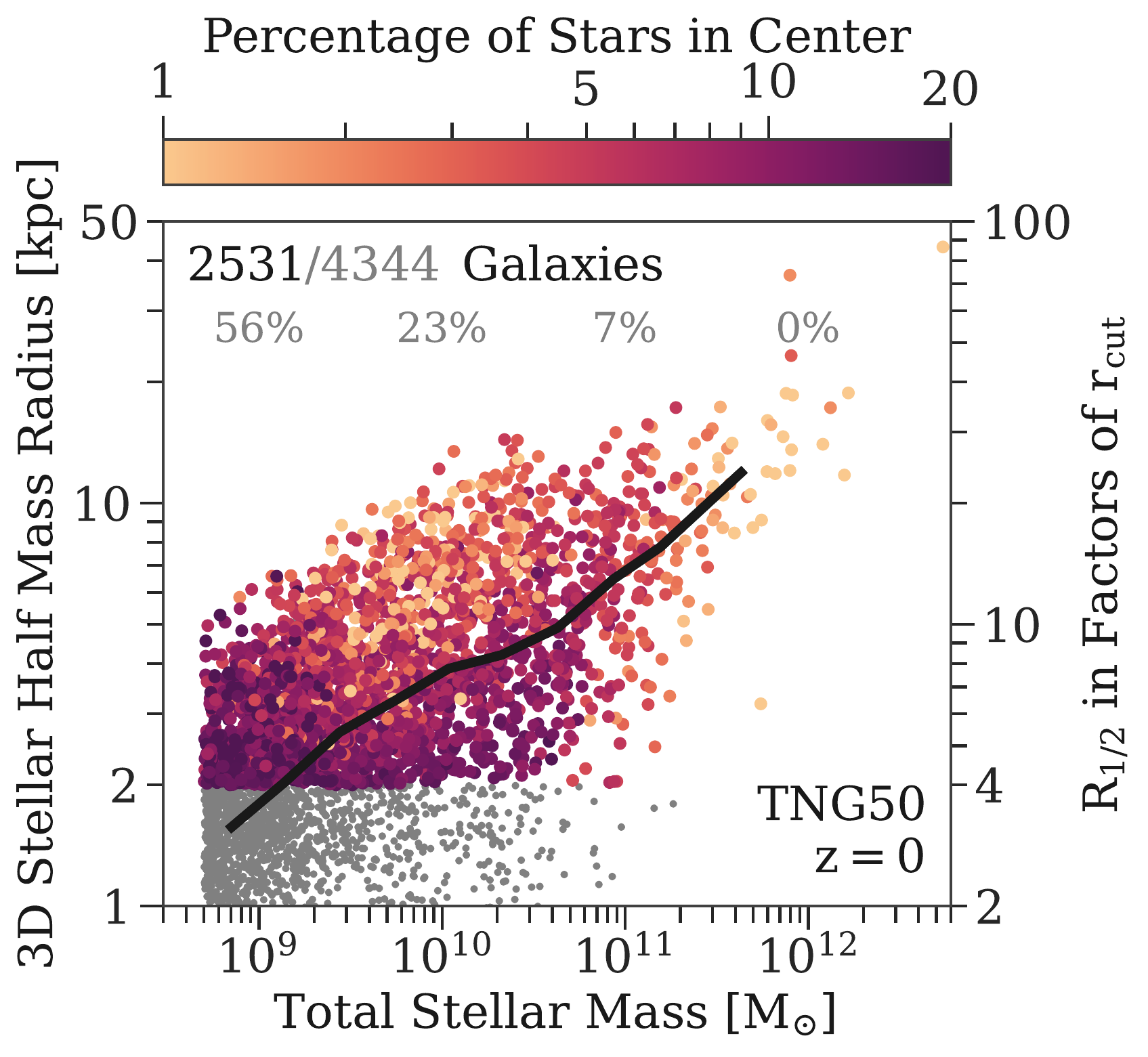}
    \caption{\textbf{Sample selection of TNG50 galaxies at} $\mathbf{z=0.}$ Stellar mass-size relation of TNG50 galaxies at $\mathrm{z}=0$ considered in this analysis colored coded according to their percentage of stars in the center relative to their total number of stellar particles. We employ a lower total stellar mass cut of $5\times 10^{8}\,\mathrm{M}_{\odot}$ leaving 4344 Galaxies. We additionally impose a minimum 3D stellar half mass radius $\mathrm{R}_{1/2}$ of 2\,kpc (i.e., $\mathrm{R}_{1/2}/\mathrm{r}_{\mathrm{cut}}\geq 4$ with $\mathrm{r}_{\mathrm{cut}}=500\,\mathrm{pc}$) resulting in a sample size of 2531 galaxies. Excluded galaxies are shown with grey points and percentages show their number fractions in bins of 0.5\,dex. The median stellar mass-size relation of \emph{all} TNG50 galaxies is shown as the black line.}
    \label{fig: sample}
\end{figure}

\begin{figure}
    \centering
    \includegraphics[width=\columnwidth]{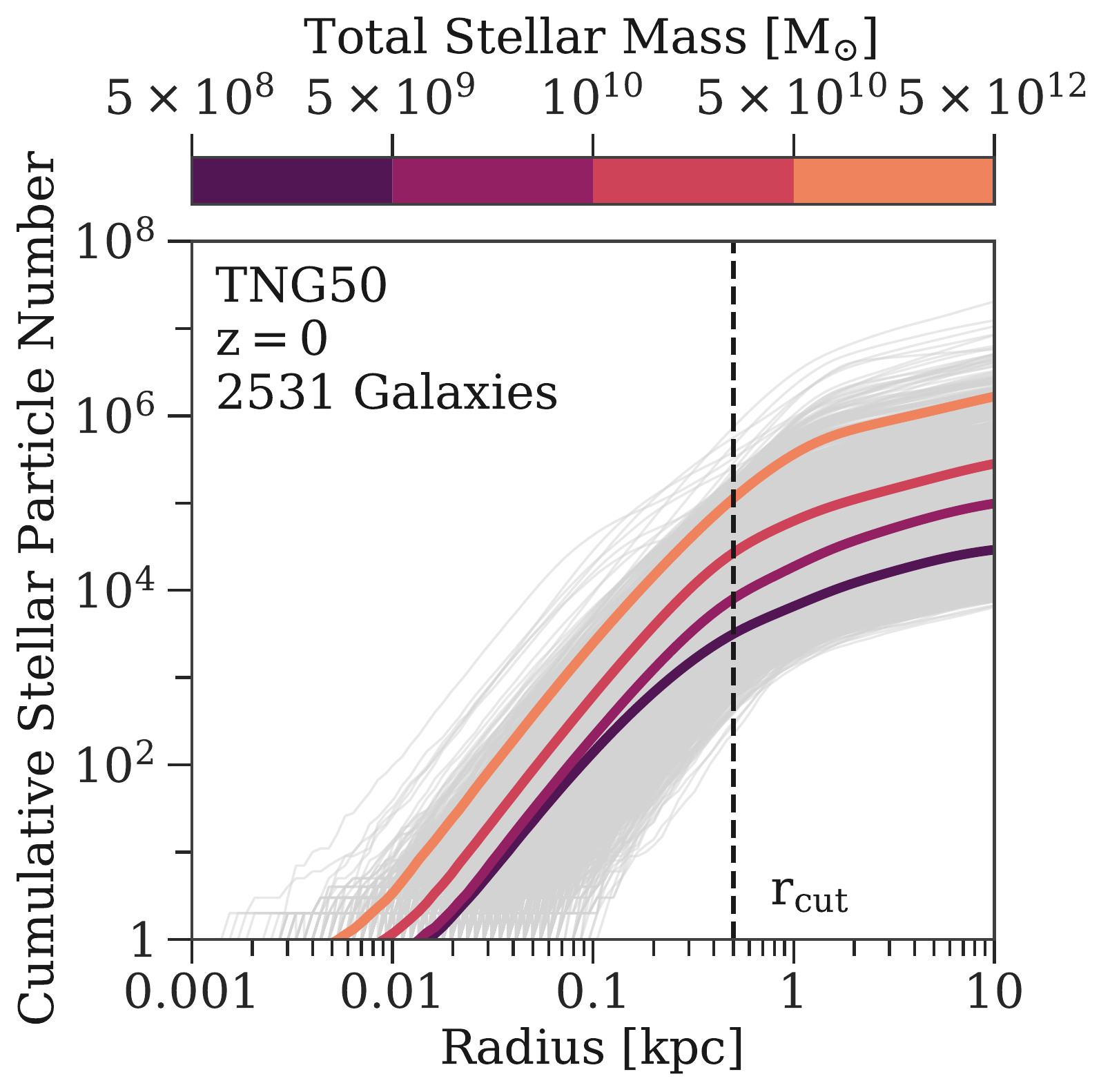}
    \caption{\textbf{Number of stellar particles in the central 500\,pc of our TNG50 galaxy sample at} $\mathbf{z=0.}$ Cumulative number of stellar particles as a function of radius per individual galaxy are shown as thin gray lines. The thick colored lines show the average per galaxy stellar mass bin as depicted by the colorbar. The dashed black line shows our adopted $\mathrm{r}_{\mathrm{cut}}$ value of 500\,pc. The average number of stellar particles within $\mathrm{r}_{\mathrm{cut}}$ lies between $10^3$ and $10^5$ for the lowest and highest galaxy mass bins respectively. No individual galaxy in our sample has less than 100 stellar particles in the center.}
    \label{fig: numberpar}
\end{figure}

\subsubsection{Galaxy sample selection}\label{sec: sample_choice}

Due to the choice of a fixed central aperture of 500\,pc we have to make some selection for our galaxy sample considered in this analysis. \par
Generally, sizes of TNG50 galaxies are numerically well converged above a stellar mass of $\sim 3\times 10^{8}\,\mathrm{M}_{\odot}$ \citep[see][]{pillepich19} at $\mathrm{z}=0$, but we employ a slightly higher lower mass cut of $5\times 10^{8}\,\mathrm{M}_{\odot}$ ensuring that the galaxies have a sufficient number of stellar particles for our analysis. We also only consider subhalos/galaxies that are of cosmological origin (i.e. \texttt{SubhaloFlag} is true). Any scaling relations used in this analysis such as the galaxy mass-size or the stellar-mass-black-hole-mass relation (to determine for example if galaxies lie above or below the median at fixed stellar mass) are always computed with respect to this galaxy sample, which contains 4344 galaxies. \par
Furthermore, for our main analysis of the centers of TNG galaxies, we enforce that the ratio of the 3D stellar half mass radius $\mathrm{R}_{1/2}$ and the central aperture ($\mathrm{r}_{\mathrm{cut}}=500\,\mathrm{pc}$) is greater than four, i.e. $\mathrm{R}_{1/2} \geq 2\,\mathrm{kpc}$. Otherwise, galaxies are too compact for our selected central aperture and about half of the entire galaxy will be classified as ``central''. Additionally, we make sure that at least a hundred stellar particles are within the central 500\,pc according to Section \ref{sec: center}, otherwise the galaxy is disregarded. \par
Our final galaxy sample selection yields 2531 TNG50 galaxies and their masses and sizes are visualized in Figure \ref{fig: sample}. The data points are color-coded by the percentage of stars inside the central 500\,pc compared to the total amounts of stars. The color trend is neither uniform in the direction of increasing stellar mass nor size. This hints at different density profiles for galaxies across their stellar mass-size plane. We note that our subsequent results do not show any strong differential trends, even though our constrain of $\mathrm{R}_{1/2} \geq 2\,\mathrm{kpc}$ disregards half the galaxies with stellar masses $<10^{9.5}\,\mathrm{M}_{\odot}$ (see Section \ref{sec: moresims} for a further discussion). \par
In Figure \ref{fig: numberpar} we show the cumulative number of stellar particles as a function of their instantaneous radius at $\mathrm{z}=0$ for our galaxy sample. The average number of stellar particles in the center, i.e. within 500\,pc, is around $10^3$ for galaxy stellar masses between $5\times10^8\,\mathrm{M}_{\odot}$ and $5\times10^9\,\mathrm{M}_{\odot}$ and increases towards $10^5$ for the highest mass bin\footnote{A synonymous measure can be achieved with the \texttt{StellarHsml} field, which gives an approximation for the spatial extent a single stellar particle samples from the underlying stellar density field. The spherical radius of stars within 500\,pc ranges between $10-100\,\mathrm{pc}$ for the highest to lowest mass galaxies respectively.}. Hence, our choice for $r_{\mathrm{cut}}$ ensures that we have enough stellar particles in the center to reliably study their properties. We can also observe a turn-over in the stellar particle number profile at radii around $0.5-1\,\mathrm{kpc}$ confirming that we are indeed probing the densest (central) region of TNG50 galaxies.

\section{The different origins of stars in the center of TNG50 galaxies}\label{sec: results2}

After selecting stars in the center of TNG50 galaxies at $\mathrm{z}=0$, we investigate their different origins. We find three general populations of stars in the central region of galaxies, which we describe in Section \ref{sec: origindef} in detail. We also present the distribution of their birth origin for stacks in galaxy stellar mass in Section \ref{sec: birth}.

\subsection{Definition of different origins}\label{sec: origindef}

We define the following different origins of stars in the center of TNG50 galaxies at $\mathrm{z}=0$:

\begin{itemize}
    \item \textit{\color{insitu}in-situ}: stars were born inside the host galaxy's center and are still found there at $\mathrm{z}=0$.
    \item \textit{\color{migrated}migrated}: stars were born gravitationally bound inside the host galaxy but outside its center. At $\mathrm{z}=0$ they reside in the host galaxy's center.
    \item \textit{\color{exsitu}ex-situ}: stars were born inside other galaxies, which merged with the host and are ultimately found inside the host's center at $\mathrm{z}=0$.
\end{itemize}

\subsubsection{Born inside or outside the host galaxy}\label{sec: exsitu}

To determine whether a star is born inside a galaxy or was brought in through merger events, we use the stellar assembly catalogue\footnote{This particular catalogue has not been publicly released.} produced by methods of \citet{rodriguesgomez16} for TNG50. This classifies stellar particles that formed along the main progenitor branch of a galaxy, i.e. the galaxy with the most massive history behind it, as in-situ ($\mathtt{InSitu}=1$) and otherwise as ex-situ ($\mathtt{InSitu}=0$). The ex-situ stars generally have two possible origins: they either came from galaxies that completely merged with the main galaxy, i.e. they are present in the merger tree of the host, or were stripped from galaxies that do not belong to the host's merger tree, e.g. flybys. \par 
Additionally, we treat subhalos/satellites that directly merged onto the main progenitor branch of a galaxy but are flagged as not being of cosmological origin (i.e. $\mathtt{SubhaloFlag}=0$ in the subhalo catalogue) differently in this study. These subhalos are often formed within another galaxy as e.g. a fragment of the baryonic disk, contain little dark matter and hence are not thought of as galaxies \citep[see also][Section 5.2]{publictng}. Because the construction of the stellar assembly catalogue involves the use of merger trees, which only track stellar particles and star-forming gas cells of subhalos, these spurious galaxies are counted as of ex-situ origin. Here, we change their labelling back to in-situ (i.e. their \texttt{InSitu} flag in the stellar assembly catalogue becomes true again) for now (see Section \ref{sec: clump} for the implications of this), because we only consider ex-situ particles coming from true external galaxies. We verify with Figure \ref{fig: tot_exsitu} in Appendix \ref{appendix: validate2} that this change does not alter the overall \emph{total} ex-situ stellar mass fraction of TNG50 galaxies significantly. We note that spurious galaxies brought to the main progenitor branch of the host galaxy through prior merging with a real galaxy are continued to be counted as ex-situ. \par

\subsubsection{Born in-situ or migrated to the center}\label{sec: migrated}

To address whether a stellar particle is born inside the center of the host galaxy or migrated to the center from elsewhere inside the host galaxy, we need to determine its birth radius. A stellar particle with a birth radius smaller than $\mathrm{r}_{\mathrm{cut}}=500\,\mathrm{pc}$ is then consequently born in-situ and otherwise counts as migrated\footnote{We here apply a simple cut in the birth radius instead of calculating $E_{\mathrm{cut}}$ (i.e. following Section \ref{sec: center}), as the potential is not recorded for every snapshot.}. \par
In TNG, two new fields (\texttt{BirthPos} and \texttt{BirthVel}) for their stellar particles were added. These represent the spatial position and velocity of the star-forming gas cell that parented the stellar particle at its exact time of birth (i.e. \texttt{GFM\_StellarFormationTime}). In theory, this provides us with knowledge of the \emph{exact} birth condition of a stellar particle at the original time step resolution of the simulation; and not only at the output time steps of the snapshots. \par
Because these quantities are provided in the reference frame of the simulation box, we need to center them on the reference frame of the galaxy of interest. This however becomes an impossible task to do to the precision needed for our analysis, as we only know the center position of subhalos at the one hundred output snapshots, but the information of its trajectory in-between is lost. We find that even interpolating the subhalos' position with a higher order spline to the exact birth times of stars can lead to centering offsets of several kpc, especially when there is a merger in process or a pericenter passage around another galaxy (see Figure \ref{fig: subbox_centering}). As we are interested in typical scales of one \,kpc or less in this study, this problem is severe and will result in a strong bias towards stars being classified as migrated even though they where formed inside our selected spherical aperture. \par
We therefore define the birth position of stellar particles as the position they have in the snapshot they first appear in. Practically this is done by matching particles at $\mathrm{z}=0$ to their birth snapshot through their unique \texttt{ParticleIDs}. The caveat of this approach is that the stellar particles have already moved since their exact formation time, which can also lead to a wrong classification of migrated and in-situ stars. However, the error created by this approach is much smaller than the incorrect centering described above (see Figure \ref{fig: subbox_migrated}). \par
We verify this approach by looking at two subhalos that reside in the subboxes of TNG50. The subbox has 3600 snapshot outputs, which makes it possible to track the center position of galaxies across a much finer time resolution of a few Myr. The reader is referred to Appendix \ref{appendix: validate3} for details.

\subsubsection{Clumpy or smoothly migrated}\label{sec: clump}

Because we have changed the \texttt{InSitu} flag from the stellar assembly catalogue for spurious galaxies, in Section \ref{sec: exsitu}, we now find two types of migrated stellar particles in the center of galaxies. Stars either travelled individually (`smoothly' migrated) or together in clumps (`clumpy' migrated) to their galaxy's center. Smoothly migrated stars are genuinely born on the main progenitor branch of the subhalo/galaxy in question and the clumpy migrated stars originate from these spurious galaxies, i.e. stellar clumps. \par
Generally, these clumps are ubiquitous in TNG50 galaxies, about 36\% of all galaxies considered in this work have at least one throughout their life time. In stellar or gas surface mass density maps they look like massive star cluster like objects (see Figure \ref{fig: clumps_sum} for an example) that form within spiral arms or gaseous (disk) fragments during galaxy interactions. However, we want to be extremely cautious here, as it is unclear, if their formation is physical or due to some numerical artifact, even though measures against artificial fragmentation are in place. In fact, their sizes (i.e. 3D stellar half mass radii) lie mostly below the gravitational softening length of TNG50.\par
Once these clumps are formed, however, their dynamical evolution within the host galaxy is determined by gravity, which we believe is well captured in TNG50 (modulo the softening). Hence, depending on their density and the exerted tidal forces on the clumps, they are either completely disrupted or travel to the center of their host galaxy due to dynamical friction and deposit their stellar particles there. Their typical stellar masses are $\sim10^8\,\mathrm{M}_{\odot}$. We point the interested reader to Appendix \ref{appendix: validate4} for more statistics on the clumps and their properties. We provide an extensive discussion on the existence and formation of stellar clumps in simulations and observation in Section \ref{sec: clumpdiscuss}. \par
For the rest of the paper, we sometimes make the distinction between migrated particles coming from the `smooth' or `clumpy' migration, if it is explicitly stated. Otherwise, all general references to migrated properties always include both types.

\begin{figure*}
    \centering
    \includegraphics[width=\textwidth]{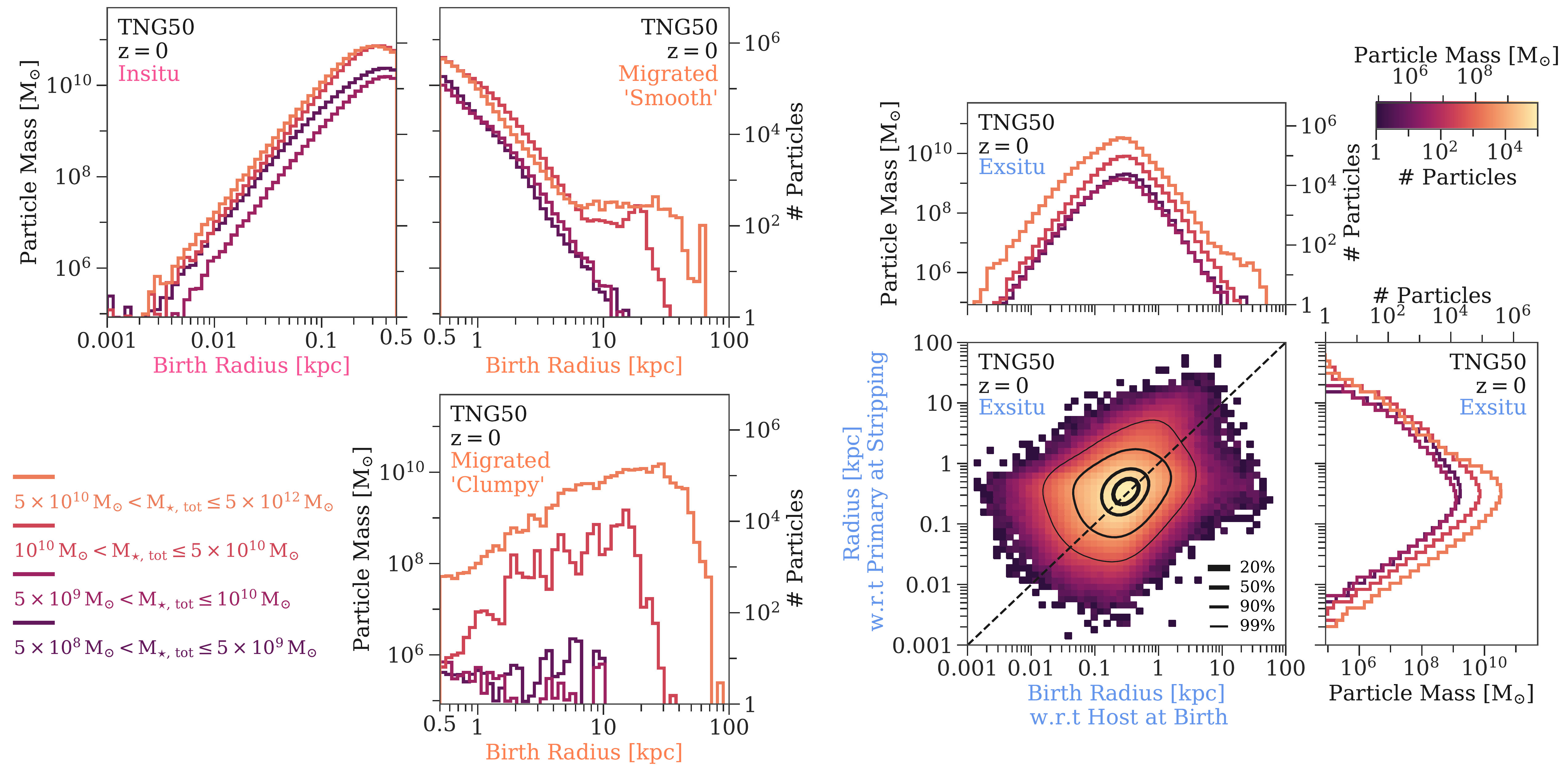}
    \caption{\textbf{Distribution of birth radii of in-situ, migrated and ex-situ stellar populations in the central 500\,pc of TNG50 galaxies at} $\mathbf{z=0.}$ \textit{Left}: Histograms of birth radii of the in-situ, `smooth' migrated and `clumpy' migrated stars colored according to stacks of galaxy stellar mass. The left hand side of the y-axis shows stacked particle mass, whereas the right hand side shows the number of stellar particles. The migrated stars can originate from large radii ($>10\,\mathrm{kpc}$) and the smoothly and clumpy migrated stars show different distributions. \textit{Right}: 2D histogram of the birth radius with respect to the host galaxy at birth and the radius with respect to the primary galaxy (i.e. final $\mathrm{z}=0$ host) at the time of stripping for all central ex-situ stars in our galaxy sample. The contours (from thicker to thinner lines) include 20\%, 50\%, 90\% and 99\% of all ex-situ particles. The respective 1D histograms for stacks in galaxy stellar mass are also shown. Most ex-situ stars are born in the central $\sim1\,\mathrm{kpc}$ of their birth galaxy and stay together until they are deposited also within the central $\sim1\,\mathrm{kpc}$ of their $\mathrm{z}=0$ host galaxy.}
    \label{fig: birthloc}
\end{figure*}

\subsection{Birth locations of the central stars}\label{sec: birth}

The distributions of birth radii of in-situ, migrated and ex-situ central stars are illustrated in Figure \ref{fig: birthloc} in stacks of galaxy stellar mass. \par
The in-situ stars are born (by definition) in the center of the host galaxy at $\mathrm{z}=0$. The peak of the birth radii distribution is around $200-300\,\mathrm{pc}$ for galaxies larger than $10^{10}\,\mathrm{M}_{\odot}$ and shifts slightly towards larger radii for the lower mass galaxy bins. We also see that higher mass galaxies birth more in-situ stars at all radii and hence are more centrally concentrated (see also Figure \ref{fig: numberpar}). \par

\subsubsection{Individually migrated stars originate close to the galaxy's center}

Most of the smoothly migrated stellar particles were also born close to the center with radii between 500\,pc and 1\,kpc, which is partly due to how we have defined them (i.e. purely based on their birth radius) and partly a consequence of the typical density profile of galaxies (i.e. more stars reside in the center of galaxies). For galaxies below $10^{10}\,\mathrm{M}_{\odot}$ the distribution of birth radii declines exponentially reaching the highest values of about 10\,kpc. The lowest mass galaxies in our sample ($\leq5\times10^{9}\,\mathrm{M}_{\odot}$) have 11\% of smoothly migrated stars, which are born in the range of $1-10\,\mathrm{kpc}$, whereas this increases slightly to 17\% for the next higher mass bin ($\leq10^{10}\,\mathrm{M}_{\odot}$). \par
For galaxies above $10^{10}\,\mathrm{M}_{\odot}$ we observe a plateau for the distribution of birth radii starting at $\sim10\,\mathrm{kpc}$, which stops at around 30\,kpc and 60\,kpc for galaxies with $\leq5\times10^{9}\,\mathrm{M}_{\odot}$ and $>5\times10^{9}\,\mathrm{M}_{\odot}$ respectively. The migrated stars originating from these large distances likely come from gas that was stripped during a merger, but which was already attributed to be gravitationally bound to the primary galaxy according to the \textsc{Subfind} algorithm and hence was counted as being born in-situ to the primary host. The percentage of smoothly migrated stars with birth radii larger than 1\,kpc is around 20\% and 14\% for the two highest stellar mass bins respectively. \par

\subsubsection{Stars migrated in clumps originate from the outskirts of galaxies}

The clumpy migrated stars show a distinctively different distribution than the smoothly migrated ones. For galaxies below $10^{10}\,\mathrm{M}_{\odot}$ their contribution is negligible. For galaxies between $10^{10}\,\mathrm{M}_{\odot}$ and $5\times10^{10}\,\mathrm{M}_{\odot}$ the clumpy migrated stars are only 3\% of the total migrated stars, whereas for galaxies above $5\times10^{10}\,\mathrm{M}_{\odot}$ the contribution rises to almost 50\%. Therefore, clump migration is only important for high mass galaxies, where it becomes the dominant driver for contributing migrated stars in the centers (see also Section \ref{appendix: validate4}). \par
Furthermore, the peak of the birth radii distribution of clumpy migrated stars is above $10\,\mathrm{kpc}$ for the high mass galaxies. This is in agreement with the fact that the gaseous disk of galaxies is much more extended than the stellar one \citep[e.g.][]{nelson12}. Stars travelling in stellar clumps are therefore able to migrate to the center of galaxies from much farther distances compared to when they travel individually.
\par

\subsubsection{Central ex-situ stars originate from the nuclei of their birth galaxies}

Regarding the ex-situ stars, we investigate two different locations: 1) their birth place with respect to their \emph{birth} host galaxy and 2) the location they were deposited inside their $\mathrm{z}=0$ \emph{host} (primary). The latter is defined as the radius the stellar particles have with respect to the primary at stripping time, i.e. the time they last switched galaxies. \par
We show the distribution of these two quantities also in Figure \ref{fig: birthloc} in the same stacks of galaxy stellar mass, as well as the 2D distribution of all ex-situ stars for these two radii. About half of all ex-situ stars that reside in the center of galaxies at $\mathrm{z}=0$ exhibit values between 100\,pc and 1\,kpc for both radii respectively. This means that the ex-situ stars are also born in the center of their respective birth galaxies as well as remain in said center until they are deposited right in the center of the primary galaxy during the merger process. Hence, the central, most bound cores of galaxies are more likely to stay together during accretion events until they arrive close to the center of the primary galaxy and ultimately deposit a large quantity of stars there. This is a consequence of mergers preserving the rank order of the particles' binding energy \citep{barnes88,hopkins09}. \par
We also find two other cases of ex-situ stars, albeit much lower in number. Firstly, TNG predicts a slight excess of ex-situ stars that are born at larger radii ($1-100\,\mathrm{kpc}$), but are still deposited close to the primary galaxy at stripping time, i.e. within $\sim1\,\mathrm{kpc}$. These stars represent a second generation of `migrated' stars; or likely in the case of ex-situ stars with birth radii of $\geq10\,\mathrm{kpc}$, stars that were formed from stripped gas during secondary mergers, which only appear for the most massive $\mathrm{z}=0$ hosts. Consequently, these ex-situ stars were born at large radii in their respective host galaxies (i.e. which will become the secondary galaxy during the merger process onto the $\mathrm{z}=0$ host), then migrated to the center of said galaxy in order to be deposited close to the center of the primary host during accretion. We confirm this by explicitly checking that their radii are indeed central ($\lesssim1\,\mathrm{kpc}$) with respect to the merging host galaxy one snapshot before the merger coalesces. \par
The second case represents ex-situ stars that were deposited at larger radii from the primary ($>1\,\mathrm{kpc}$), but born within the central 1\,kpc of their birth galaxy. Despite being stripped outside the center of the $\mathrm{z}=0$ host galaxy, these stars still were able to migrate such that they are found in the center of their respective galaxy at $\mathrm{z}=0$. There is a possibility that these stars were stripped earlier, i.e. before the merger coalesces, but their dynamics were still following the orbit of the galaxy undergoing the merger and hence they could arrive at the center of the final host galaxy.

\section{The central in-situ, migrated and ex-situ populations across TNG50 galaxies}\label{sec: results}

In this section we present our results of \textcolor{insitu}{in-situ}, \textcolor{migrated}{migrated} and \textcolor{exsitu}{ex-situ} populations within the central $\sim 500\,\mathrm{pc}$ of TNG50 galaxies. \par
We study their contributions across different galaxy properties (Section \ref{sec: demo}) and examine differences in their stellar population and dynamical properties (Section \ref{sec: pop}).

\subsection{Galaxy population demographics}\label{sec: demo}

Below we depict the contribution of the central stellar mass of the different origins as an overall trend with galaxy mass (Section \ref{sec: masstrend}), in correlation to each other (Sec \ref{sec: diversity}) and for different galaxy types (Section \ref{sec: types}).

\subsubsection{Galaxy mass trends}\label{sec: masstrend}

\begin{figure*}
    \centering
    \includegraphics[width=1.5\columnwidth]{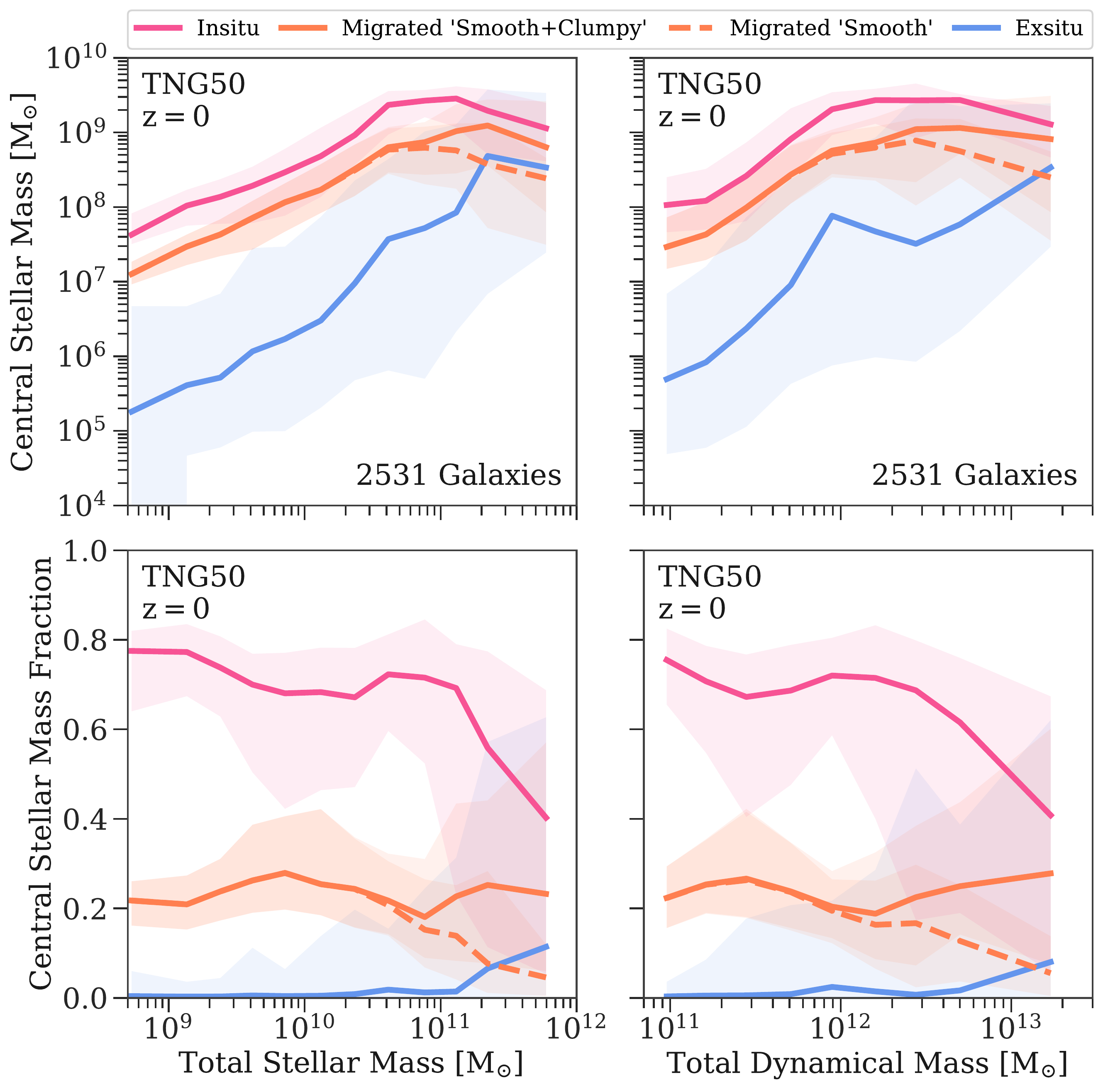}
    \caption{\textbf{Central (500\,pc) stellar mass of the in-situ, migrated and ex-situ populations of TNG50 galaxies at} $\mathbf{z=0.}$ Median trends of central stellar mass (\textit{top panels}) and central stellar mass fraction (\textit{bottom panels}) as a function of the galaxies' total stellar mass (\textit{left}) and total dynamical mass (\textit{right}) divided in the three origins: in-situ (\textit{pink}), migrated (\textit{orange}) and ex-situ (\textit{blue}). The migrated population is shown for both `smooth+clumpy' (\textit{solid line}) and just `smooth' (\textit{dashed line}) migration (see \protect\ref{sec: clump} for details). Shaded areas show the 16th and 84th percentiles. Overall the in-situ population dominates on average across all galaxy masses with the migrated population contributing around 20\% to the total central stellar mass. Only above galaxy masses of $10^{11}\,\mathrm{M}_{\odot}$ the ex-situ population starts to significantly contribute to the central mass build-up.}
    \label{fig: mass}
\end{figure*}

\begin{figure*}
    \centering
    \includegraphics[width=\textwidth]{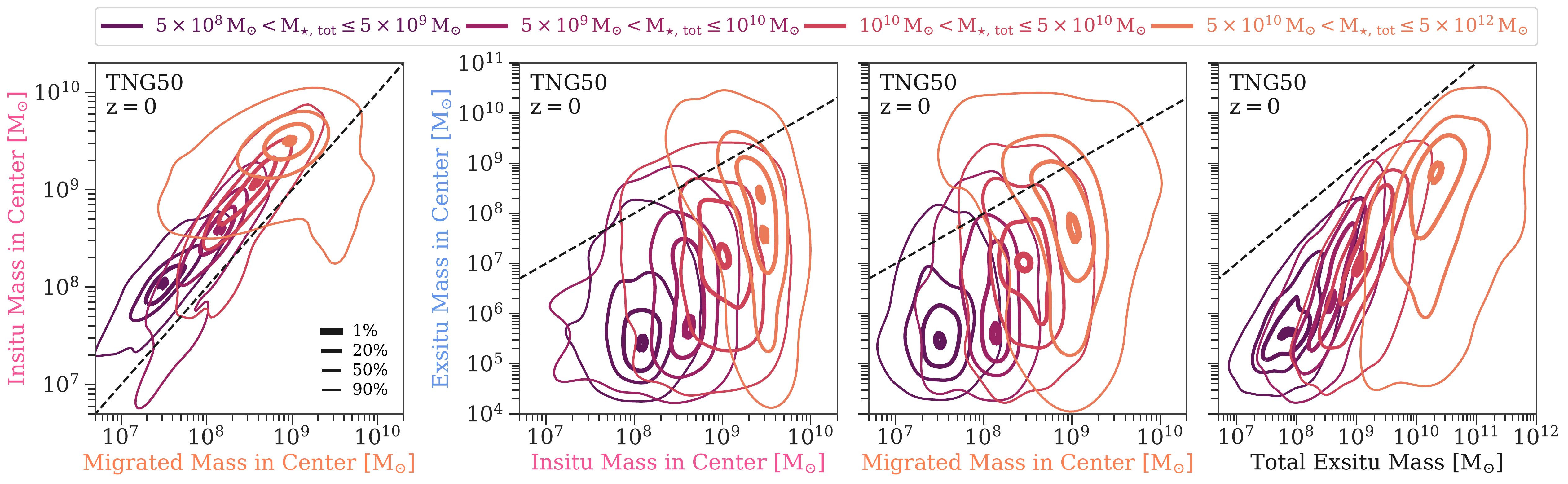}
    \caption{\textbf{2D distributions of the central (500\,pc) in-situ, migrated, ex-situ as well as the \emph{total} ex-situ stellar mass of TNG50 galaxies at} $\mathbf{z=0.}$ Gaussian kernel density estimates for different combinations of in-situ, migrated and ex-situ mass in the center as well as total ex-situ mass color-coded according to four different total stellar mass bins. The contours show different percentiles encompassing 1\%, 20\%, 50\% and 90\% of all data points (\textit{thickest to thinnest line}). The black dashed line shows the one-to-one relation in all panels. The correlation between the central mass of the in-situ and migrated population follow the one-to-one relation closely with a fixed offset, whereas the central ex-situ population exhibits a large scatter across the other central populations.}
    \label{fig: mass2}
\end{figure*}

In Figure \ref{fig: mass} we give an overview of the absolute and relative contribution of the ex-situ, migrated and in-situ population across galaxy masses (both stellar and dynamical) in TNG50.\par 
For all three populations the central stellar mass increases with increasing galaxy mass with the in-situ population dominating at all galaxy masses. Whereas the relation for the in-situ and the smoothly migrated stars have the same shape, the slope for the ex-situ population is steeper. The latter also shows a larger overall scatter due to the stochasticity of merger events contributing stars to the center. \par
Even though the fractional mass of the ex-situ population in the center is negligible for galaxy stellar masses below $10^{11}\,\mathrm{M}_{\odot}$, there are only 227 (9\%) galaxies in our total sample that have no central ex-situ mass, i.e. they do not possess a single stellar particle of ex-situ origin in their central 500\,pc, or in other words they possess less than $\sim8\times10^4\,\mathrm{M}_{\odot}$, the mass of a stellar particle, in ex-situ stars. Above $10^{11}\,\mathrm{M}_{\odot}$ the ex-situ mass becomes of the same order as the in-situ and migrated population, which is a consequence of mergers contributing a significant amount stellar mass to build up of these galaxies. The ex-situ mass reaches about 10\% of the total central stellar mass at the highest galaxy masses, albeit with a large scatter of up to 60\%. \par
Around galaxy stellar masses of about $5\times10^{10}\,\mathrm{M}_{\odot}$, the relation flattens for the in-situ and smoothly migrated stars, with the in-situ population reaching about 4\% of the total galaxy stellar mass. Although we have low number statistics of galaxies in this regime within the TNG50 volume (there are 18 galaxies with stellar masses above $5\times10^{11}\,\mathrm{M}_{\odot}$), it is reasonable that the in-situ mass goes down, because the ex-situ mass increases in addition to galaxies being quenched by AGN feedback. The consequential increased stochasticity is also seen by the larger scatter in the in-situ and migrated population at the highest galaxy stellar masses\footnote{Another possibility for the large scatter at high galaxy stellar masses for the in-situ and migrated central stellar mass could be stars formed from accreted gas, which was brought in by gas-rich mergers. We do not quantify this further as it is beyond the scope of this study.}. \par
The contribution of clumpy migrated stars to the overall central migrated population only starts to significantly affect galaxies with stellar masses higher than $5\times10^{10}\,\mathrm{M}_{\odot}$. For galaxies higher than $2\times10^{11}\,\mathrm{M}_{\odot}$ the clumps are responsible for roughly quadrupling the mass of migrated stars, or, in fractional terms, increasing the contribution of migrated stars to the total central mass from below 10\% to slightly above 20\%. Hence, the clumps are important driver to bring in stars from the outskirts of galaxies in TNG50. \par
Taking into account the entire migrated population (`smooth+clumpy'), we find a contribution of around 20\% to the total stellar mass in the center across all TNG50 galaxies. Interestingly, the total central migrated fraction around galaxy stellar masses of $\sim10^{10}\,\mathrm{M}_{\odot}$ slightly increases, with the 84th percentile reaching almost 40\%. We explicitly confirm that this is \emph{not} due to mixing galaxies with different sizes and hence different total central stellar masses (see also Figure \ref{fig: sample}). \par
The statements made so far also apply when correlating the central stellar masses of the three populations with the total dynamical mass of TNG50 galaxies. The larger scatter in all three relations is due to the scatter in the stellar-to-halo mass relation. \par

\subsubsection{The diversity of central stellar mass at fixed galaxy mass}\label{sec: diversity}

In Figure \ref{fig: mass2} three correlations between the central ex-situ, migrated and in-situ stellar mass as 2D Gaussian kernel estimates in bins of total galaxy stellar masses are shown. The bins were specifically chosen based on the change of the average migrated fraction as a function of galaxy stellar mass and, in case of the highest mass bin, to ensure enough galaxies per bin to reliably perform the kernel density estimate. \\ \par
\textbf{\textit{In-situ vs. migrated mass (Figure \ref{fig: mass2}, first panel)}:} At all stellar masses, the mass of migrated stars correlates strongly with the in-situ stellar mass with three times as much in-situ than migrated mass. This reflects our previous statements that the shape of the median relation of the central in-situ and migrated mass versus the total stellar galaxy mass is similar. \par
Nevertheless, for some galaxies the migrated mass is larger than the in-situ mass in the center as seen by the 90\% contours for the three highest galaxy mass bins in the range of $5\times10^9-5\times10^{12}\,\mathrm{M}_{\odot}$. Galaxies in this regime are dominated by clumpy migration. Hence, the mass contributed to the center by clumps can be significant enough to break the otherwise tight one-to-one relation of migrated and in-situ mass. \par
Lastly, we find for the 90\% contour in the highest mass bin for galaxies above $5\times10^{10}\,\mathrm{M}_{\odot}$ that there is a larger tail of galaxies with lower in-situ and migrated mass in the center. Most galaxies situated in this space have a high ex-situ central mass fraction of 40\% or higher. \\ \par
\textbf{\textit{Ex-situ vs. in-situ and migrated mass (Figure \ref{fig: mass2}, second and third panels)}:} At roughly fixed central in-situ or migrated masses there is a large variety of ex-situ mass that is deposited in the center of galaxies. The scatter of the ex-situ mass in the center increases roughly from four to six dex the from smallest to largest galaxy stellar mass bin. Compared to that the scatter in the in-situ and migrated mass direction is rather small, being roughly one dex across all galaxy stellar mass bins. However, some galaxies in mass bins below $5\times10^{10}\,\mathrm{M}_{\odot}$ have lower in-situ masses of up to one dex below the majority of the other galaxies in their respective bins. Galaxies lying in this region have above average migrated fractions of 40\% or more with some reaching extreme values of above 80\%. \par
The spread in migrated mass compared to the in-situ mass is larger for the highest mass galaxies. This is mainly due to the increased stochasticity in the total \emph{central} stellar mass for the 18 galaxies above $5\times10^{11}\,\mathrm{M}_{\odot}$ in galaxy stellar mass, which almost spans one dex as opposed to only a quarter dex for galaxies between $5\times10^{10}$ and $5\times10^{11}\,\mathrm{M}_{\odot}$. These 18 galaxies lie between the 50\% and 90\% contour and have ex-situ masses spanning from $10^{7}$ to $10^{11}\,\mathrm{M}_{\odot}$. Their in-situ masses are exclusively below the 50\% contour, whereas their respective migrated masses can lie towards lower or higher values. \par
The peak of the central ex-situ mass distributions begins to rise for galaxies above $10^{10}\,\mathrm{M}_{\odot}$ in total stellar mass, going from about three dex below the one-to-one relation to one dex. This break point roughly translates to $4\times10^8\,\mathrm{M}_{\odot}$ in central in-situ mass and $1-2\times10^8\,\mathrm{M}_{\odot}$ in central migrated mass. The former roughly coincides with the critical mass needed for the SMBH to be in the kinetic feedback mode \citep[e.g.][Figure 1]{zinger20}. The \emph{total} ex-situ mass also begins to rise for galaxies with total stellar masses above a few $10^{10}\,\mathrm{M}_{\odot}$ (see Figure \ref{fig: tot_exsitu}). \\ \par
\textbf{\textit{Central ex-situ vs. total ex-situ mass (Figure \ref{fig: mass2}, fourth panel)}:} Lastly, we also show the correlation between the \emph{central} ex-situ mass and the \emph{total} ex-situ mass, i.e. all stars that were ever accreted onto the $\mathrm{z}=0$ host galaxy. For fixed total galaxy stellar mass the slope of the contours depict that a higher total ex-situ mass generally also implies a higher central ex-situ mass. The slope of this correlation is rather steep. While the total ex-situ mass spans approximately two dex per galaxy stellar mass bin, the ex-situ mass in the center spans four to six dex from the lowest to highest galaxy stellar masses. Consequently, it is quite stochastic which merging satellite galaxies deposit stellar mass in the center. \par
Furthermore, the central density contours shift closer to the one-to-one relation with increasing galaxy stellar mass. This means that more galaxies in the highest mass bin have mergers that are more effective in bringing a larger fraction of their total ex-situ mass into their center as opposed to lower mass galaxies. Nevertheless, the 90\% contours for galaxy stellar masses above $5\times10^9\,\mathrm{M}_{\odot}$ extend right up to the one-to-one relation, meaning that there some galaxies that have almost all their ex-situ mass in the central 500\,pc. \par

\begin{figure*}
    \centering
    \includegraphics[width=1.75\columnwidth]{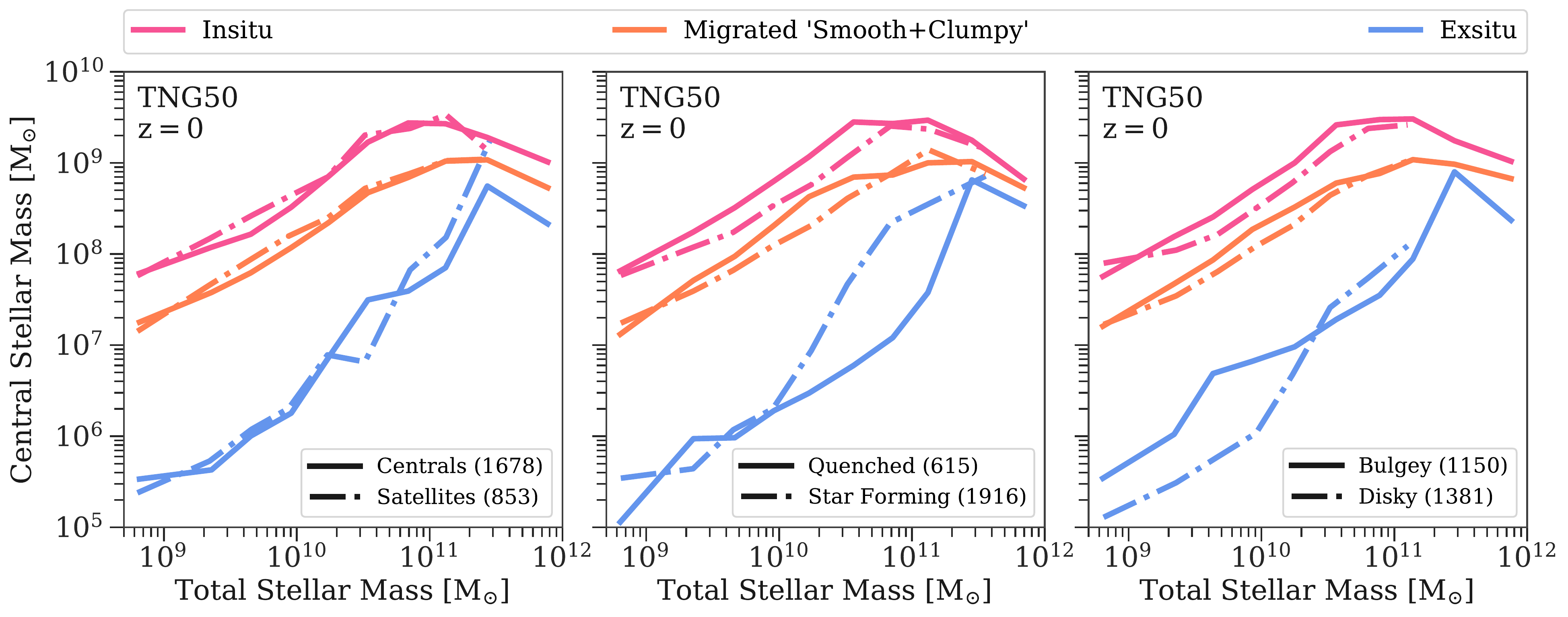}\\
    \vspace*{12pt}
    \includegraphics[width=1.75\columnwidth]{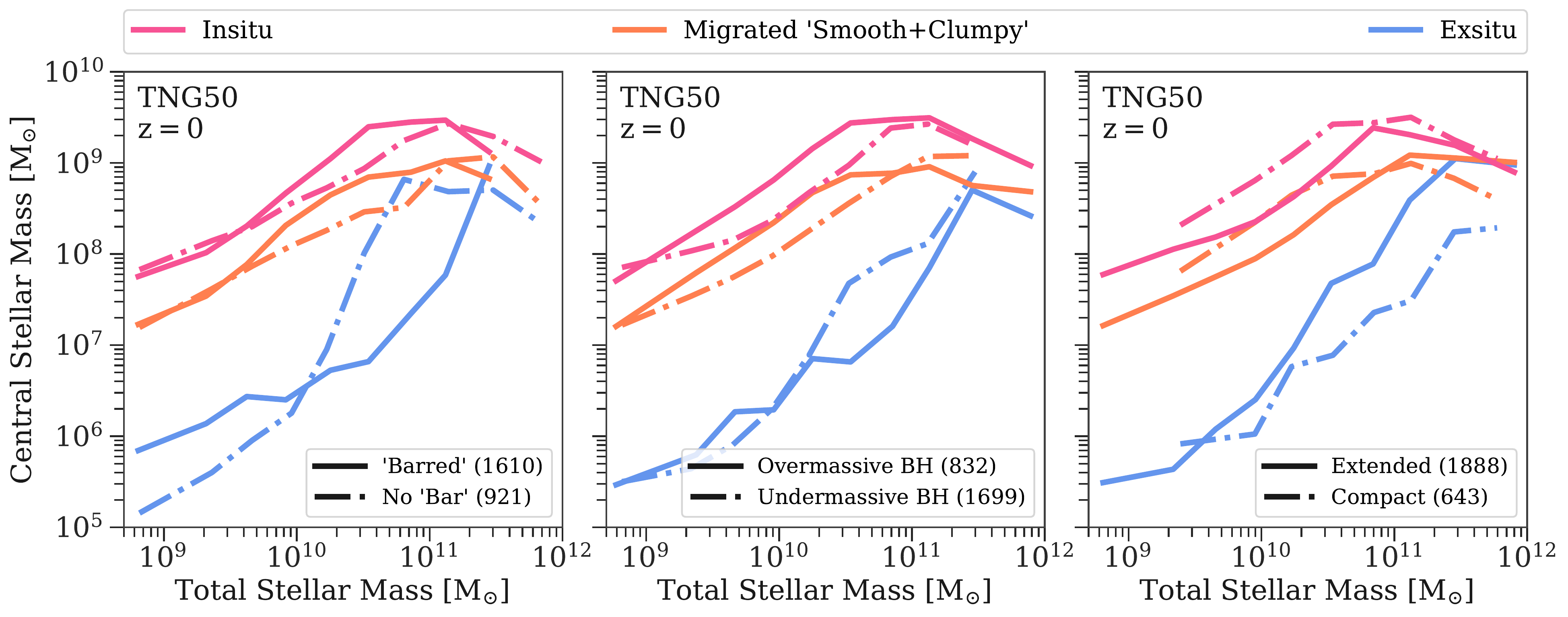}
    \caption{\textbf{Differences in the central (500\,pc) in-situ, migrated and ex-situ populations for different galaxy properties of TNG50 galaxies at} $\mathbf{z=0.}$ Each panel shows median trends of central stellar mass divided in the three origins: in-situ (\textit{pink}), migrated (\textit{orange}) and ex-situ (\textit{blue}) as a function of the galaxies' total stellar mass. The dashed and solid lines split the TNG50 galaxy population according to different properties, which are (\textit{form left to right}): central vs. satellite, quenched vs. star forming, bulgey vs. disky, `barred' vs. `non barred', overmassive vs. undermassive black holes (BH) and extended vs. compact galaxies. The bracketed numbers show the total amount of galaxies in each category.}
    \label{fig: props}
\end{figure*}

\subsubsection{Trends for different galaxy types}\label{sec: types}

Galaxies with different present-day properties are thought to have undergone different formation pathways. Is this reflected in different contributions of in-situ, migrated and ex-situ stars building up the center of these galaxies? \par 
In Figure \ref{fig: props} we show the running median of the central stellar mass of the three origins as a function of total galaxy stellar mass split into six different galaxy properties. The definitions of the different galaxy properties are summarized in Table \ref{tab: proptab} and described in detail in Appendix \ref{sec: galprops}. \par
All in all, the most significant differences are seen in the central ex-situ population across various galaxy properties. This significance manifests in separation between the median relations including the scatter (which we do not show, however, in favour of clarity). Small differences for the in-situ and migrated populations are not significant with respect to the scatter around the median relations. \\ \par

\textit{\textbf{Centrals vs. Satellites (Figure \ref{fig: props}, top left panel)}}: On average, centrals and satellites contain the same amount of central in-situ, migrated and ex-situ mass, showing that their central 500\,pc is unaffected by their environment at $\mathrm{z}=0$. This is sensible considering that galaxy centers likely assemble before the galaxy becomes a satellites. Additionally, most environmental effects should first take effect in the outskirts of galaxies. Similarly, we find no significant difference in the central mass of the three populations, when the galaxies are divided by the mass of their host halo. \\ \par

\textit{\textbf{Quenched vs. Star Forming (Figure \ref{fig: props}, top middle panel)}}: Quenched galaxies between $5\times10^{9}$ and $5\times10^{10}\,\mathrm{M}_{\odot}$ have slightly higher central in-situ and migrated mass than for star forming ones. This difference primarily arises because the star forming galaxies tend to have lower central densities on average than quenched ones. \par
A larger difference is seen in the ex-situ. For galaxy stellar masses above $10^{10}\,\mathrm{M}_{\odot}$ the average ex-situ mass starts to rise more rapidly for star forming galaxies than for quenched ones. For galaxies around $5\times10^{10}\,\mathrm{M}_{\odot}$ this difference becomes largest, with the median central ex-situ mass of star forming galaxies being higher by more than one dex. \par
While this trend may seemingly be counter-intuitive for the current consensus of galaxy evolution, the difference is also true when considering the \emph{total} ex-situ stellar mass in TNG50 (and also TNG100) as seen in Figure \ref{fig: tot_exsitu} in Appendix \ref{appendix: validate3}. This could be an indication that today's star forming galaxies had more or larger mass ratio mergers with galaxies with high gas content at later cosmic times (see Section \ref{sec: buildup} for a further discussion). We obtain a consistent picture when galaxies are divided according to their $g-i$ colour or total gas mass at $\mathrm{z}=0$. \\ \par

\textit{\textbf{Bulgey vs. Disky (Figure \ref{fig: props}, top right panel)}}: The in-situ and migrated central mass for disky and bulgey galaxies show a similar trend as for the star forming and quenched population. However, the trend for the central ex-situ mass is distinct. Bulgey galaxies below $10^{10}\,\mathrm{M}_{\odot}$ in stellar mass have higher ex-situ masses (by roughly half a dex) in their centers than their disky counterparts. This difference disappears for galaxy stellar masses above $10^{10}\,\mathrm{M}_{\odot}$. \par
We have checked the median relation for the \emph{total} ex-situ mass and find that bulgey galaxies have a constant higher offset of about 0.25 dex compared to disk galaxies across the whole galaxy mass range. Hence, disky galaxies below $10^{10}\,\mathrm{M}_{\odot}$ have not only lower absolute central and total ex-situ masses, but also a lower central-to-total ex-situ fraction of about 0.4\% as compared to 1\% for bulge dominated galaxies. Thus the relative amount of ex-situ mass that is deposited in the center might be an important driver for morphological transformation in these lower galaxy mass regimes. \par
For galaxies above $10^{10}\,\mathrm{M}_{\odot}$ in stellar mass, the central-to-total ex-situ fraction decreases strongly as a function of galaxy stellar mass, with disk galaxies having consequently slightly higher values. This could be an indication that once a massive rational support exists in the stellar component it is hard to destroy it through mergers. Similar relations are found when adopting other definitions for disky and bulgy galaxies, such as the ratio of the kinetic energy in ordered motion compared to the total kinetic energy \citep[see][]{rodriguesgomez17}. \\ \par

\textit{\textbf{Barred vs. No Bar (Figure \ref{fig: props}, bottom left panel)}}: For galaxies below $10^{10}\,\mathrm{M}_{\odot}$ TNG50 predicts no difference in the central in-situ and migrated mass, however the galaxies with bar-like features have higher ex-situ masses than galaxies with no bar-like features. This trend is similar for the bulgey vs. disky galaxies. We have explicitly checked that indeed high ex-situ masses in the center of galaxies within this mass regime mainly occur in bulgey and barred galaxies, whereas bulgey and unbarred galaxies as well as disky galaxies, both barred and unbarred, have lower central ex-situ masses by approximately one dex. \par
For galaxies above $10^{10}\,\mathrm{M}_{\odot}$ in total stellar mass, this relation for the central ex-situ mass swaps. In this regime unbarred galaxies have higher ex-situ masses in the center regardless whether they are disky or bulgey. We find that the same statements for barred and unbarred galaxies across the entire mass range are true when correlating the \emph{total} ex-situ mass of galaxies. \par
Lastly, the in-situ and migrated mass in the center is higher for barred galaxies between $10^{10}\,\mathrm{M}_{\odot}$ and $10^{11}\,\mathrm{M}_{\odot}$. Hence, barred galaxies in this mass regime have higher central densities than unbarred galaxies in TNG50, which is consistent with observations \citep[see][]{barconcen}. \\ \par

\textit{\textbf{Over- vs. Undermassive Black Holes (Figure \ref{fig: props}, bottom middle panel)}}: We could expect that AGN feedback has an influence on the stellar mass growth in the center of galaxies. We therefore split our TNG50 sample according to whether the galaxies have an over- or undermassive black hole at $\mathrm{z}=0$. Identical relations are found when the galaxy population is split according to the cumulative energy injection of each feedback mode or both. \par 
On average, galaxies between $10^{10}\,\mathrm{M}_{\odot}$ and $10^{11}\,\mathrm{M}_{\odot}$ in stellar mass with an undermassive black hole have a higher central ex-situ mass by about one dex than galaxies with an overmassive black hole at the same stellar masses. For galaxies with total stellar masses in the range of $5\times10^{9}-5\times10^{10}\,\mathrm{M}_{\odot}$, the ones with an overmassive black hole have in-situ and migrated masses in the center that are about half a dex higher than for galaxies with an undermassive black hole. Consequently, galaxies with overmassive black holes in this mass regime have higher central densities. \par
We find that mainly all of these differences in the in-situ, migrated and ex-situ mass for galaxies with over- and undermassive black holes emerge because galaxies at fixed stellar mass with overmassive black holes tend to be more compact in TNG50 and vice versa. Therefore, a similar behaviour of the central stellar mass in the three populations with total galaxy stellar mass is found when the galaxy population is split into compact and extended galaxies (see below). This connection between black hole masses, central densities and sizes of galaxies at fixed galaxy stellar mass is also found in observations \citep[][]{chen20}. \\ \par

\begin{figure*}
    \centering
    \includegraphics[width=\textwidth]{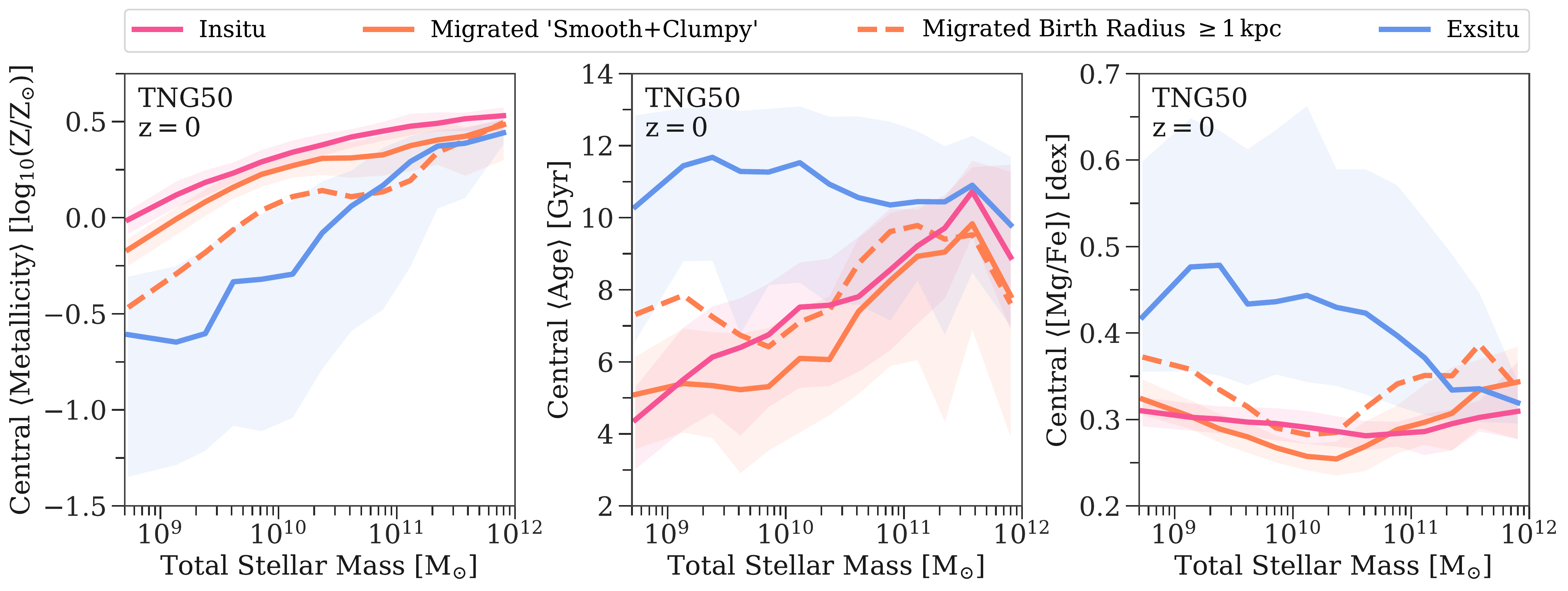}
    \caption{\textbf{Average stellar population properties of in-situ, migrated and ex-situ stars in the central 500\,pc of TNG50 galaxies at} $\mathbf{z=0}.$ \textit{From left to right}: The central mass-weighted average metallicity, age and magnesium-to-iron abundance [Mg/Fe] as a function of total galaxy stellar mass for the in-situ (\textit{pink}), migrated (\textit{orange}) and ex-situ (\textit{blue}) stars. The shaded bands depict the 16th and 84th percentile. The dashed orange line only shows the quantities for migrated stars that have birth radii larger than 1\,kpc. Overall, the stellar population properties of in-situ and migrated stars are similar, whereas the ex-situ stars are more metal-poor, older and have higher [Mg/Fe] across the galaxy mass range.}
    \label{fig: stellpop}
\end{figure*}

\textit{\textbf{Extended vs. Compact (Figure \ref{fig: props}, bottom right panel)}}: Extended galaxies tend to have on average more ex-situ mass in the center than compact galaxies at the same total stellar mass. The difference is around one dex for galaxies above $10^{10}\,\mathrm{M}_{\odot}$ in stellar mass. \par 
When we correlate with the \emph{total} ex-situ mass, we find an opposite behaviour in TNG50. Galaxies $\lesssim5\times 10^{10}\,\mathrm{M}_{\odot}$ and with higher total ex-situ fractions are on average more extended (see Figure \ref{fig: exsitu_rad} in Appendix \ref{appendix: validate2}). \par
Compact galaxies between $5\times10^{9}\,\mathrm{M}_{\odot}$ and $5\times10^{10}\,\mathrm{M}_{\odot}$ have more in-situ and migrated mass in the center, and therefore higher central densities (and black hole masses, see above). As a matter of fact, this difference is also seen for quenched vs. star forming and bulgey vs. disky galaxies, even though to a lesser extent. This stems from the fact that generally star forming galaxies tend to be more disky and hence more extended and vice versa. \\ \par

\subsection{Stellar population and dynamical properties}\label{sec: pop}

Are there distinguishable features in the stellar population and dynamical properties of the in-situ, migrated and ex-situ stars? The short answer is yes, especially for galaxies below $\lesssim10^{11}\,\mathrm{M}_{\odot}$, where the majority of ex-situ stars originate from lower mass satellites. 

\subsubsection{Average age, metallicity and [Mg/Fe] of central stars}

Metallicities, ages and magnesium-to-iron abundances [Mg/Fe] of stars encode information about their birth places. Figure \ref{fig: stellpop} illustrates average quantities of stellar populations belonging to the in-situ, migrated and ex-situ origin as a function of their galaxy's stellar mass. We also show separate relations for migrated stars that have birth radii larger than 1\,kpc to exclude the majority of migrated stars that were born close to the center, which dominate the average stellar population properties (see smoothly migrated stars in Figure \ref{fig: birthloc}). \\ \par

\textit{\textbf{Metallicity (Figure \ref{fig: stellpop}, left panel)}}: Stars in the central 500\,pc follow a mass-metallicity relation, where galaxies at the lowest mass end ($5\times10^8\,\mathrm{M}_{\odot}$) have on average solar metallicity and galaxies at the highest mass end ($\sim10^{12}\,\mathrm{M}_{\odot}$) have metallicities of around 0.5\,dex. The total mass-metallicity relation of all the stars in the center is very close to the one for the in-situ population only, as they dominate the mass in the center of galaxies on average (see Figure \ref{fig: mass}). \par
Furthermore, the average metallicity for central stars is consistently offset by about 0.3 dex towards higher metallicities across the whole galaxy mass range compared to the mass-metallicity relation which takes into account all the stars belonging to a given galaxy. This emphasizes the self-similarity of galactic chemical enrichment. \par
On top of that, the total central mass-metallicity relation is tight having a scatter of around 0.1\,dex, which also holds when only in-situ or only migrated stars are considered. Hence, there is little galaxy-to-galaxy variation at fixed stellar mass regarding in-situ star formation. \par
The average metallicity of the in-situ population is the highest, followed by the migrated stars, which is less than a quarter dex lower across the whole galaxy mass range. This small difference is expected as most of the migrated stars are born very close to the center ($0.5-1\,\mathrm{kpc}$). When including only migrated stars with large birth radii ($>1\,\mathrm{kpc}$), the difference becomes larger to about half a dex due to internal metallicity gradients present in galaxies, which is in turn caused by less efficient star formation in the galactic outskirts. Above galaxy stellar masses of $2\times10^{11}\,\mathrm{M}_{\odot}$ the average metallicity of all migrated stars and only those with birth radii larger than 1\,kpc becomes similar again. This is because migrated stars from clumps are dominating at these galaxy masses, which originate from larger distances (median distance is $30\,\mathrm{kpc}$) and have high metallicities (median metallicity is 0.2\,dex). \par
The mass-metallicity relation for the central ex-situ stars follows a steeper slope than the one for the in-situ and migrated stars, because we are showing the mass of the $\mathrm{z}=0$ host galaxy and not of the galaxy they were born in. The average metallicity of ex-situ stars is around 0.5 dex lower at the lowest galaxy masses compared to the metallicity for the in-situ stars. At the highest mass end the average metallicity of the ex-situ stars becomes close to the one for the migrated stars, which is around 0.25\,dex. This steeper slope emphasizes that ex-situ stars in the center of low mass galaxies originate from galaxies of even lower mass, while most of the central ex-situ stars in high mass galaxies originate from galaxies of more similar stellar mass. \par
Lastly, the galaxy-to-galaxy variation at fixed galaxy stellar mass for the average metallicity of ex-situ stars is much larger compared to the in-situ and migrated population. The scatter varies from around one dex at the low mass galaxy end to close to a quarter dex for the highest galaxy masses. This emphasizes that at lower host galaxy stellar mass, a larger variety of satellite galaxies (i.e. with different stellar masses) can deposit stars in the center of their respective $\mathrm{z}=0$ hosts. \\ \par

\textit{\textbf{Age (Figure \ref{fig: stellpop}, middle panel)}}: The ex-situ stars have a rather constant, old age of around 10\,Gyr across the whole galaxy mass range, albeit with a large scatter of around 2\,Gyr. This is not surprising as most mergers happen before the redshift of one, which corresponds to a lookback time of around 8\,Gyr. The flat relation for the average age of the ex-situ stars is not in conflict with their corresponding mass-metallicity relation. Because high mass galaxies are more efficient in chemical enrichment than low mass galaxies, they will consequently have higher metallicities at fixed stellar age. \par
The median relations for the average age for the in-situ and migrated stars are again similar to each other with the in-situ stars being slightly older by around 1\,Gyr or less at fixed galaxy stellar mass. Overall, in-situ and migrated stars are younger, with average ages between 3 and 6\,Gyr, in the lowest mass galaxies ($\sim10^9\,\mathrm{M}_{\odot}$), and become increasingly older with average ages of around $8-10\,\mathrm{Gyr}$ at the highest mass end ($\sim10^{12}\,\mathrm{M}_{\odot}$). \par
The scatter of the average ages for the in-situ and migrated stars is much larger than their corresponding variations in metallicity. This could have multiple reasons, for example: different pathways in star formation histories (i.e. star formation rate as a function of time) can result in the same metallicity but different average ages, or the metallicity enrichment starts to saturate once a metallicity above solar is reached and therefore it does not matter, if star formation continues for another few Gyr. \par
Galaxies above $10^{11}\,\mathrm{M}_{\odot}$ in stellar mass exhibit a larger scatter of the average age of their migrated population compared to their in-situ stars. This arises because migration to the center in this regime is dominated by clumps, which have a rather flat formation time distribution with the majority forming between 4 and 10\,Gyr ago (see Figure \ref{fig: clumps_sum}). \par
Below galaxy stellar mass of $10^{11}\,\mathrm{M}_{\odot}$, the migrated stars, which were born at distances larger than 1\,kpc, have a running median of averages ages that are around $1-2\,\mathrm{Gyr}$ older than the total migrated population. As these stars need significantly more time to arrive in the center, their ages are consequently older. \\ \par

\textit{\textbf{[Mg/Fe] (Figure \ref{fig: stellpop}, right panel)}}: In extragalactic studies magnesium is the predominant $\alpha$-element present in optical spectra \citep[see e.g.][]{nacho18b,nacho19,nacho21,gallazzi20}. We therefore show the running median of the mass-weighted average magnesium-to-iron abundance as a function of galaxy stellar mass as a proxy for the total $\alpha$-to-iron abundance. This abundance ratio provides to first hand information about the star formation time scale before supernovae type Ia significantly enrich the interstellar medium with iron peak elements\footnote{Influences on [$\alpha$/Fe] due to IMF (initial mass function) changes are not captured in the simulation, as a Chabrier IMF \citep{chabrier03} is assumed for every stellar particle}. \par
The average central [Mg/Fe] is almost constant for the in-situ population across the whole galaxy mass range with a value of about 0.3\,dex. For galaxies between $2\times10^9\,\mathrm{M}_{\odot}$ and $6\times10^{10}\,\mathrm{M}_{\odot}$, the migrated stars have slightly lower values. The lowest average [Mg/Fe] of around 0.25\,dex is reached for galaxies around $2\times10^{10}\,\mathrm{M}_{\odot}$. This directly maps to the increased difference of the average age between in-situ and migrated stars of around 1\,Gyr in the same mass regime. Hence, in-situ stars of these galaxies form on average earlier and more rapidly as opposed to their migrated stars. \par
Above $6\times10^{10}\,\mathrm{M}_{\odot}$, the average [Mg/Fe] for the migrated populations rises above the one for the in-situ stars to around 0.35\,dex at the highest mass end. This cross-over is not seen in the average ages. An explanation for this could be that in the high galaxy mass regime, an increasing number of migrated stars can originate from larger distances and possibly formed from stripped gas of merging lower mass systems (see Section \ref{sec: birth}), which have larger [Mg/Fe] values due to lesser efficiency in chemical enrichment. \par
When only including migrated stars originating from distances farther than 1\,kpc away from the center, the average [Mg/Fe] becomes larger by around 0.1\,dex across all galaxy stellar masses. For galaxies below $\sim10^{11}\,\mathrm{M}_{\odot}$ the corresponding ages become older, which is thus consistent in having formed from true in-situ gas of their respective host galaxies. \par
The age for migrated stars with birth radii $>1\,\mathrm{kpc}$ in galaxies above $10^{10}\,\mathrm{M}_{\odot}$ does not increase even though their [Mg/Fe] increase as well. This could indeed provide evidence for some migrated stars having formed from stripped gas for galaxies in this mass regime. \par
The ex-situ stars have overall higher average [Mg/Fe] values of around 0.45\,dex, which decreases to around 0.35\,dex for galaxies above $10^{11}\,\mathrm{M}_{\odot}$ in stellar mass. This is consistent with their old ages and being formed in lower mass satellite galaxies that produced stars less efficiently than their respective $\mathrm{z}=0$ hosts. \par
The scatter in average [Mg/Fe] for the ex-situ population is significantly larger than for the in-situ and migrated population across all galaxy masses, but especially $\lesssim 10^{11}\,\mathrm{M}_{\odot}$. The onset of type Ia supernovae creates probably more stochasticity in lower mass galaxies as single supernovae events can significantly enrich the interstellar medium of the entire host galaxy. \par

\begin{figure*}
    \centering
    \includegraphics[width=\textwidth]{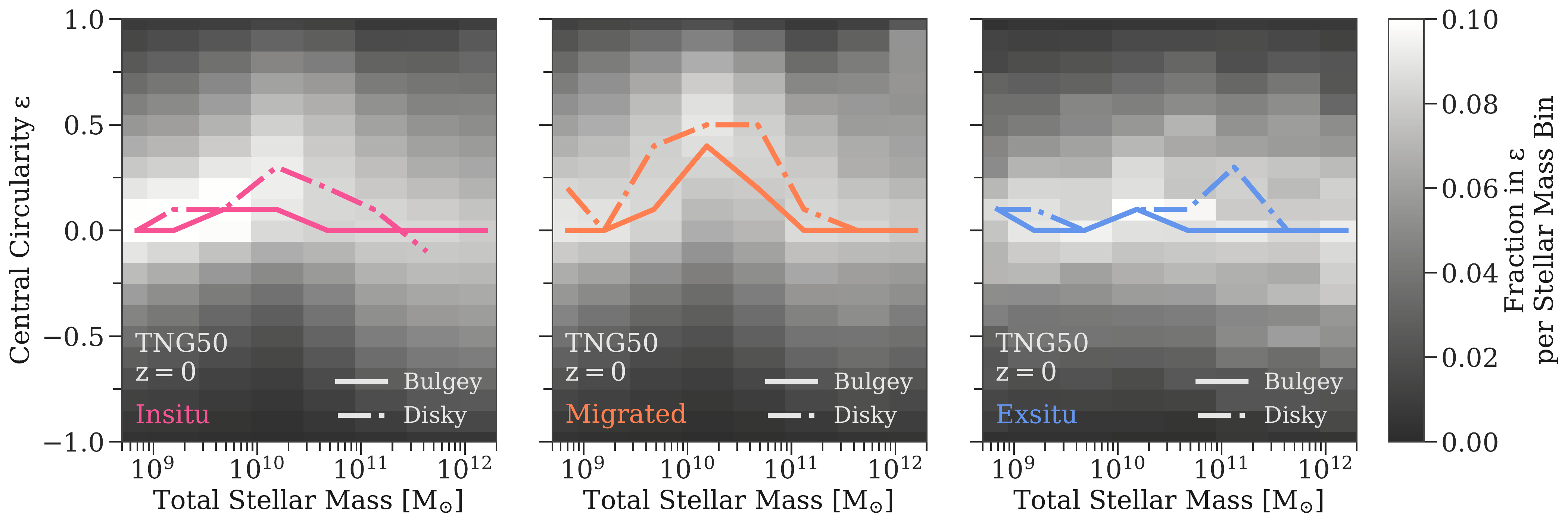}
    \caption{\textbf{Average differences in the dynamical properties of in-situ, migrated and ex-situ stars in the central 500\,pc of TNG50 galaxies at} $\mathbf{z=0}.$ \textit{From left to right}: The central circularity distributions in stacks of total galaxy stellar mass for in-situ (\textit{pink}), migrated (\textit{orange}) and ex-situ (\textit{blue}) stars. Per galaxy, stellar particles in the center belonging to either the in-situ, migrated or ex-situ population are binned according to their circularity $\epsilon$ and normalized to unity respectively. They are then stacked together according to the displayed galaxy stellar mass bins and then normalized again. The lines trace the circularity bin with the maximum fractional mass across galaxy stellar masses divided by bulgey (\textit{solid}) and disky (\textit{dashed-dotted}) galaxies respectively. Clearly the migrated population has the most rotational support for galaxies around $10^{10}\,\mathrm{M}_{\odot}$ in total stellar mass, regardless of the host being disk or bulge dominated.}
    \label{fig: circ}
\end{figure*}

\subsubsection{Stacked circularity distributions}

Different birth origins also leave imprints on the stars' dynamics, which can still be visible until the present-day. We investigate such imprints by quantifying the instantaneous circularity $\epsilon$ of stars \citep[see][]{zhu21}. Circularities close to one indicate circular orbits, values around zero indicate random motion dominated orbits and negative ones show counter-rotating orbits. We then compute the normalized circularity distribution for each galaxy with a bin size of 0.1 for $\epsilon$. Circularity distributions are than stacked together according to the galaxy's total stellar mass in bins of approximate 0.5\,dex and are re-normalized. \par
The results for the different in-situ, migrated and ex-situ populations are displayed in Figure \ref{fig: circ}. The lines in Figure \ref{fig: circ} trace the peak of the circularity distributions across galaxy stellar mass, separately for disky and bulgey galaxies. \par
The circularity distribution of the in-situ population is centered on random motion dominated orbits for galaxies with stellar masses smaller than $3\times10^{9}\,\mathrm{M}_{\odot}$ and larger than $10^{11}\,\mathrm{M}_{\odot}$. Galaxies with stellar masses in between have a circularity distribution with a peak shifted towards slightly higher circularities of around 0.25. We see that this shift is caused by galaxies that are overall disky, as the bulge dominated galaxies have a circularity peak that stays around zero. Nevertheless, the in-situ stars are in summary on warm to hot orbits even for disk dominated galaxies, which is not surprising as the velocity dispersion generally rises towards the center of galaxies. \par
For galaxies below $10^{10}\,\mathrm{M}_{\odot}$ the stacked circularity distributions for in-situ stars have a sharper peak, whereas galaxies of higher masses have an overall broader distribution in $\epsilon$. This could be an indication of a smaller galaxy-to-galaxy variation of the circularity distribution in the center of the smallest galaxies, regardless of whether they are disky or bulgey, as in this mass regime the absolute numbers of those two galaxy types are approximately the same in TNG50. At the high mass end on the other hand, a broader circularity distribution could indicate that in-situ stars become redistributed in their orbits due to the increased influence of mergers and contribution of ex-situ stars. \par
For the migrated population the circularity distribution is again centered on random motion orbits for galaxies $<3\times10^{9}\,\mathrm{M}_{\odot}$ and $>10^{11}\,\mathrm{M}_{\odot}$, although now the distribution is also overall broader for the low mass galaxies. Migrated stars in intermediate mass galaxies are on even higher circularity orbits than their corresponding in-situ stars reaching a peak of around 0.5 for galaxy stellar masses of $\sim10^{10}\,\mathrm{M}_{\odot}$. This peak is seen in disk and bulge dominated galaxies alike. Hence, migrated stars tend to have the most rotational support for galaxies in the intermediate mass regime, which could be an indication for migration being caused by different mechanisms across the galaxy mass range in TNG50. However, migrated stars are also on average younger than the in-situ stars in these galaxies, which might be the reason why they are still more on circular orbits (see Figure \ref{fig: stellpop}). We also point out that some galaxies above $\sim3\times10^{10}\,\mathrm{M}_{\odot}$ have a very double peaked (i.e. one around zero and around 0.5 or higher) circularity distribution, which is washed out in Figure \ref{fig: circ} due to the stacking. These stars originate from (recently) migrated clumps. \par
Ex-situ stars have circularities centered around zero across the entire galaxy mass range in TNG50 and also for both disk and bulge dominated galaxies. Because they originate from stochastic merger events, stars are put on average on hot, random motion dominated orbits. Nevertheless, we see a a large scatter throughout the circularity distributions for the ex-situ stars in the different galaxy stellar mass bins indicating a lot of individual galaxy-to-galaxy variation. Depending on the exact time the merger occurred and how the orbits between the host and merging satellite were configured, ex-situ stars can very well retain some rotational support and often be on counter rotating orbits. \par

\begin{figure*}
    \centering
    \includegraphics[width=\textwidth]{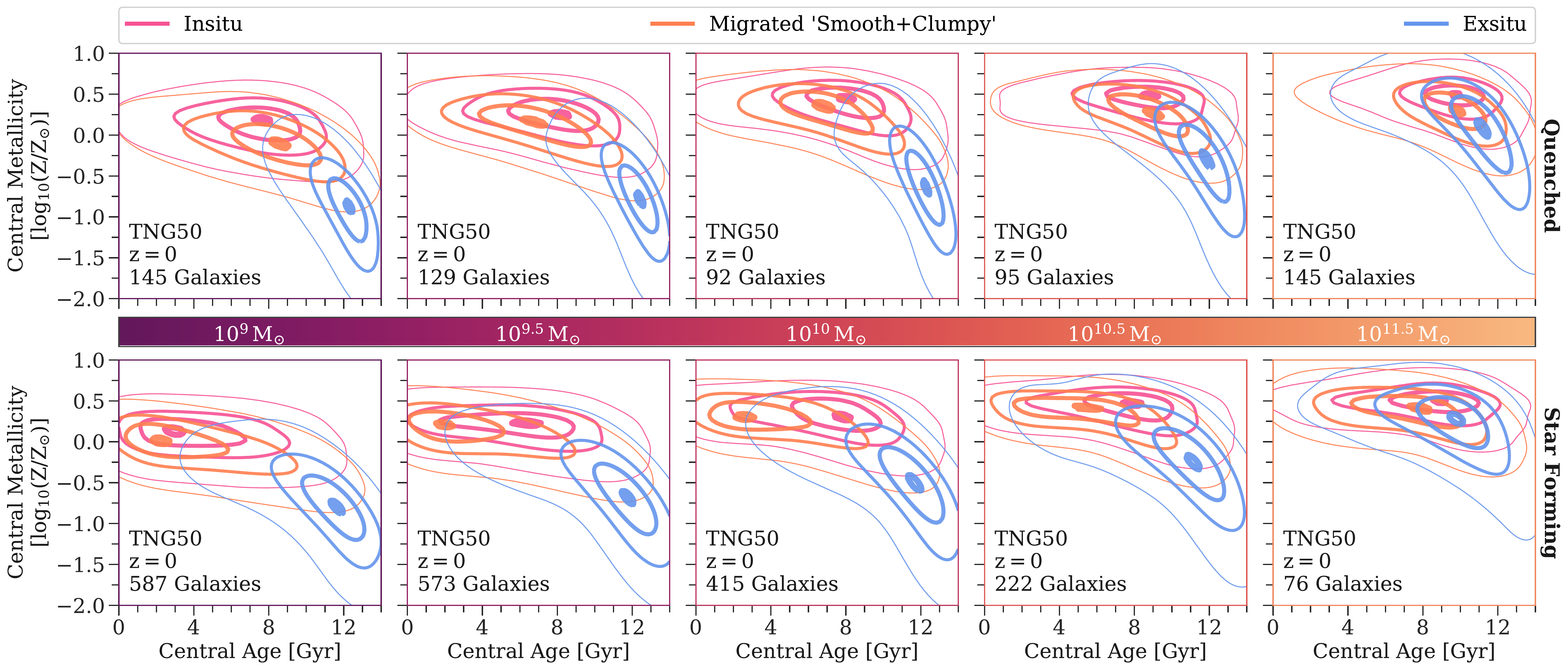}
    \caption{\textbf{Age-metallicity distributions of central (500\,pc) stars of TNG50 galaxies in stacks of stellar mass at} $\mathbf{z=0}.$ Gaussian kernel density estimates for in-situ \textit{(pink)}, migrated \textit{(orange)} and ex-situ \textit{(blue)} stars encompassing 1\%, 20\%, 50\% and 90\% of all central stellar mass \textit{(thickest to thinnest)} are shown in the respective galaxy stellar mass bin increasing from left to right as depicted by the colorbar. The galaxy mass bins are centered on the indicated stellar mass and are 0.5 dex wide, except for the last panel, which is approximately one dex wide. Prior to stacking the age-metallicity distribution of each galaxy is normalized. The top row shows quenched and bottom row shows star forming galaxies respectively. In each panel the number of galaxies in the corresponding stellar mass bin are indicated. The galaxy-averaged age-metallicity distribution of the three origins becomes best separated around galaxies with stellar masses of $10^{10}\,\mathrm{M}_{\odot}$.}
    \label{fig: agemet}
\end{figure*}

\subsubsection{2D distributions of ages, metallicity and circularities}

\begin{figure*}
    \centering
    \includegraphics[width=\textwidth]{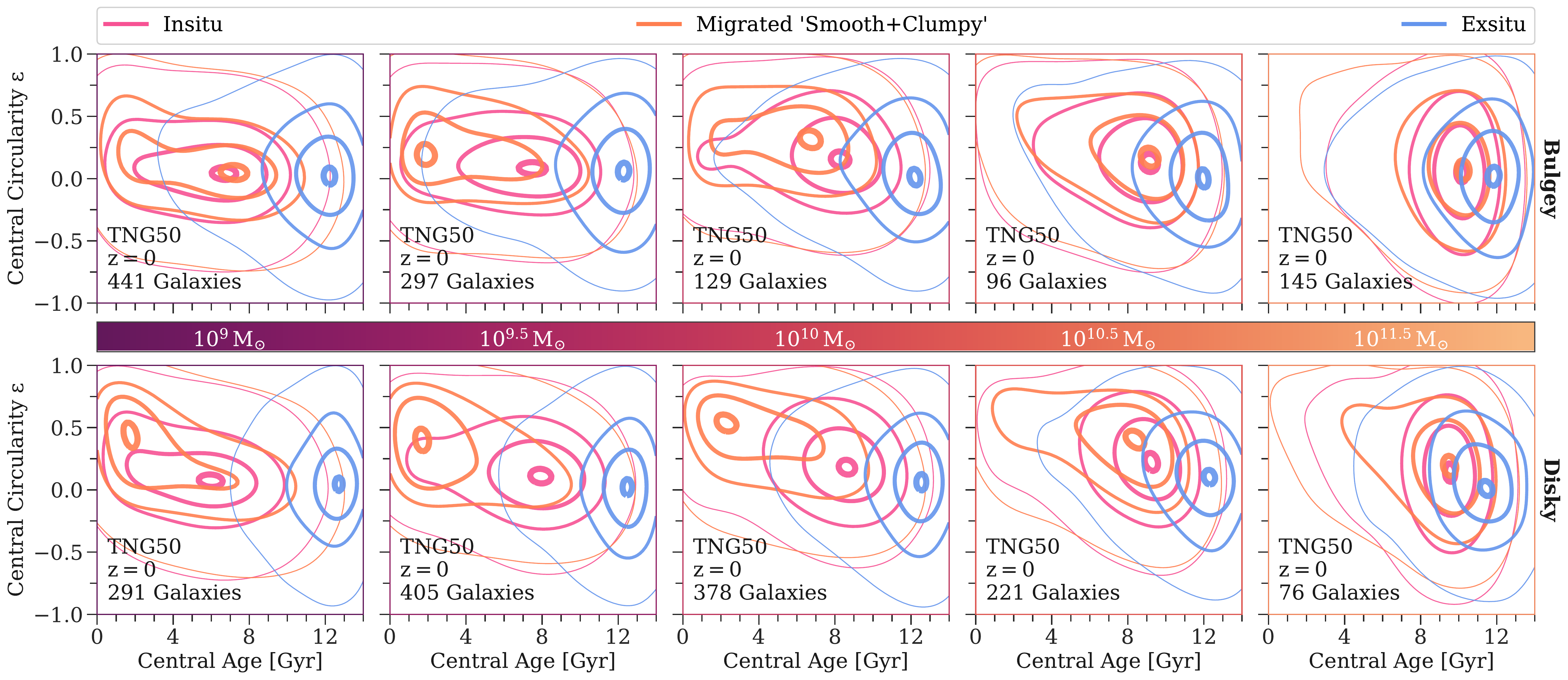}
    \caption{\textbf{Age-circularity distributions of central (500\,pc) stars of TNG50 galaxies in stacks of stellar mass at} $\mathbf{z=0}.$ Gaussian kernel density estimates for in-situ \textit{(pink)}, migrated \textit{(orange)} and ex-situ \textit{(blue)} stars encompassing 1\%, 20\%, 50\% and 90\% of all stellar mass \textit{(thickest to thinnest)} are shown in the respective galaxy stellar mass bin increasing from left to right as depicted by the colorbar. The galaxy mass bins are centered on the indicated stellar mass and are 0.5 dex wide, except for the last panel, which is approximately one dex wide. Prior to stacking the age-circularity distribution of each galaxy is normalized. The top row shows bulge dominated and bottom row shows disk dominated galaxies respectively. In each panel the number of galaxies in the corresponding stellar mass bin are indicated. The galaxy-averaged age-circularity distribution of the three origins becomes best separated around galaxies with stellar masses of $10^{10}\,\mathrm{M}_{\odot}$.}
    \label{fig: agecirc}
\end{figure*}

In Figures \ref{fig: agemet} and \ref{fig: agecirc} we show the 2D distributions of age and metallicity and age and circularity of the central in-situ, migrated and ex-situ stars respectively in stacks of galaxy stellar mass. For each galaxy we first compute the mass-weighted and normalized 2D histogram of the respective quantities with bin sizes of 0.5 Gyr for age, 0.25 dex for metallicity and 0.1 for circularity. We then stack those according to the total stellar mass bin of the galaxies, normalize again and then compute the Gaussian kernel density estimate. The galaxy stellar mass bins are 0.5 dex wide for galaxies between $10^{8.75}\,\mathrm{M}_{\odot}$ and $10^{10.75}\,\mathrm{M}_{\odot}$. We stack all galaxies with stellar masses between $10^{10.75}\,\mathrm{M}_{\odot}$ and $10^{12}\,\mathrm{M}_{\odot}$ together as a finer binning did not reveal any mass-dependent trends and also became stochastic due to low number statistics in this mass regime. The five galaxies with stellar masses above $10^{12}\,\mathrm{M}_{\odot}$ are not included. Additionally, we show the stacked age-metallicity distributions for quenched and star forming galaxies separately to avoid averaging over too many dissimilar galaxies in this parameter space. Similarly, we divide between bulge and disk dominated galaxies for the age-circularity distributions. \par
With the 2D distributions we can observe a couple of new trends that are not necessarily apparent from the average stellar population properties in Figure \ref{fig: stellpop} and the 1D circularity distributions of Figure \ref{fig: circ}. \\ \par

\textit{\textbf{Age-metallicity (Figure \ref{fig: agemet})}}: For star forming galaxies (bottom row of Figure \ref{fig: agemet}), the average distribution of the migrated stars changes very little in shape and position, apart from shifting towards higher metallicities, from the lowest mass galaxies until the $10^{10.5}\,\mathrm{M}_{\odot}$ galaxy stellar mass bin. They are centered between $2-4\,\mathrm{Gyr}$. The ex-situ stars behave similarly and are centered around 12\,Gyr. However, the average distribution of the in-situ stars shows an entirely different mass trend. While the in-situ stars are almost entirely coinciding with the migrated stars in age-metallicity space for the lowest mass bin, the peak of the in-situ distribution gradually shifts towards older ages from 2\,Gyr to 8\,Gyr for increasing galaxy mass. In the process the in-situ average age-metallicity distribution becomes more elongated around $10^{9.5}\,\mathrm{M}_{\odot}$ in the age direction when focusing on the 20\% contour. \par
For $10^{10}\,\mathrm{M}_{\odot}$ galaxies the in-situ distribution becomes more centrally concentrated again. Furthermore, the average age-metallicity distributions of the three origins are maximally separated in this mass regime, with the migrated stars being the youngest ($1-6\,\mathrm{Gyr}$ for the 20\% contour), followed by the in-situ stars ($6-10\,\mathrm{Gyr}$ for the 20\% contour) at similar metallicity and the ex-situ stars populating the oldest ($10-13\,\mathrm{Gyr}$ for the 20\% contour) and most metal-poor tail. Thus, there must be a mechanism for these galaxies that halts in-situ star formation in their centers, while it continues outside of it in order to be able to produce young migrated stars. It is likely that this is connected to the (kinetic) AGN feedback implement in TNG, which quenches galaxies from inside-out \citep[][see also \ref{sec: mecha} and Figure \ref{fig: assembly}]{nelson21}. \par
Starting at $10^{10.5}\,\mathrm{M}_{\odot}$ the average age-metallicity distribution of the migrated stars also becomes more elongated towards older ages and above $10^{10.5}\,\mathrm{M}_{\odot}$ coincides again with the one of the in-situ stars. The peak of the ex-situ distribution increases towards metallicities similar to those of the in-situ and migrated stars. For galaxies between $10^{11}\,\mathrm{M}_{\odot}$ and $10^{12}\,\mathrm{M}_{\odot}$ in stellar mass the average distributions for the in-situ, migrated and ex-situ stars become almost indistinguishable in age-metallicity space. \par
For quenched galaxies (top row of Figure \ref{fig: agemet}) the behaviour for the age-metallicity distributions of the in-situ and migrated stars across galaxy stellar mass is different. They are not clearly separated in any galaxy stellar mass bin as was the case for the star forming galaxies. Both the in-situ and migrated average age-metallicity distributions are more centrally concentrated than for star forming galaxies, their shapes are very similar to each other and their peaks are both at old ages (around 8\,Gyr) exhibiting little galaxy mass dependence. The peak of the age-metallicity distribution for the migrated stars seem to be slightly younger for galaxies in mass bins between $10^{9.5}\,\mathrm{M}_{\odot}$ and $10^{10}\,\mathrm{M}_{\odot}$ and slightly older otherwise. Interestingly at both the low and high mass end, the separation in metallicity between the in-situ and migrated stars is larger for the quenched galaxies as for the star forming ones. For galaxies between $10^{11}\,\mathrm{M}_{\odot}$ and $10^{12}\,\mathrm{M}_{\odot}$ the three distributions are again indistinguishable. \par

\textit{\textbf{Age-circularity  (Figure \ref{fig: agecirc})}}: Stars with higher circularities are usually younger. The distribution for migrated stars of disky galaxies (bottom row of Figure \ref{fig: agecirc}) in the lowest galaxy stellar peaks at around 2\,Gyr with high circularity values of around 0.5, whereas the distributions for the in-situ stars peaks at older ages (6\,Gyr) centered on circularity values of zero. Nevertheless, the 20\% contour for the in-situ stars still has a tails towards younger ages and slightly above zero circularities. In the next higher galaxy stellar mass bin the 20\% contour of the distribution for the migrated stars looses its tail of older ages ($4-8\,\mathrm{Gyr}$) and zero circularities. The 20\% contour for the in-situ stars in now centered on even older ages (8\,Gyr). Beginning around galaxy stellar masses of $10^{10}\,\mathrm{M}_{\odot}$ the 20\% contour for the migrated stars elongates towards older ages spanning now $1-8\,\mathrm{Gyr}$, while roughly maintaining the high circularity. The distribution for the in-situ stars becomes broader and slightly shifts towards above zero circularities. In the $10^{10.5}\,\mathrm{M}_{\odot}$ galaxy stellar mass bin the peak of the migrated stars shifts from young (2\,Gyr) to old (8\,Gyr) ages with just a slight decrease in circularity. Above galaxies with $10^{10.5}\,\mathrm{M}_{\odot}$ in stellar mass the migrated stars switch from a rotationally supported distribution to random motion dominated one until they coincide with the age-circularity distributions of the in-situ and ex-situ stars at the highest galaxies. \par
The peak of the age-circularity distribution for the in-situ stars, albeit having the same young age as for the migrated stars in the lowest stellar mass bin, is near zero circularity. With increasing galaxy mass the age-circularity distribution for in-situ stars shifts towards old ages and becomes broader in the circularity direction, but stays mostly centered around zero circularity with perhaps a slight shift towards higher circularities around the $10^{10}\,\mathrm{M}_{\odot}$ galaxy stellar mass bin as already observed in Figure \ref{fig: circ}. The age-circularity distributions for the ex-situ stars show practically no galaxy mass dependence; they are centered on random motion dominated orbits and the oldest ages. \par
The centers of bulge dominated galaxies (top row of Figure \ref{fig: agecirc}) above $10^{9}\,\mathrm{M}_{\odot}$ have overall similar age-circularity distributions as disk dominated galaxies. However, the absolute values of the migrated distribution do not reach the same high circularities as for the disky galaxies and its peak transitions quicker to old ages (8\,Gyr) between mass bins of 9.5 and 10\,dex. Below $10^{9}\,\mathrm{M}_{\odot}$ both the migrated and in-situ distrbition are centered on zero circularities and old ages; distinct to the disky galaxies. \par
For both bulge and disk dominated galaxies the average age-circularity distribution of the in-situ, migrated and ex-situ stars are well separated in mass ranges between $10^{9.5}\,\mathrm{M}_{\odot}$ and $10^{10}\,\mathrm{M}_{\odot}$. This dependence of increasing circularity for younger ages, especially prominent for the migrated stars, gives an indication that recently (i.e. young) migrated stars travel to the center of their host galaxies by loosing their angular momentum \citep[``churning''; see e.g.][for the Milky Way disk]{frankel20} and then, once they have arrived in the center, become dynamically heated over time. 

\section{Discussion, implications and outlooks}\label{sec: discuss}

In this section we discuss the implications of the studied mass assembly of the central 500\,pc in TNG50 galaxies on the formation scenarios of central galaxy components. We also discuss the clumps found in TNG50 as well as the robustness of our results within the TNG modelling framework. In addition, we assess how our results on the stellar population and dynamical properties can be compared to observations and used to understand the mass build-up of galaxies in general.

\subsection[The build-up of galaxy centers in a $\Lambda$CDM cosmology]{The build-up of galaxy centers in a $\boldsymbol{\Lambda}$CDM cosmology}\label{sec: buildup}

Throughout this paper, we have unravelled a set of relations between the properties of the stellar centers of galaxies at $\mathrm{z}=0$. Galaxy centers are dominated by in-situ stars (see Figure \ref{fig: mass}) and follow well established relations \citep[e.g.][]{gallazzi05} that correlate their increasing stellar masses with increasing average ages and metallicities (see Figure \ref{fig: stellpop}). Stars that migrated to the center are second most abundant. They follow the trends for the in-situ stars, however are often distinctively younger and on more rotation supported orbits. Ex-situ stars in the center become only significant (in mass) at high galaxy stellar masses ($>10^{11}\,\mathrm{M}_{\odot}$) (see Figure \ref{fig: mass}). The majority of ex-situ stars originate from the centers of the accreted galaxies \citep[see also][]{gao04}. Moreover, they are amongst the oldest and most metal-poor and random motion dominated stars (see Figures \ref{fig: stellpop} and \ref{fig: circ} as well as), which is in agreement with \citet[e.g.][]{elbadry18}, who studied three MW-like galaxies from the Latte project \citep{wetzel16}. \par
While these trends are consistent with our general understanding of galaxy formation in a $\Lambda$CDM cosmology, we find others that may be more surprising. For example, there seems to be no average difference between the central mass assembly of central and satellite galaxies (see Figure \ref{fig: props}). Generally, central galaxies are thought to have accreted more satellite galaxies. We have checked this relation also for the total accreted mass within TNG50 and also found no significant difference between centrals and satellites on average. Thus, perhaps TNG50 is not probing enough very high mass central galaxies around stellar masses of $10^{12}\,\mathrm{M}_{\odot}$, where this trend might become apparent. \par
Another, rather unexpected result compared to usual assumptions, is that star forming galaxies above $10^{10}\,\mathrm{M}_{\odot}$ possess on average more ex-situ mass in their centers compared to quenched ones (see Figure \ref{fig: props}). Again, this difference, even though much less significant remains when considering the total amount of ex-situ mass (see Figure \ref{fig: tot_exsitu}). This trend also exists for the larger box of TNG100, thus eliminating the fact for low number statistics at the higher mass end, and is in contrast to the original Illustris simulation \citep[see][Figure 5]{rodriguesgomez16}. We have checked the median mass growth of the central ex-situ stars for star forming and quenched galaxies alike between stellar masses of $10^{10}-10^{11}\,\mathrm{M}_{\odot}$ and found that quenched galaxies stop acquiring ex-situ mass in their centers after $\mathrm{z}\sim1.7$ (lookback time $\sim 10\,\mathrm{Gyr}$). Only if we split quenched galaxies further into bulgey and disky as well as barred and non barred do we see that quenched, bulgey and non barred galaxies have a similarly high ex-situ mass in their centers as their star forming counterparts. Together, this is an indication that the time of accretion and consequently the absolute amount of stellar and gas mass of the secondary galaxy (the former will be higher at later cosmic times and the latter will influence the amount of newly formed stars during the merger process) will matter in the build-up of ex-situ mass in the center of the primary and ultimately dictate what properties it has today.\par
On top of that, the fraction of in-situ, migrated and ex-situ stars in the center of galaxies has a significant scatter at fixed galaxy stellar mass regardless of the galaxy's bulk properties at $\mathrm{z}=0$ (see Figure \ref{fig: mass2}). Hence, median trends for different galaxy populations only reveal half of the picture, as the stochasticity of galaxy mergers and interactions in a $\Lambda$CDM cosmology leads to diverse pathways in the build-up of stellar mass in the centers of galaxies. Thus, characteristic properties of galaxies at $\mathrm{z}=0$ are only a limited indicator of the exact formation history of an individual galaxy. For example, there are perfectly regular MW-like spiral galaxies in TNG50 $\mathrm{z}=0$, of which some have experienced (multiple) major mergers and of which some had a more quiet assembly \citep[see also][]{diego22}. \par
This diversity in the central 500\,pc of TNG50 galaxies potentially reflects the variety of central galaxy components seen in observations (see Section \ref{sec: intro}). Even though nuclear rings, disks and star clusters are at or below the resolution limit of TNG50, the stellar population and dynamical properties that we find for central stars of different origins might be a first indication that this would also manifest in structurally distinct components. For example, the distinctly high circularities of migrated stars in $10^{9}-10^{10}\,\mathrm{M}_{\odot}$ galaxies (see Figure \ref{fig: agecirc}) reflect that nuclear disk-like configurations are able to arise. Even more intriguing are their predominantly younger ages of $1-2\,\mathrm{Gyr}$ compared to the underlying old ($\sim8\,\mathrm{Gyr}$) in-situ population, which is in line with observational findings of nuclear disks/rings \citep{bittner20}. Typically, the formation of nuclear rings in disk galaxies is associated with bars funnelling gas towards the center \citep[see e.g.][for dedicated simulations]{seo19,tress20,sormani20}. Even though we did not explicitly investigate the inflow of gas in this study, we see that the migration of stars to the center is likely connected to temporarily induced non-axisymmetries during galaxy interactions (see Section \ref{sec: mecha}). Hence, this shows that mechanisms that are associated with producing distinct nuclear galaxy components are captured in TNG50. Follow-up zoom-in simulations of TNG50 galaxies would show if indeed nuclear components such as disks and rings form from these mechanisms (see Section \ref{sec: moresims}).

\begin{figure*}
     \centering
     \includegraphics[width=\textwidth]{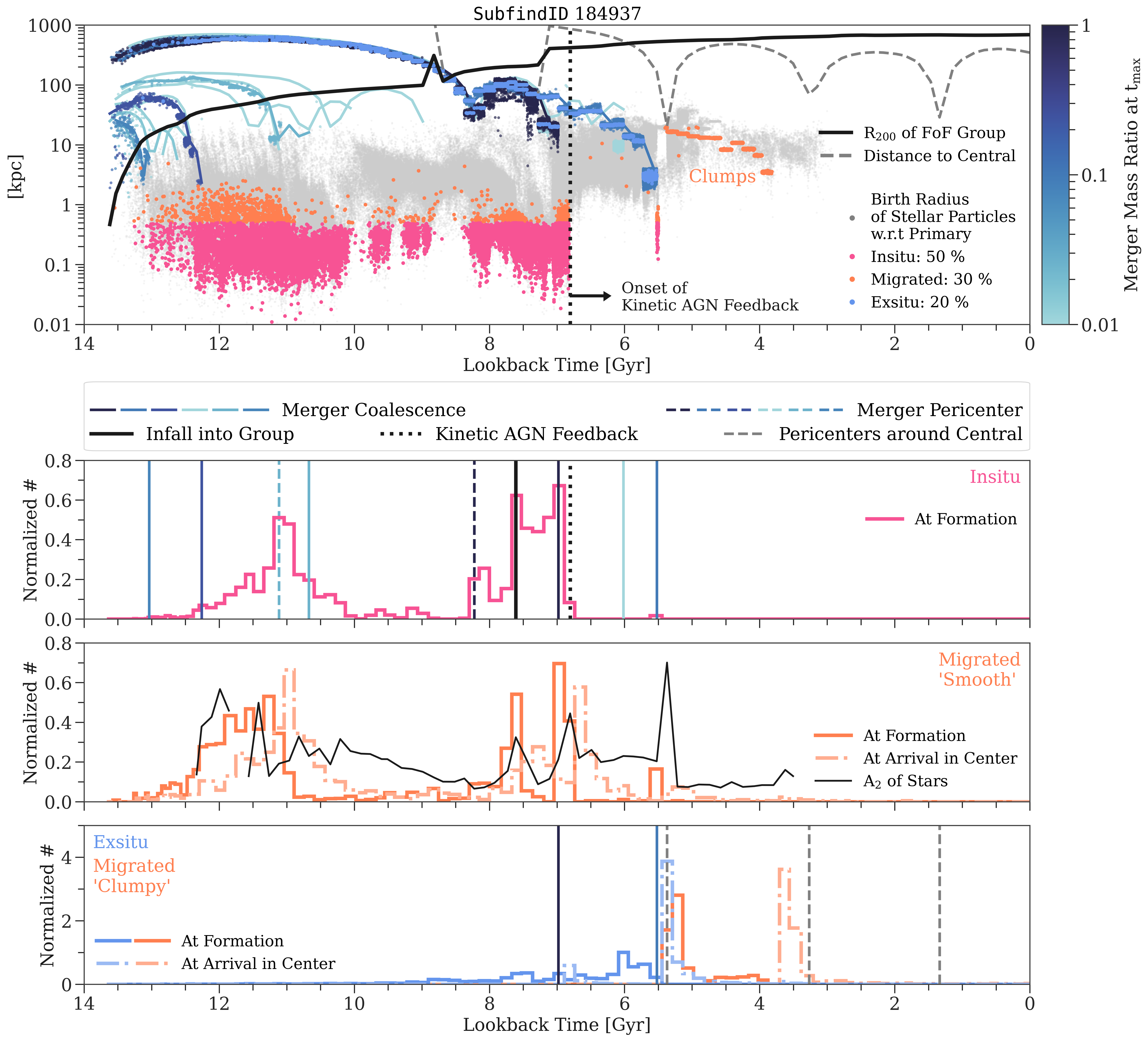}
     \caption{\textbf{Central (500\,pc) assembly history of an individual galaxy (\texttt{SubfindID} 184937) in TNG50 with a total stellar mass of} $\mathbf{10^{10.8}\,M_{\odot}}.$ \textbf{This galaxy encompasses many mechanisms that can shape the stellar mass build-up in the center of galaxies in a} $\mathbf{\Lambda}$\textbf{CDM cosmology.} \textit{Top panel}: Points show all individual stellar particles that belong to that galaxy at $\mathrm{z}=0$. Their distance at the time of birth is shown with respect to their current host in the case of in-situ formed stars (\textit{light gray}: all in-situ particles, \textit{\textcolor{insitu}{pink}}: central in-situ stars only, \textit{\textcolor{migrated}{orange}}: central migrated stars only). In the case of the ex-situ formed stars the distance is shown with respect to their future host galaxy at the time of birth (\textit{color-coded according to the colorbar}: all ex-situ stars, \textit{\textcolor{exsitu}{blue}}: central ex-situ stars only). The distance to individual satellite galaxies (only with maximum stellar masses above $10^{6}\,\mathrm{M}_{\odot}$) that will merge with the primary at some point are shown with thinner solid lines. Their coloring also follows the colorbar, which visualizes the merger mass ratio taken at the time $\mathrm{t}_{\mathrm{max}}$, when the secondary galaxy reaches maximum stellar mass. The thick black solid line shows the radius of the FoF Group the galaxy belongs to at a given lookback time (represented as $\mathrm{R}_{200}$, where the group's density is 200 times the critical density of the Universe). The thick gray dashed line shows the distance between the individual galaxy and the central galaxy of the FoF group it belongs to. Approximately 7\,Gyr ago the galaxy fell into another group and became a satellite galaxy. Before that it was the central of its own FoF group. The vertical black dotted line represents the time the kinetic AGN feedback starts to take effect, which quenches the center. This galaxy has 50\% in-situ, 30\% migrated (of which only 9\% are 'smoothly' migrated and the rest comes from migrated clumps) and 20\% ex-situ stars in its center. \textit{Bottom panel}: Histograms of formation times of \textcolor{insitu}{in-situ} (\textit{top}), formation (\textit{solid}) and arrival (\textit{dashed-dotted}) times at the center for \textcolor{migrated}{`smoothly' migrated} (\textit{middle}) as well as \textcolor{exsitu}{ex-situ} and \textcolor{migrated}{`clumpy' migrated} (\textit{bottom}) stars. Additionally, in the panel for the in-situ stars, we mark the time of coalescence for the six most massive mergers of this galaxy with thin blue colored solid lines. According to the colorbar of the top panel, a darker blue means a higher merger mass ratio. Pericenter passages for two mergers are shown by thin dashed lines following the same colorcode. The approximate time of the galaxy falling into its $\mathrm{z}=0$ FoF group is shown by the thick black solid line and the onset of the kinetic AGN feedback is shown as the black dotted line. In the panel for the `smoothly' migrated stars we also show the A$_2$ mode of the stars for a given lookback time (see Appendix \protect\ref{sec: galprops} for a definition). In the panel for the ex-situ and `clumpy' migrated stars, we show the time of coalescence of the two mergers that deposited ex-situ stars in the galaxy's center (\textit{blue solid lines}) as well as the three pericenter passages of the galaxy around its central galaxy after it became a satellite (\textit{gray dashed lines}). }
     \label{fig: assembly}
\end{figure*}

\subsection{Mechanisms for the formation and deposit of stars in the center of galaxies}\label{sec: mecha}

The cosmological framework of TNG50 produces diverse properties of galaxies and their centers. Consequently, the mechanisms that are responsible for the formation and deposit of stars in the centers of galaxies also have to be diverse. \par
To visualize possible mechanisms for the formation and deposit of stars in the center of galaxies we walk through the central assembly history of an individual galaxy as seen in Figure \ref{fig: assembly}. We picked this particular galaxy, which has a stellar mass of $10^{10.8}\,\mathrm{M}_{\odot}$ at $\mathrm{z}=0$, as it shows many of the possible mechanisms that can be present in the formation of galaxy centers. This however does not mean that all galaxies show the same amount of complexity. Most galaxies will only exhibit one or two of these mechanisms with varying impact depending on their individual formation pathway. The galaxy's center at $\mathrm{z}=0$ consists of around 50\% in-situ, 30\% migrated and 20\% ex-situ stars. \par
The main summary of the subsequent sections and Figure \ref{fig: assembly} is the following: galaxy mergers and other interactions are probably the most important driver in central stellar mass assembly, as they also strongly influence the formation of in-situ stars. Due to the diverse statistics of galaxy interactions, many of the proposed formation scenarios of central galaxy components arise naturally and in conjunction to each other, when hierarchical galaxy formation is considered. Thus, TNG50 highlights the necessity to study the central mass assembly of galaxies in a cosmological context. \par

\subsubsection{In-situ stars}

Galaxy mergers can trigger bursts of star formation as the tidal forces compress and shock gas efficiently \citep[e.g][]{mihos96,dimatteo07,cox08,dimatteo08}, even in its nuclear region \citep{powell13}. While the relative enhancement of star formation rates depend on the specific configuration of the merging galaxies, e.g. merger mass ratio, gas content, orbital infall parameters, the times of intense star formation coincide with pericenter passages and coalescence \citep[see also][]{diego22}. \par
Peaks in the formation time of central in-situ stars in Figure \ref{fig: assembly} coincide with times of pericenters and coalescence of mergers that this galaxy has experienced. \emph{Thus the formation history of the central in-situ stars is directly connected with the merger history of a galaxy.} However, it has to be further quantified whether also the \emph{bulk} of in-situ stars is formed during such events or if that actually happens in-between galaxy interactions. Nevertheless, it is clear that a variety of different mergers are able to produce peaks in the formation of central in-situ stars. \par
For example, the peak between 10\,Gyr and 12\,Gyr ago was induced by a very minor merger\footnote{We adopt the definition of \citet{rodriguesgomez16} for the calculation of merger ratios.} with a stellar merger mass ratio of around 0.02. At these high redshifts the primary still had a high gas fraction ($\sim80\%$) and thus the minor merger was enough to trigger a peak of star formation in the center. Evidently, the formation of in-situ stars in the center decreased between 10\,Gyr and 8\,Gyr ago, as the amount of available gas decreased. Thus, in order to trigger another significant peak in in-situ star formation later on, the merger between 8\,Gyr and 7\, Gyr had to bring in a large amount of gas. While the ratio of the stellar mass between the secondary and primary was around one (and therefore a major merger) at the time when the secondary reached its maximum stellar mass, the secondary still had around 20 times more gas than the primary. \par
Furthermore, this major merger, as well as two smaller ones that coalesced around 6\,Gyr ago, happened while the primary galaxy was in the process of falling into another FoF Group, i.e. transitioning from being a central galaxy of its own FoF group to being a satellite galaxy of another FoF group. This is seen by the two sharp jumps in R$_{200}$ between 9\,Gyr and 7\,Gyr ago. Evidently, this process produced another peak of in-situ star formation at around 7.5\,Gyr ago, which could stem from the new, higher density environment. We have also seen in other galaxies, that were able to retain enough gas in their centers after infalling into a group, that in-situ star formation was triggered during the pericenter passages around the central until the galaxy became quenched. In such occasions, again tidal forces are able to compress the gas efficiently. \par
Within TNG50, there are two main processes that can quench the in-situ star formation in the center of galaxies. The first one is the onset of the \emph{kinetic} AGN feedback mode implemented in TNG. Often this feedback mode switches on after a merger has been completed, as is the case for our galaxy in Figure \ref{fig: assembly} at around 7\,Gyr, shortly after the major merger coalesces. We see that only the central 1\,kpc becomes quenched, while the outskirts of the galaxy continue to form stars. This is because in TNG, AGN driven quenching proceeds from inside out \citep[see][for details]{weinberger17,nelson19,nelson21}. After this mode is switched on only occasional gas-rich mergers or migrated clumps are able to bring in new gas to the center to cause new in-situ star formation, as seen at 5.5\,Gyr in Figure \ref{fig: assembly}. Lastly, the thermal feedback mode, which is often active prior to the kinetic mode switches on and also injects relatively more energy, is not responsible for quenching the centers of galaxies in TNG50 \citep{zinger20}. \par
The second process that will shut down star formation in the center of TNG50 galaxies is when the galaxy as a whole becomes quenched, either through environmental processes, e.g. after a few pericenters after infall into a group (as is the case for the galaxy in Figure \ref{fig: assembly} around 3\,Gyr ago) or through AGN feedback, which is primarily important for the highest mass galaxies \citep[see also][]{donnari21a}.

\subsubsection{Migrated stars}

The formation times of `smoothly' migrated stars in Figure \ref{fig: assembly} is closely related to the formation times of the in-situ stars, which is not the case for the `clumpy' migrated stars. This is reasonable, because the majority of `smoothly' migrated stars are born already close to the center ($\lesssim2\,\mathrm{kpc}$), while the `clumpy' migrated stars formed predominantly in the outer disk. The `clumpy' migrated stars make up 91\% of the total mass of migrated stars in the center of this galaxy. \par
However, the star formation in the central 2\,kpc is \emph{not} a guarantee to produce a significant amount of `smoothly' migrated stars, as seen between lookback times of $9-11\,\mathrm{Gyr}$ and also between $7-7.5\,\mathrm{Gyr}$ in Figure \ref{fig: assembly}. Thus, specific conditions must be met that transport stars from around $1-2\,\mathrm{kpc}$ to the center. \par
Non-axisymmetric features, such as spiral arms and bars, are well known to be able to diffuse the angular momentum of stars and cause radial migration \citep[e.g.][]{sellwood02,minchev10}. While this effect is mainly studied in the (outer) disk of galaxies, we show here in Figure \ref{fig: assembly} that similar non-axisymmetries are likely responsible for the inward migration of stars to center of galaxies. We see that peaks in the A$_2$ mode (see Appendix \ref{sec: galprops} for a definition) of the stellar mass distribution occur \emph{before} peaks of migrated stars arriving in the galaxy center. We detect similar enhancements for the Fourier modes of the \emph{gas} mass distribution \citep[see also][]{dimatteo07}. \par
These temporary enhancements of non-axisymmetric features are clearly induced during galaxy interactions and the exerted torques on the gas and stars can produce these `smoothly' migrated stars. This also indicates that it is possible for a galaxy to have experienced migration events of stars, even if the galaxy itself does not exhibit any signs of bar- or spiral-like features today. \par
The `clumpy' migrated stars form after the first pericenter passage of the galaxy around its central around 5.5\,Gyr ago and arrive at the center shortly before the second pericenter passage around 2\,Gyr later. Similarly, we have seen qualitatively for other galaxies that clumps formed rather recently ($\mathrm{z}<1$) are mainly induced by fly-bys, as these are still able to destabilize the disk significantly after the predominant merger phase of the Universe is over. However, clumps are also able to form without any significant galaxy interactions and a follow-up study is needed to characterize this further as well as establish overall the credibility of the formation of the clumps (see Section \ref{sec: clumpdiscuss} for a further discussion).

\subsubsection{Ex-situ stars}

The two mergers that are responsible for the majority of the ex-situ stars in the center of the galaxy in Figure \ref{fig: assembly}, are the 1:1 and 1:10 merger that coalesced around 7\,Gyr and 5.5\,Gyr ago respectively. Both mergers brought in a comparable \emph{total} amount of stellar mass of around $1.2\times10^{10}\,\mathrm{M}_{\odot}$ and $8.1\times10^{9}\,\mathrm{M}_{\odot}$ respectively. However, the major merger deposited around 10 times less stars in the central 500\,pc compared to the minor merger, i.e. $0.6\%$ and $5\%$ of their respective total stellar mass arrived in the center. This highlights that the merger mass ratio cannot be the only parameter determining the amount of ex-situ stellar mass that is deposited in the center of galaxies. We expect that the spin-orbit coupling of the primary and secondary galaxy as well as other orbital parameters play a role in this, as the exerted tidal forces and the influence of dynamical friction differ for different configurations \citep[see e.g.][for a study]{renaud09}. \par
Around 67\% and 93\% of the ex-situ stars that arrived from the major and minor merger respectively were formed \emph{after} both satellite galaxies entered R$_{200}$ of the primary's FoF halo around 8.75\,Gyr ago. As also all central ex-situ stars were born in the center ($\sim500\,\mathrm{pc}$) of their respective birth galaxies, this confirms that significant nuclear star formation is also triggered in the secondary galaxy after infall. \par
Most of the ex-situ stars in the center arrive there immediately after the merger coalesces. This is the case if their distance to the center of the primary was less than 500\,pc at the time of stripping. Otherwise it can take up to 2\,Gyr. Interestingly, the arrival of the `clumpy' migrated stars at the center around 3.5\,Gyr ago induced a second peak in the arrival of ex-situ stars from the minor merger into the center, albeit being ten times lower and hence not visible in Figure \ref{fig: assembly}.

\subsection{The case of stellar clumps}\label{sec: clumpdiscuss}

Disk fragmentation can occur due to gravitational instabilities in a galaxy's gas-rich and turbulent disk \citep{toomre64,springel05,hopkins13a}. This fragmentation can form highly star forming clumps, which have been reproduced in several studies using hydrodynamical galaxy simulations, either isolated or fully cosmological ones \citep[e.g.][]{bournaud07,genel12,bournaud14,mandelker14,mandelker17,buck17}. The execution of these simulations was motivated by the discovery of the clumpy morphology in the rest-frame UV light of high redshift, star forming galaxies \citep[e.g.][]{elmegreen07,guo15}. Therefore, these simulations are tailored to focus on clump formation in massive disk galaxies $10^{10-11}\,\mathrm{M}_{\odot}$ at $\mathrm{z}\geq1$. \par
In observations, clumps have masses between $10^{7}\,\mathrm{M}_{\odot}$ and $10^{9}\,\mathrm{M}_{\odot}$, as well as sizes of 1\,kpc or less. Clumps in the simulations are usually identified via regions of enhanced gaseous surface mass density or from mock stellar light images. In TNG50, the identification of clumps is (so far) a passive byproduct of the \textsc{Subfind} algorithm, nevertheless the extracted baryonic mass distribution of the clumps peaks at $10^{8}\,\mathrm{M}_{\odot}$ exhibiting overall high gas fractions (see Figure \ref{fig: clumps_sum}) are in agreement with the other studies. However, all clumps in TNG50 have 3D baryonic half mass radii below 300\,pc and therefore seem to be much more compact compared to observations and some simulation studies \citep[see Figure 9 in][]{buck17}. The latter could be a result of the different treatments of star formation and feedback in the simulations or the clump identification, as the numerical resolution of TNG50 is largely comparable to those of the previous studies. Additionally, in TNG50 clumps seem to form continuously throughout cosmic time (see Figure \ref{fig: clumps_sum}), which has not been investigated in other studies of clump formation. \par 
These clumps can migrate to the center of their host galaxies due to dynamical friction, as well as merge with each other while doing so \citep[][]{bournaud07,dekel13,bournaud14,mandelker14,dekel21}. The migration time to the center is found to be of the order of a few hundred Myr, which is similar for clumps in TNG50, where most of the clumps arrive at their respective galaxy's center after $1-2$ snapshots ($\sim200\,\mathrm{Myr}$) (see Figure \ref{fig: clumps_sum}). This mechanism contributes to the formation of the bulge \citep[e.g.][]{bournaud07,elmegreen07,dekel09}. Consequently, the properties of the clumps need to be specific, such that they survive their own internal stellar feedback and the tidal forces on their way to the center with enough stellar mass to significantly contribute to the formation of the bulge. In TNG50, the fraction of clumpy migration stars to the total stellar mass in the center is greater than 40\% for about 12\% of galaxies with any clumpy migrated stars in their centers (all of those galaxies have stellar masses above $10^{10.5}\,\mathrm{M}_{\odot}$; see Figure \ref{fig: clumps_sum}). Thus, according to TNG50, the stellar mass transported by the migration of clumps is likely not very important for bulge formation for the majority of high mass galaxies. Nevertheless, the clumps often retain a large amount of gas until the center, or drag gas along (see clump closest to the galaxy's center in the top right of Figure \ref{fig: clumps_sum}), from which stars might form. We have not checked explicitly if this increases the contribution of stellar mass in the center significantly, or if such gas is lost by directly funneling into the SMBH. \par
In contrast to that, some simulations report clump formation, but then no migration due to almost immediate dissolution or disruption \citep[][]{hopkins12,hopkins13b,mayer16,oklop17,buck17}. This is likely due to the different simulation set ups, as well as the exact implementation of stellar feedback, or simply because not enough galaxy diversity is probed with isolated or zoom-in galaxy simulations. For example, in Figure \ref{fig: clumps_sum}, we see that galaxies above $10^{10}\,\mathrm{M}_{\odot}$ in stellar mass start to exhibit more than one clump on average, however significant amounts of clumps that are able to migrate to the center reside in galaxies with stellar masses around $10^{11}\,\mathrm{M}_{\odot}$ and above. Hence, an investigation of clump formation and migration in a fully cosmological box is necessary to not only capture galaxies of different masses but also different galaxy assembly histories. In TNG50 mergers and other galaxy interactions, such as flybys, can trigger a significant amount of clump formation \citep[although not exclusively; see also][for similar reports]{dimatteo08,hopkins13b,calabro19}. \par
Still, within TNG50 we want to exercise caution when it comes to the trustworthiness of clump formation and their exact properties. Only follow-up zoom-in simulations with higher resolution for different galaxies, as well as different treatment of star forming gas and stellar feedback \citep[see e.g.][for the influence of highly resolved star formation and different stellar feedback schemes in galaxy simulations]{hopkins13c,smith18,smith21a,smith21b}, will allow for a more robust quantification of clumps in TNG50. Nevertheless, the clump formation in TNG50 is unlikely to be a numerical artifact in its entirety, as the adaptive mesh refinement naturally allows for smaller cell sizes in areas of high gas density.

\subsection{The predictive power of TNG50 at small(er) scales}\label{sec: moresims}

\begin{figure*}
    \centering
    \includegraphics[width=\textwidth]{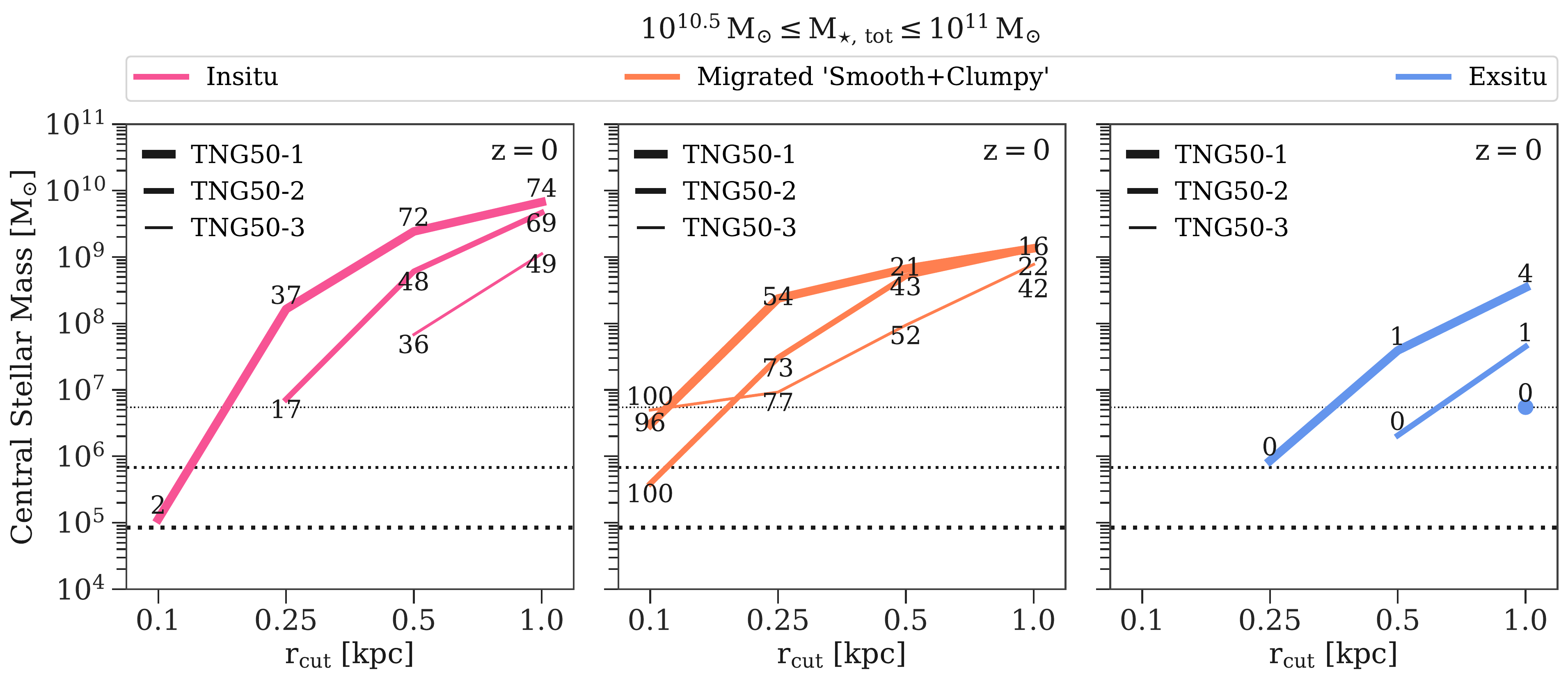}
    \caption{\textbf{Effects of numerical resolution and aperture size on the central stellar mass for in-situ, migrated and ex-situ stars in TNG50 galaxies with} $\mathbf{10^{10.5-11}\,M_{\odot}}$ \textbf{in stellar mass at} $\mathbf{z=0.}$ Lines show the median central stellar mass for in-situ (\textit{pink}), migrated (\textit{orange}) and ex-situ (\textit{blue}) stars for four choices of $\mathrm{r}_{\mathrm{cut}}=0.1,\ 0.25,\ 0.5\ \mathrm{and}\ 1 \, \mathrm{kpc}$ and three resolution realizations of TNG50 (thicker lines indicate better resolution). TNG50-1 is the highest resolution (flagship), followed by TNG50-2 and -3, which have 2 and 4 times lower spatial resolution. The mass resolution is 8 and 64 times lower respectively and indicated by the dotted horizontal lines. The numbers indicate the respective central stellar mass \emph{fraction} in percent. Decreasing size of the center means decreasing stellar mass. However the fraction of migrated mass increases, while the in-situ fraction consequentially decreases. At a hypothetical higher resolution (TNG50-0) the latter effect would be lessened as more stellar mass is formed within a given aperture size. Similar trends are recovered for other galaxy masses.}
    \label{fig: morecentral}
\end{figure*}

While the TNG modelling framework is extremely successful in reproducing key observational results of galaxy populations, numerical resolution and the implementation of sub-grid physics are insuperable limitations of the physical model of galaxy formation. Regarding the former we demonstrate in Figure \ref{fig: convergence} in Appendix \ref{appendix: validate5} that the total stellar mass within the central 500\,pc of galaxies in TNG50 is converging \cite[see also][]{pillepich19}. When splitting the central mass into the contribution of in-situ, migrated and ex-situ stars, the start of convergence is more difficult to assess due to the fixed size of $\mathrm{r}_{\mathrm{cut}}$ (see Appendix \ref{appendix: validate5} for a more detailed discussion) as well as the overall influence of resolution on the amount of accreted stellar mass \citep[which should overall increase with resolution, see also][]{grand21}. Thus higher resolution runs are needed to fully determine the amount of convergence. \par
Higher resolution zoom-in simulations of some TNG50 galaxies - additionally with models variation of stellar feedback and a better resolved cold gas phase of the star forming gas - are certainly interesting and needed to properly evaluate the convergence of the central stellar mass, the formation of stellar clumps and observed nuclear galaxy components. Nevertheless, our study of TNG50 shows that the cosmological context plays a major role in the assembly of galaxy centers, which is unlikely to become less significant with numerical resolution and other modelling aspects. Already at the resolution of TNG50 it is \emph{rare} to find a galaxy with \emph{no} ex-situ stars in its central 500\,pc, which is only the case for around 9\% of all galaxies in our sample spanning a range between $5\times10^{8}-5\times10^{12}\,\mathrm{M}_{\odot}$. We note that this percentage remains the same, even if we do not impose a constraint on the physical size of galaxies included in our analysis. In fact, our selection of galaxies with half-mass radii $>2\,\mathrm{kpc}$ does not impose any strong differential effects on our results. We have explicitly checked Figures \ref{fig: mass} and \ref{fig: props}, which show the same trends when galaxies with $\mathrm{R}_{1/2}<2\,\mathrm{kpc}$ are included. Thus, we do not expect that any other results of our study are significantly effected by our choice. \par
Therefore, our findings highlight that the high density, nuclear regions of galaxies can survive tidal forces and contribute to the build-up of the centers of other galaxies, \emph{including} low-mass galaxies. Additionally, if the two merging galaxies are massive enough to both host a black hole, the merger of the central regions of galaxies will contribute to the growth of SMBHs \citep[see e.g.][for a recent observational confirmation of such a system]{schweizer18,voggel21}. \par
Currently there are no large-box, cosmological, hydrodynamical simulations (including TNG50) that can resolve smaller central galaxy components, such as nuclear star clusters (NSCs), until $\mathrm{z}=0$ \citep[see][that investigate NSC formation in a cosmological set-up until $\mathrm{z}=1.5$]{brown18}. However, by extending the trends in our study to even smaller scales, it is not impossible to think that the formation and evolution of NSCs are also governed by galaxy interactions. Even though the relative fraction of ex-situ stars on tens of pc scales is likely very small for the majority of galaxies, it is clear that the in-situ star formation and the migration of stars to the center is closely connected to the formation pathway of the entire galaxy (see Figure \ref{fig: assembly}), because galaxy interactions are able to create the conditions needed to funnel gas and stars to the center. Therefore, it is important to treat NSC formation in the context of the hierarchical build-up of galaxies in a $\Lambda$CDM cosmology \citep[][]{brown18} and consider the influence of galaxy interactions in (semi-)analytical models \citep{leaman21}. \par
We have explicitly checked within TNG50 if we can make predictions at smaller scales than 500\,pc, by repeating our entire analysis for $\mathrm{r}_{\mathrm{cut}}$ of 250\,pc (approximately the softening length of TNG50) and 100\,pc, as well as 1\,kpc for a consistency check. In addition to TNG50-1 (the highest resolution), we repeated this for the two lower resolution realizations, which are TNG50-2 and TNG50-3 respectively. The results are shown in Figure \ref{fig: morecentral} for galaxies between $10^{10.5}$ and $10^{11}\,\mathrm{M}_{\odot}$. \par
With decreasing size of the center ($\mathrm{r}_{\mathrm{cut}}$) the absolute mass decreases for all three origins. However, the \emph{fraction} of the in-situ population decreases with decreasing central size, while the migrated fraction increases; a consequence of the smaller volume that is proved. At the same time, this behaviour is also affected by the resolution, which not only sets the absolute normalization of the mass fraction, but also the spatial size at which the relative contribution of the in-situ and migrated fraction swaps. We therefore conclude that a hypothetical higher resolution (TNG50-0) would increase (decrease) the in-situ (migrated) fraction below 250\,pc due to the convergence behaviour of the absolute stellar mass at fixed aperture size. Similarly, the contribution of ex-situ stars will increase at a given aperture and also likely reach scales smaller than 500\,pc (see Appendix \ref{appendix: validate5} for more details). \par
This behaviour emphasizes that the contribution of all three origins will likely remain relevant on scales of 100\,pc.

\subsection{Galaxy centers as tracers of overall galaxy assembly}

Unveiling the merger history of galaxies proves difficult to tackle outside our own Galaxy due to many reasons. Perhaps the most severe one is the fact that accreted material is not necessarily visually apparent in the forms of streams and shells (or any other form of irregularity), especially when the merger coalesced many Gyr ago. \par
Since deep photometry of galaxies (initially stacked for many galaxies) revealed the need for an additional S\'ersic component to accurately fit their surface brightness profiles beyond tens of kpc \citep[e.g.][]{zibetti04,tal11,dsouza14}, the focus of quantifying accreted material has primarily been on the outskirts of galaxies, i.e. their stellar halos \citep[e.g.][]{monachesi16,merritt16,spavone17,huang18,spavone20}. It is understood that the excess of light at large galactic radii should mark the transition from the in-situ to ex-situ dominated areas of a galaxy, as (significant) stellar mass can be build-up through minor merging there. \par
However, the new era of cosmological hydrodynamical simulations suggest that such a transition does not necessarily exist for every galaxy, as especially high mass galaxies can be dominated by ex-situ stars at all radii \citep[][]{tacchella19,pulsoni21}. Furthermore, \citet[][]{remus21} showed that the transition in surface brightness profiles traced by two S\'ersic fits does not correspond to the true in-situ and ex-situ dominated regions. Similarly, changes in kinematic profiles at large radii, which can, for example, be obtained with globular clusters \citep[e.g.][]{arnold14} or planetary nebulae \citep[e.g.][]{pulsoni18}, do not, in general, correspond to transitions between in-situ and ex-situ dominated regions \citep[][]{schulze20,pulsoni21}. \par
While detailed studies of stellar halos are certainly important and necessary, our study suggests that there lies potential in using the centers of galaxies to study their accretion history (see Figure \ref{fig: mass2}). Not only are the centers the brightest region of a galaxy and hence deliver the highest quality data, but they are also increasingly covered in numbers by IFU surveys, which provide detailed kinematic and stellar population information (e.g. SAMI: \citealp[][]{sami}, MaNGA: \citealp[][]{bundy15}). In particular, our results in Figures \ref{fig: agemet} and \ref{fig: agecirc} show that in-situ and ex-situ stars in the center are (on average) well separated in age-metallicity-circularity space for
galaxies with stellar masses $\leq10^{10.5}\,\mathrm{M}_{\odot}$. Newly developed techniques are able to extract such \emph{distributions} in ages and metallicity as well as circularities from IFU measurements, and already have been proven to be able to estimate the true underlying accreted stellar material much more realistically \citep{boecker19,boecker20a,zhu20,davison21,zhu21}. \par
Even though in-situ and ex-situ stars separate better in their stellar population and dynamical properties for these lower mass galaxies, the ex-situ stellar mass fraction in the central 500\,pc is \emph{on average} tenth of percent, thus picking up accreted signatures in the very centers will still be challenging even with these new techniques. However, low redshift IFU observations easily cover $1-2$ half-light radii, which extend beyond 500\,pc and hence should encompass more accreted material. More follow up work will be needed to quantify the optimal extent needed from a galaxy's center to reliably pick up ex-situ fractions in observations. \par
On top of that, the large spread seen in the central ex-situ mass \emph{at fixed total ex-situ mass} (see Figure \ref{fig: mass2}) points towards significant spatial variation of ex-situ stars in the host galaxy, regardless of whether the galaxy has accreted a lot of stellar material or not. It is likely that measuring these spatial variations, will inform us about the types of mergers that have happened. Typical characteristics could be the merger ratio, for example major mergers will have the ability to deposit more of their stars in the center of galaxies, but also the gas content or orbital infall parameters. We plan to exploit this in future work. \par

\begin{figure*}
     \centering
     \includegraphics[width=1.75\columnwidth]{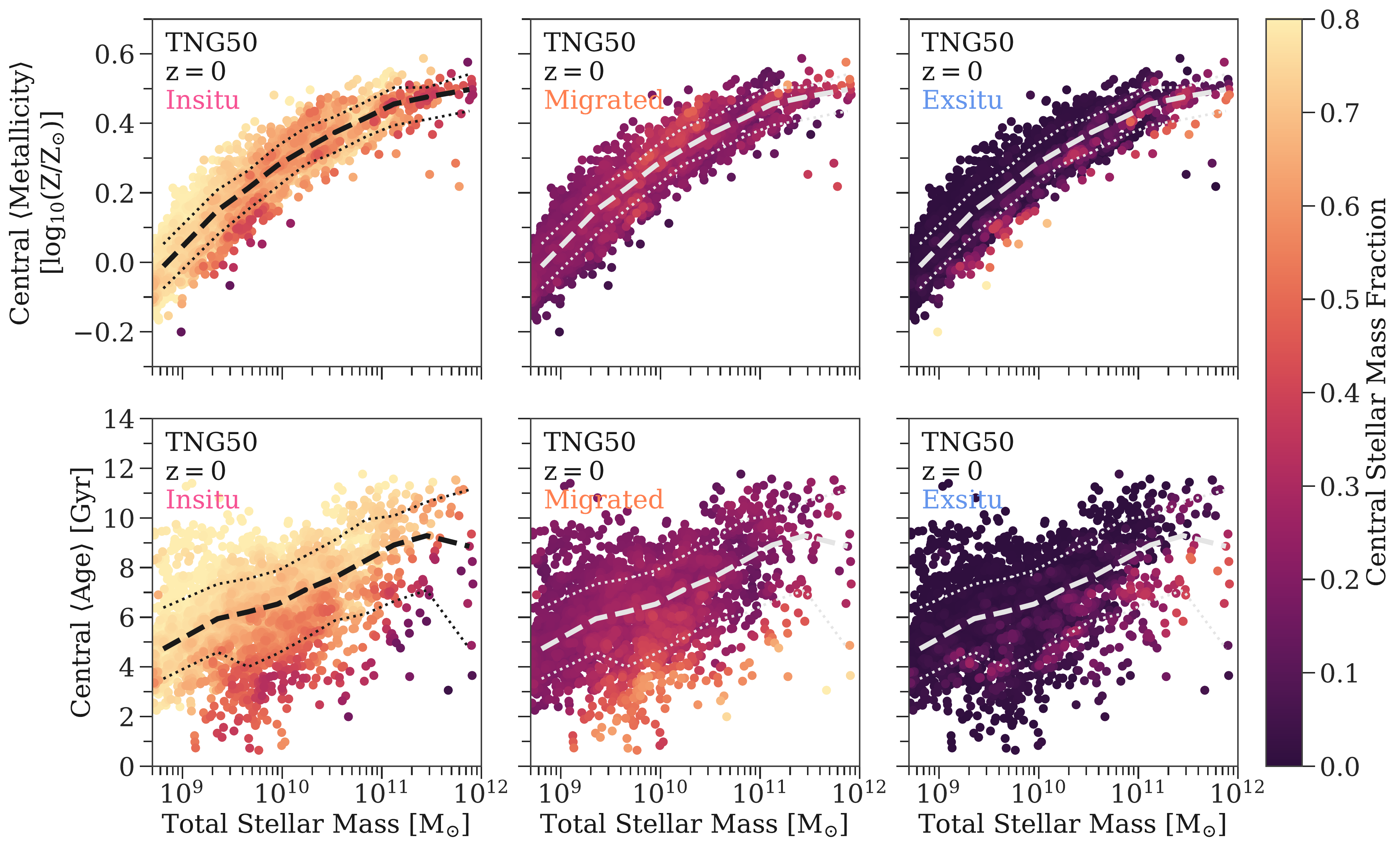}
     \caption{\textbf{Information about the central (500\,pc) fractional mass associated with in-situ, migrated and ex-situ stars contained in average age and metallicity measurements from TNG50 galaxies at} $\mathbf{z=0}.$ \textit{Top row}: Mass-weighted average metallicities for \emph{all} central stars as a function of the galaxy's total stellar mass, but color-coded in each panel \textit{(from left to right)} according to their fraction of in-situ, migrated and ex-situ stars. The colors are LOESS \protect\citep{loess} smoothed to show the average around neighbouring galaxies. The thick dashed line shows the median relation in each panel, and the thin dotted lines show the 16th and 84th percentile respectively. \textit{Bottom row}: The same but for the mass-weighted average age. Galaxies that are more metal-poor than the 16th percentile for their corresponding stellar mass are more likely to have high ex-situ fractions in their centers, while galaxies with high central migrated fractions are younger than the 16th percentile.} 
     \label{fig: obs}
\end{figure*}

\subsection{Hints for (SDSS-like) observations}

TNG50 predicts a diverse mass build-up of galaxy centers. What are the prospects to learn about a galaxy's central in-situ, migrated and ex-situ fraction from more ``traditional'' observations? For example, from SDSS DR7 \citep{sdss}, that provides single $3^{\prime\prime}$ fiber spectra for centers of hundred of thousand of galaxies, \emph{average} ages, metallicities and [$\alpha$/Fe] abundances can be determined \citep{gallazzi20}. How much information do such measurements contain about the contribution of stellar populations of different origins to a galaxy's center? \par
In Figure \ref{fig: obs} we show the mass-weighted average central age and metallicity for our sample of TNG50 galaxies color-coded by the mass fraction attributed to each origin (LOESS smoothed; \citealp{loess}). \par
If the measured average central age and metallicity of a galaxy lies on the respective mass-age and mass-metallicity relation, the galaxy has likely a high fraction of in-situ stars in its center, except if its larger than $10^{11}\,\mathrm{M}_{\odot}$ in stellar mass. If the measured average metallicity is \emph{below} the 16th percentile at fixed stellar mass, it is more likely that the galaxy's center is dominated by ex-situ stars. Similarly, if the galaxy has an average age below the 16th percentile, it has likely a high amount of migrated stars in its center. High mass galaxies above $10^{11}\,\mathrm{M}_{\odot}$ with significant amounts of ex-situ stars in their centers, also have a slightly younger age (between the 16th percentile and the median) than the typical average galaxy in that mass regime. \par
Naturally, a proper mocking of observed average stellar population properties from TNG50 is needed to compare accurately to measurements from \citet{gallazzi20}. However, Figure \ref{fig: obs} seems to acknowledge that such measurements provide some leverage in determining the fraction of stars with different origins in the centers of galaxies. \par
With respect to comparisons to the whole SDSS galaxy sample, it would be necessary to repeat the analysis of this paper for different spatial apertures, The fixed  $3^{\prime\prime}$ diameter of the SDSS fibers will already encompass larger physical sizes than 1\,kpc for galaxies with $\mathrm{z}>0.02$. It would be interesting to understand how the relative contribution of stars from the different origins change with greater spatial extent, especially for the ex-situ stars.

\section{Summary and Conclusions}\label{sec: summary}

Galaxies growth hierarchically in a $\Lambda$CDM universe. Their centers are the regions where usually the highest quality observations are available. What information about the hierarchical growth of galaxy formation is encoded in this observationally favourable region? To answer this, we investigated the central 500\,pc mass assembly of galaxies ranging from $5\times10^{8}\,\mathrm{M}_{\odot}$ to $5\times10^{12}\,\mathrm{M}_{\odot}$ with half-mass radii $>2\,\mathrm{kpc}$ in the TNG50 simulation. \par
Stars that are found at the center of TNG50 galaxies at $\mathrm{z}=0$ originate from one of the three possibilities: \textcolor{insitu}{in-situ} (formed inside the center), \textcolor{migrated}{migrated} (formed inside the host galaxy, but outside the center), \textcolor{exsitu}{ex-situ} (brought in by mergers). Stars can migrate to the center either as as continuous distribution of individuals (smooth) or in clumps. \par
For each origin we characterized their radius with respect to their host galaxy at birth to understand the travelling distances for the migrated stars as well as the spatial environment of ex-situ stars at the time of birth and deposit into the $\mathrm{z}=0$ host. \par
We then investigated the amount of the central stellar mass contained in each of three origins and their relative contribution as well as their correlation to each other across the entire TNG50 galaxy mass range. Additionally, we studied differences in central in-situ, migrated and ex-situ for different galaxy types at $\mathrm{z}=0$. \par
To address whether the different origins of central stars leave a discernible imprint on their (observable) features, we characterized and correlated their ages, metallicities, [Mg/Fe] abundances as well as dynamical properties with their distributions in circularity as a function of the galaxy's total stellar mass. We summarize our most important findings below:

\begin{itemize}
    \item In-situ stars are on average the dominant component in stellar mass in the central 500\,pc of TNG50 galaxies across the entire mass range of $5\times10^{8-12}\,\mathrm{M}_{\odot}$. Migrated stars contribute on average 20\% to the total stellar mass in the center, where below (above) $5\times10^{10}\,\mathrm{M}_{\odot}$ in galaxy stellar mass smoothly (clumpy) migrated stars encompass their majority. The central stellar mass fraction of ex-situ stars becomes \emph{on average} non negligible above galaxy masses of $5\times10^{10}\,\mathrm{M}_{\odot}$ with a large scatter of up to 80\%. However, it is the \emph{exception} to find a galaxy without \emph{any} ex-situ stellar mass in its central 500\,pc, which is only the case for about 9\% of galaxies in our total sample. (Figure \ref{fig: mass})
    \item The majority of smoothly migrated stars originate close to the center (radii between 500\,pc and 1\,kpc), whereas $\sim15\%$ come from larger distances up until 10\,kpc. Compared to that, clumpy migrated stars possess a distinctively different distribution of birth radii which peaks around $20-30\,\mathrm{kpc}$ for galaxies with stellar masses greater than $5\times10^{10}\,\mathrm{M}_{\odot}$. Most of the ex-situ stars originate in the central 1\,kpc of their birth galaxies, where they remain until they are deposited inside their $\mathrm{z=0}$ host galaxy. (Figure \ref{fig: birthloc})
    \item \emph{At fixed galaxy stellar mass} the amount of central ex-situ stellar mass exhibits a significant scatter between $4-6\,\mathrm{dex}$, reflecting the stochasticity of the merger history of individual galaxies. In some cases, close to the \emph{entire} total amount of ex-situ stellar material ever deposited inside the host galaxy resides within the central 500\,pc. (Figure \ref{fig: mass2})
    \item In TNG at $\mathrm{z}=0$, star forming galaxies with stellar masses above $10^{10}\,\mathrm{M}_{\odot}$ have on average \emph{larger} ex-situ central stellar masses than their quenched counterparts. Only quenched galaxies that are additionally bulgey and have no bar signature show a rise of central ex-situ stellar mass above $10^{10}\,\mathrm{M}_{\odot}$ similar to the star forming galaxies. Galaxies between $5\times10^{9}-5\times10^{10}\,\mathrm{M}_{\odot}$ with an overmassive (undermassive) SMBH in the center are more compact (extended) and show on average a higher (lower) in-situ and migrated central stellar mass. There is \emph{no} difference in neither in-situ, migrated nor ex-situ central stellar masses for central or satellite galaxies. (Figure \ref{fig: props})
    \item Central ex-situ stars have on average the lowest metallicities, the oldest ages and the highest [Mg/Fe] abundances. The slope of their mass-metallicity relation is slightly steeper than that of the in-situ and migrated stars, and their mass-age relation is flat compared to the positive correlation between central age and galaxy stellar mass for the in-situ and migrated stars. Overall, the average stellar populations properties of in-situ and migrated stars are very similar, with in-situ stars being slightly more metal-rich and older. (Figure \ref{fig: stellpop})
    \item The majority of central stars for galaxies with stellar masses below $10^{9}\,\mathrm{M}_{\odot}$ and above $10^{11}\,\mathrm{M}_{\odot}$ regardless of their origin are on random motion dominated orbits. For galaxies in between those stellar masses, the peak of the circularity distribution shifts by 0.5 (0.25) towards rotationally supported orbits for migrated (in-situ) stars \emph{for both} disk and bulge dominated galaxies, whereas ex-situ stars remain random motion dominated at all galaxy masses. (Figure \ref{fig: circ})
    \item For star forming galaxies around $10^{10}\,\mathrm{M}_{\odot}$ in stellar mass, in-situ, migrated and ex-situ stars clearly separate in age-metallicity space, while the distinction becomes less clear for star forming galaxies outside that mass range and quenched galaxies in general. (Figure \ref{fig: agemet})
    \item For both disk and bulge dominated galaxies between $10^{9.5-10}\,\mathrm{M}_{\odot}$ in stellar mass, in-situ, migrated and ex-situ stars clearly separate in age-circularity space. The migrated stars are the youngest with the highest amount of rotational support and the ex-situ stars are the oldest and purely random motion supported, whereas the in-situ stars are situated in between. (Figure \ref{fig: agecirc})
\end{itemize}

Furthermore, we have demonstrated the diversity of the central 500\,pc of galaxies as governed by the hierarchical mass build-up in a $\Lambda$CDM universe. Galaxy interactions are an important driver in not only contributing ex-situ stars to the center of galaxies, but also in dictating the formation of in-situ stars and the migration of stars to the center. This leads to an entanglement of different mechanisms that influence the formation history of stars in the center of galaxies. In Figure \ref{fig: assembly} we have qualitatively identified these mechanisms that are present in TNG50, which includes episodes of in-situ star formation and stellar migration to the center during times of pericenter passages and/or coalescence of mergers or flybys, infall into galaxy groups/clusters as well as depletion of the central gas reservoir through kinetic AGN feedback and environmental effects. \par
In the future, higher resolution simulations (not only spatially but also concerning star formation and stellar feedback prescriptions) will be needed to fully address the formation and migration of stellar clumps and to study the formation of nuclear galaxy structures, such as nuclear disks and star clusters, in a fully cosmological context. \par
Bright galaxy centers have the potential to be used in observations as tracers of the overall galaxy assembly history. TNG50 predicts distinct stellar populations and dynamical properties for the stars of different origins in the center of galaxies, which can be observed with today's IFU capabilities. Figure \ref{fig: obs} demonstrates that there is even promise to deduce the fractional contribution of central in-situ, migrated and ex-situ stars from SDSS-like observations in a galaxy population averaged sense.\par
In summary, TNG50 is a tremendous advancement in predicting the stellar build-up of sub-kpc scales in a fully cosmological context. Its predictive power is valuable to consider new pathways in modelling formation scenarios of central stellar components as well as to push forward novel observational techniques to unveil the formation history of galaxies.

\section*{Data availability}

The IllustrisTNG simulations, including TNG50, are publicly available and accessible at \url{www.tng-project.org/data} \citep[][]{publictng}. Data directly related to this publication and its figures are available upon reasonable request from the corresponding author.

\section*{Acknowledgements}

AB is grateful to Ignacio Mart\'in-Navarro and Jes\'us Falc\'on-Barroso for their support and scientific discussion throughout this project. AB also likes to thank Glenn van de Ven and Francisco Aros for useful discussions. We thank the anonymous referee for helpful comments on the manuscript. This work was funded by the Deutsche Forschungsgemeinschaft (DFG, German Research Foundation) -- Project-ID 138713538 -- SFB 881 (``The Milky Way System'', subproject B08). NF was supported by the Natural Sciences and Engineering Research Council of Canada (NSERC), [funding reference number CITA 490888-16] through the CITA postdoctoral fellowship and acknowledges partial support from a Arts \& Sciences Postdoctoral Fellowship at the University of Toronto. RR acknowledges funding from the Deutsche Forschungsgemeinschaft (DFG) through an Emmy Noether Research Group (grant number NE 2441/1-1). The IllustrisTNG simulations were undertaken with compute time awarded by the Gauss Centre for Supercomputing (GCS) under GCS Large-Scale Projects GCS-ILLU and GCS-DWAR on the GCS share of the supercomputer Hazel Hen at the High Performance Computing Center Stuttgart (HLRS), as well as on the machines of the Max Planck Computing and Data Facility (MPCDF) in Garching, Germany. The computations for this work were performed on the ISAAC cluster of the Max Planck Institute for Astronomy at the Rechenzentrum in Garching.





\begin{thebibliography}{}
\makeatletter
\relax
\def\mn@urlcharsother{\let\do\@makeother \do\$\do\&\do\#\do\^\do\_\do\%\do\~}
\def\mn@doi{\begingroup\mn@urlcharsother \@ifnextchar [ {\mn@doi@}
  {\mn@doi@[]}}
\def\mn@doi@[#1]#2{\def\@tempa{#1}\ifx\@tempa\@empty \href
  {http://dx.doi.org/#2} {doi:#2}\else \href {http://dx.doi.org/#2} {#1}\fi
  \endgroup}
\def\mn@eprint#1#2{\mn@eprint@#1:#2::\@nil}
\def\mn@eprint@arXiv#1{\href {http://arxiv.org/abs/#1} {{\tt arXiv:#1}}}
\def\mn@eprint@dblp#1{\href {http://dblp.uni-trier.de/rec/bibtex/#1.xml}
  {dblp:#1}}
\def\mn@eprint@#1:#2:#3:#4\@nil{\def\@tempa {#1}\def\@tempb {#2}\def\@tempc
  {#3}\ifx \@tempc \@empty \let \@tempc \@tempb \let \@tempb \@tempa \fi \ifx
  \@tempb \@empty \def\@tempb {arXiv}\fi \@ifundefined
  {mn@eprint@\@tempb}{\@tempb:\@tempc}{\expandafter \expandafter \csname
  mn@eprint@\@tempb\endcsname \expandafter{\@tempc}}}

\bibitem[\protect\citeauthoryear{{Abadi}, {Navarro}, {Steinmetz}  \&
  {Eke}}{{Abadi} et~al.}{2003}]{abadi03}
{Abadi} M.~G.,  {Navarro} J.~F.,  {Steinmetz} M.,   {Eke} V.~R.,  2003, \mn@doi
  [\apj] {10.1086/378316}, \href
  {https://ui.adsabs.harvard.edu/abs/2003ApJ...597...21A} {597, 21}

\bibitem[\protect\citeauthoryear{{Abazajian} et~al.,}{{Abazajian}
  et~al.}{2009}]{sdss}
{Abazajian} K.~N.,  et~al., 2009, \mn@doi [\apjs]
  {10.1088/0067-0049/182/2/543}, \href
  {https://ui.adsabs.harvard.edu/abs/2009ApJS..182..543A} {182, 543}

\bibitem[\protect\citeauthoryear{{Agarwal} \& {Milosavljevi{\'c}}}{{Agarwal} \&
  {Milosavljevi{\'c}}}{2011}]{agarwal11}
{Agarwal} M.,  {Milosavljevi{\'c}} M.,  2011, \mn@doi [\apj]
  {10.1088/0004-637X/729/1/35}, \href
  {https://ui.adsabs.harvard.edu/abs/2011ApJ...729...35A} {729, 35}

\bibitem[\protect\citeauthoryear{{Aharon} \& {Perets}}{{Aharon} \&
  {Perets}}{2015}]{aharon15}
{Aharon} D.,  {Perets} H.~B.,  2015, \mn@doi [\apj]
  {10.1088/0004-637X/799/2/185}, \href
  {https://ui.adsabs.harvard.edu/abs/2015ApJ...799..185A} {799, 185}

\bibitem[\protect\citeauthoryear{{Antonini}}{{Antonini}}{2013}]{antonini13}
{Antonini} F.,  2013, \mn@doi [\apj] {10.1088/0004-637X/763/1/62}, \href
  {https://ui.adsabs.harvard.edu/abs/2013ApJ...763...62A} {763, 62}

\bibitem[\protect\citeauthoryear{{Antonini}, {Capuzzo-Dolcetta},
  {Mastrobuono-Battisti}  \& {Merritt}}{{Antonini} et~al.}{2012}]{antonini12}
{Antonini} F.,  {Capuzzo-Dolcetta} R.,  {Mastrobuono-Battisti} A.,   {Merritt}
  D.,  2012, \mn@doi [\apj] {10.1088/0004-637X/750/2/111}, \href
  {https://ui.adsabs.harvard.edu/abs/2012ApJ...750..111A} {750, 111}

\bibitem[\protect\citeauthoryear{{Antonini}, {Barausse}  \& {Silk}}{{Antonini}
  et~al.}{2015}]{antonini15}
{Antonini} F.,  {Barausse} E.,   {Silk} J.,  2015, \mn@doi [\apj]
  {10.1088/0004-637X/812/1/72}, \href
  {https://ui.adsabs.harvard.edu/abs/2015ApJ...812...72A} {812, 72}

\bibitem[\protect\citeauthoryear{{Arentsen} et~al.,}{{Arentsen}
  et~al.}{2020}]{arentsen20}
{Arentsen} A.,  et~al., 2020, \mn@doi [\mnras] {10.1093/mnrasl/slz156}, \href
  {https://ui.adsabs.harvard.edu/abs/2020MNRAS.491L..11A} {491, L11}

\bibitem[\protect\citeauthoryear{{Arnold} et~al.,}{{Arnold}
  et~al.}{2014}]{arnold14}
{Arnold} J.~A.,  et~al., 2014, \mn@doi [\apj] {10.1088/0004-637X/791/2/80},
  \href {https://ui.adsabs.harvard.edu/abs/2014ApJ...791...80A} {791, 80}

\bibitem[\protect\citeauthoryear{{Asplund}, {Grevesse}, {Sauval}  \&
  {Scott}}{{Asplund} et~al.}{2009}]{asplund}
{Asplund} M.,  {Grevesse} N.,  {Sauval} A.~J.,   {Scott} P.,  2009, \mn@doi
  [\araa] {10.1146/annurev.astro.46.060407.145222}, \href
  {https://ui.adsabs.harvard.edu/abs/2009ARA&A..47..481A} {47, 481}

\bibitem[\protect\citeauthoryear{{Athanassoula}}{{Athanassoula}}{2005}]{athanassoula05}
{Athanassoula} E.,  2005, \mn@doi [\mnras] {10.1111/j.1365-2966.2005.08872.x},
  \href {https://ui.adsabs.harvard.edu/abs/2005MNRAS.358.1477A} {358, 1477}

\bibitem[\protect\citeauthoryear{{Athanassoula}, {Machado}  \&
  {Rodionov}}{{Athanassoula} et~al.}{2013}]{athanassoula13}
{Athanassoula} E.,  {Machado} R. E.~G.,   {Rodionov} S.~A.,  2013, \mn@doi
  [\mnras] {10.1093/mnras/sts452}, \href
  {https://ui.adsabs.harvard.edu/abs/2013MNRAS.429.1949A} {429, 1949}

\bibitem[\protect\citeauthoryear{{Barbuy}, {Chiappini}  \& {Gerhard}}{{Barbuy}
  et~al.}{2018}]{barbuy18}
{Barbuy} B.,  {Chiappini} C.,   {Gerhard} O.,  2018, \mn@doi [\araa]
  {10.1146/annurev-astro-081817-051826}, \href
  {https://ui.adsabs.harvard.edu/abs/2018ARA&A..56..223B} {56, 223}

\bibitem[\protect\citeauthoryear{{Barnes}}{{Barnes}}{1988}]{barnes88}
{Barnes} J.~E.,  1988, \mn@doi [\apj] {10.1086/166593}, \href
  {https://ui.adsabs.harvard.edu/abs/1988ApJ...331..699B} {331, 699}

\bibitem[\protect\citeauthoryear{{Bittner} et~al.,}{{Bittner}
  et~al.}{2020}]{bittner20}
{Bittner} A.,  et~al., 2020, \mn@doi [\aap] {10.1051/0004-6361/202038450},
  \href {https://ui.adsabs.harvard.edu/abs/2020A&A...643A..65B} {643, A65}

\bibitem[\protect\citeauthoryear{{Bocquet}, {Saro}, {Dolag}  \&
  {Mohr}}{{Bocquet} et~al.}{2016}]{bocquet16}
{Bocquet} S.,  {Saro} A.,  {Dolag} K.,   {Mohr} J.~J.,  2016, \mn@doi [\mnras]
  {10.1093/mnras/stv2657}, \href
  {https://ui.adsabs.harvard.edu/abs/2016MNRAS.456.2361B} {456, 2361}

\bibitem[\protect\citeauthoryear{{Boecker}, {Leaman}, {van de Ven}, {Norris},
  {Mackereth}  \& {Crain}}{{Boecker} et~al.}{2020a}]{boecker19}
{Boecker} A.,  {Leaman} R.,  {van de Ven} G.,  {Norris} M.~A.,  {Mackereth}
  J.~T.,   {Crain} R.~A.,  2020a, \mn@doi [\mnras] {10.1093/mnras/stz3077},
  \href {https://ui.adsabs.harvard.edu/abs/2020MNRAS.491..823B} {491, 823}

\bibitem[\protect\citeauthoryear{{Boecker}, {Alfaro-Cuello}, {Neumayer},
  {Mart{\'\i}n-Navarro}  \& {Leaman}}{{Boecker} et~al.}{2020b}]{boecker20a}
{Boecker} A.,  {Alfaro-Cuello} M.,  {Neumayer} N.,  {Mart{\'\i}n-Navarro} I.,
  {Leaman} R.,  2020b, \mn@doi [\apj] {10.3847/1538-4357/ab919d}, \href
  {https://ui.adsabs.harvard.edu/abs/2020ApJ...896...13B} {896, 13}

\bibitem[\protect\citeauthoryear{{B{\"o}ker}, {Sarzi}, {McLaughlin}, {van der
  Marel}, {Rix}, {Ho}  \& {Shields}}{{B{\"o}ker} et~al.}{2004}]{boker04}
{B{\"o}ker} T.,  {Sarzi} M.,  {McLaughlin} D.~E.,  {van der Marel} R.~P.,
  {Rix} H.-W.,  {Ho} L.~C.,   {Shields} J.~C.,  2004, \mn@doi [\aj]
  {10.1086/380231}, \href
  {https://ui.adsabs.harvard.edu/abs/2004AJ....127..105B} {127, 105}

\bibitem[\protect\citeauthoryear{{Bottema}}{{Bottema}}{2003}]{bottema03}
{Bottema} R.,  2003, \mn@doi [\mnras] {10.1046/j.1365-8711.2003.06613.x}, \href
  {https://ui.adsabs.harvard.edu/abs/2003MNRAS.344..358B} {344, 358}

\bibitem[\protect\citeauthoryear{{Bournaud}, {Elmegreen}  \&
  {Elmegreen}}{{Bournaud} et~al.}{2007}]{bournaud07}
{Bournaud} F.,  {Elmegreen} B.~G.,   {Elmegreen} D.~M.,  2007, \mn@doi [\apj]
  {10.1086/522077}, \href
  {https://ui.adsabs.harvard.edu/abs/2007ApJ...670..237B} {670, 237}

\bibitem[\protect\citeauthoryear{{Bournaud} et~al.,}{{Bournaud}
  et~al.}{2014}]{bournaud14}
{Bournaud} F.,  et~al., 2014, \mn@doi [\apj] {10.1088/0004-637X/780/1/57},
  \href {https://ui.adsabs.harvard.edu/abs/2014ApJ...780...57B} {780, 57}

\bibitem[\protect\citeauthoryear{{Brown}, {Gnedin}  \& {Li}}{{Brown}
  et~al.}{2018}]{brown18}
{Brown} G.,  {Gnedin} O.~Y.,   {Li} H.,  2018, \mn@doi [\apj]
  {10.3847/1538-4357/aad595}, \href
  {https://ui.adsabs.harvard.edu/abs/2018ApJ...864...94B} {864, 94}

\bibitem[\protect\citeauthoryear{{Bryant} et~al.,}{{Bryant}
  et~al.}{2015}]{sami}
{Bryant} J.~J.,  et~al., 2015, \mn@doi [\mnras] {10.1093/mnras/stu2635}, \href
  {https://ui.adsabs.harvard.edu/abs/2015MNRAS.447.2857B} {447, 2857}

\bibitem[\protect\citeauthoryear{{Buck}, {Macci{\`o}}, {Obreja}, {Dutton},
  {Dom{\'\i}nguez-Tenreiro}  \& {Granato}}{{Buck} et~al.}{2017}]{buck17}
{Buck} T.,  {Macci{\`o}} A.~V.,  {Obreja} A.,  {Dutton} A.~A.,
  {Dom{\'\i}nguez-Tenreiro} R.,   {Granato} G.~L.,  2017, \mn@doi [\mnras]
  {10.1093/mnras/stx685}, \href
  {https://ui.adsabs.harvard.edu/abs/2017MNRAS.468.3628B} {468, 3628}

\bibitem[\protect\citeauthoryear{{Buck}, {Obreja}, {Macci{\`o}}, {Minchev},
  {Dutton}  \& {Ostriker}}{{Buck} et~al.}{2020}]{buck20}
{Buck} T.,  {Obreja} A.,  {Macci{\`o}} A.~V.,  {Minchev} I.,  {Dutton} A.~A.,
  {Ostriker} J.~P.,  2020, \mn@doi [\mnras] {10.1093/mnras/stz3241}, \href
  {https://ui.adsabs.harvard.edu/abs/2020MNRAS.491.3461B} {491, 3461}

\bibitem[\protect\citeauthoryear{{Bundy} et~al.,}{{Bundy}
  et~al.}{2015}]{bundy15}
{Bundy} K.,  et~al., 2015, \mn@doi [\apj] {10.1088/0004-637X/798/1/7}, \href
  {https://ui.adsabs.harvard.edu/abs/2015ApJ...798....7B} {798, 7}

\bibitem[\protect\citeauthoryear{{Calabr{\`o}} et~al.,}{{Calabr{\`o}}
  et~al.}{2019}]{calabro19}
{Calabr{\`o}} A.,  et~al., 2019, \mn@doi [\aap] {10.1051/0004-6361/201935778},
  \href {https://ui.adsabs.harvard.edu/abs/2019A&A...632A..98C} {632, A98}

\bibitem[\protect\citeauthoryear{{Cappellari} et~al.,}{{Cappellari}
  et~al.}{2013}]{loess}
{Cappellari} M.,  et~al., 2013, \mn@doi [\mnras] {10.1093/mnras/stt644}, \href
  {https://ui.adsabs.harvard.edu/abs/2013MNRAS.432.1862C} {432, 1862}

\bibitem[\protect\citeauthoryear{{Chabrier}}{{Chabrier}}{2003}]{chabrier03}
{Chabrier} G.,  2003, \mn@doi [\pasp] {10.1086/376392}, \href
  {https://ui.adsabs.harvard.edu/abs/2003PASP..115..763C} {115, 763}

\bibitem[\protect\citeauthoryear{{Chen} et~al.,}{{Chen} et~al.}{2020}]{chen20}
{Chen} Z.,  et~al., 2020, \mn@doi [\apj] {10.3847/1538-4357/ab9633}, \href
  {https://ui.adsabs.harvard.edu/abs/2020ApJ...897..102C} {897, 102}

\bibitem[\protect\citeauthoryear{{C{\^o}t{\'e}} et~al.,}{{C{\^o}t{\'e}}
  et~al.}{2006}]{cote06}
{C{\^o}t{\'e}} P.,  et~al., 2006, \mn@doi [\apjs] {10.1086/504042}, \href
  {https://ui.adsabs.harvard.edu/abs/2006ApJS..165...57C} {165, 57}

\bibitem[\protect\citeauthoryear{{Cox}, {Jonsson}, {Somerville}, {Primack}  \&
  {Dekel}}{{Cox} et~al.}{2008}]{cox08}
{Cox} T.~J.,  {Jonsson} P.,  {Somerville} R.~S.,  {Primack} J.~R.,   {Dekel}
  A.,  2008, \mn@doi [\mnras] {10.1111/j.1365-2966.2007.12730.x}, \href
  {https://ui.adsabs.harvard.edu/abs/2008MNRAS.384..386C} {384, 386}

\bibitem[\protect\citeauthoryear{{Crain} et~al.,}{{Crain}
  et~al.}{2015}]{crain15}
{Crain} R.~A.,  et~al., 2015, \mn@doi [\mnras] {10.1093/mnras/stv725}, \href
  {https://ui.adsabs.harvard.edu/abs/2015MNRAS.450.1937C} {450, 1937}

\bibitem[\protect\citeauthoryear{{Croton} et~al.,}{{Croton}
  et~al.}{2006}]{croton06}
{Croton} D.~J.,  et~al., 2006, \mn@doi [\mnras]
  {10.1111/j.1365-2966.2005.09675.x}, \href
  {https://ui.adsabs.harvard.edu/abs/2006MNRAS.365...11C} {365, 11}

\bibitem[\protect\citeauthoryear{{D'Souza}, {Kauffman}, {Wang}  \&
  {Vegetti}}{{D'Souza} et~al.}{2014}]{dsouza14}
{D'Souza} R.,  {Kauffman} G.,  {Wang} J.,   {Vegetti} S.,  2014, \mn@doi
  [\mnras] {10.1093/mnras/stu1194}, \href
  {https://ui.adsabs.harvard.edu/abs/2014MNRAS.443.1433D} {443, 1433}

\bibitem[\protect\citeauthoryear{{Dav{\'e}}, {Angl{\'e}s-Alc{\'a}zar},
  {Narayanan}, {Li}, {Rafieferantsoa}  \& {Appleby}}{{Dav{\'e}}
  et~al.}{2019}]{dave19}
{Dav{\'e}} R.,  {Angl{\'e}s-Alc{\'a}zar} D.,  {Narayanan} D.,  {Li} Q.,
  {Rafieferantsoa} M.~H.,   {Appleby} S.,  2019, \mn@doi [\mnras]
  {10.1093/mnras/stz937}, \href
  {https://ui.adsabs.harvard.edu/abs/2019MNRAS.486.2827D} {486, 2827}

\bibitem[\protect\citeauthoryear{{Davis}, {Efstathiou}, {Frenk}  \&
  {White}}{{Davis} et~al.}{1985}]{fof}
{Davis} M.,  {Efstathiou} G.,  {Frenk} C.~S.,   {White} S.~D.~M.,  1985,
  \mn@doi [\apj] {10.1086/163168}, \href
  {http://adsabs.harvard.edu/abs/1985ApJ...292..371D} {292, 371}

\bibitem[\protect\citeauthoryear{{Davis}, {Graham}  \& {Cameron}}{{Davis}
  et~al.}{2019}]{davis19}
{Davis} B.~L.,  {Graham} A.~W.,   {Cameron} E.,  2019, \mn@doi [\apj]
  {10.3847/1538-4357/aaf3b8}, \href
  {https://ui.adsabs.harvard.edu/abs/2019ApJ...873...85D} {873, 85}

\bibitem[\protect\citeauthoryear{{Davison}, {Norris}, {Pfeffer}, {Davies}  \&
  {Crain}}{{Davison} et~al.}{2020}]{davison20}
{Davison} T.~A.,  {Norris} M.~A.,  {Pfeffer} J.~L.,  {Davies} J.~J.,   {Crain}
  R.~A.,  2020, \mn@doi [\mnras] {10.1093/mnras/staa1816}, \href
  {https://ui.adsabs.harvard.edu/abs/2020MNRAS.497...81D} {497, 81}

\bibitem[\protect\citeauthoryear{{Davison}, {Norris}, {Leaman}, {Kuntschner},
  {Boecker}  \& {van de Ven}}{{Davison} et~al.}{2021}]{davison21}
{Davison} T.~A.,  {Norris} M.~A.,  {Leaman} R.,  {Kuntschner} H.,  {Boecker}
  A.,   {van de Ven} G.,  2021, \mn@doi [\mnras] {10.1093/mnras/stab2362},
  \href {https://ui.adsabs.harvard.edu/abs/2021MNRAS.507.3089D} {507, 3089}

\bibitem[\protect\citeauthoryear{{Dekel} \& {Krumholz}}{{Dekel} \&
  {Krumholz}}{2013}]{dekel13}
{Dekel} A.,  {Krumholz} M.~R.,  2013, \mn@doi [\mnras] {10.1093/mnras/stt480},
  \href {https://ui.adsabs.harvard.edu/abs/2013MNRAS.432..455D} {432, 455}

\bibitem[\protect\citeauthoryear{{Dekel}, {Sari}  \& {Ceverino}}{{Dekel}
  et~al.}{2009}]{dekel09}
{Dekel} A.,  {Sari} R.,   {Ceverino} D.,  2009, \mn@doi [\apj]
  {10.1088/0004-637X/703/1/785}, \href
  {https://ui.adsabs.harvard.edu/abs/2009ApJ...703..785D} {703, 785}

\bibitem[\protect\citeauthoryear{{Dekel}, {Mandelker}, {Bournaud}, {Ceverino},
  {Guo}  \& {primack}}{{Dekel} et~al.}{2021}]{dekel21}
{Dekel} A.,  {Mandelker} N.,  {Bournaud} F.,  {Ceverino} D.,  {Guo} Y.,
  {primack} J.,  2021, arXiv e-prints, \href
  {https://ui.adsabs.harvard.edu/abs/2021arXiv210713561D} {p. arXiv:2107.13561}

\bibitem[\protect\citeauthoryear{{Di Matteo}, {Combes}, {Melchior}  \&
  {Semelin}}{{Di Matteo} et~al.}{2007}]{dimatteo07}
{Di Matteo} P.,  {Combes} F.,  {Melchior} A.~L.,   {Semelin} B.,  2007, \mn@doi
  [\aap] {10.1051/0004-6361:20066959}, \href
  {https://ui.adsabs.harvard.edu/abs/2007A&A...468...61D} {468, 61}

\bibitem[\protect\citeauthoryear{{Di Matteo}, {Bournaud}, {Martig}, {Combes},
  {Melchior}  \& {Semelin}}{{Di Matteo} et~al.}{2008}]{dimatteo08}
{Di Matteo} P.,  {Bournaud} F.,  {Martig} M.,  {Combes} F.,  {Melchior} A.~L.,
   {Semelin} B.,  2008, \mn@doi [\aap] {10.1051/0004-6361:200809480}, \href
  {https://ui.adsabs.harvard.edu/abs/2008A&A...492...31D} {492, 31}

\bibitem[\protect\citeauthoryear{{D{\'\i}az-Garc{\'\i}a}, {Salo}, {Laurikainen}
   \& {Herrera-Endoqui}}{{D{\'\i}az-Garc{\'\i}a} et~al.}{2016a}]{bars3}
{D{\'\i}az-Garc{\'\i}a} S.,  {Salo} H.,  {Laurikainen} E.,   {Herrera-Endoqui}
  M.,  2016a, \mn@doi [\aap] {10.1051/0004-6361/201526161}, \href
  {https://ui.adsabs.harvard.edu/abs/2016A&A...587A.160D} {587, A160}

\bibitem[\protect\citeauthoryear{{D{\'\i}az-Garc{\'\i}a}, {Salo}  \&
  {Laurikainen}}{{D{\'\i}az-Garc{\'\i}a} et~al.}{2016b}]{barconcen}
{D{\'\i}az-Garc{\'\i}a} S.,  {Salo} H.,   {Laurikainen} E.,  2016b, \mn@doi
  [\aap] {10.1051/0004-6361/201628683}, \href
  {https://ui.adsabs.harvard.edu/abs/2016A&A...596A..84D} {596, A84}

\bibitem[\protect\citeauthoryear{{Do}, {David Martinez}, {Kerzendorf},
  {Feldmeier-Krause}, {Arca Sedda}, {Neumayer}  \& {Gualandris}}{{Do}
  et~al.}{2020}]{do20}
{Do} T.,  {David Martinez} G.,  {Kerzendorf} W.,  {Feldmeier-Krause} A.,  {Arca
  Sedda} M.,  {Neumayer} N.,   {Gualandris} A.,  2020, \mn@doi [\apjl]
  {10.3847/2041-8213/abb246}, \href
  {https://ui.adsabs.harvard.edu/abs/2020ApJ...901L..28D} {901, L28}

\bibitem[\protect\citeauthoryear{{Donnari} et~al.,}{{Donnari}
  et~al.}{2019}]{donnari19}
{Donnari} M.,  et~al., 2019, \mn@doi [\mnras] {10.1093/mnras/stz712}, \href
  {https://ui.adsabs.harvard.edu/abs/2019MNRAS.485.4817D} {485, 4817}

\bibitem[\protect\citeauthoryear{{Donnari} et~al.,}{{Donnari}
  et~al.}{2021a}]{donnari21a}
{Donnari} M.,  et~al., 2021a, \mn@doi [\mnras] {10.1093/mnras/staa3006}, \href
  {https://ui.adsabs.harvard.edu/abs/2021MNRAS.500.4004D} {500, 4004}

\bibitem[\protect\citeauthoryear{{Donnari}, {Pillepich}, {Nelson}, {Marinacci},
  {Vogelsberger}  \& {Hernquist}}{{Donnari} et~al.}{2021b}]{donnari21b}
{Donnari} M.,  {Pillepich} A.,  {Nelson} D.,  {Marinacci} F.,  {Vogelsberger}
  M.,   {Hernquist} L.,  2021b, \mn@doi [\mnras] {10.1093/mnras/stab1950},
  \href {https://ui.adsabs.harvard.edu/abs/2021MNRAS.506.4760D} {506, 4760}

\bibitem[\protect\citeauthoryear{{Dubois} et~al.,}{{Dubois}
  et~al.}{2014}]{dubois14}
{Dubois} Y.,  et~al., 2014, \mn@doi [\mnras] {10.1093/mnras/stu1227}, \href
  {https://ui.adsabs.harvard.edu/abs/2014MNRAS.444.1453D} {444, 1453}

\bibitem[\protect\citeauthoryear{{Dubois}, {Peirani}, {Pichon}, {Devriendt},
  {Gavazzi}, {Welker}  \& {Volonteri}}{{Dubois} et~al.}{2016}]{dubois16}
{Dubois} Y.,  {Peirani} S.,  {Pichon} C.,  {Devriendt} J.,  {Gavazzi} R.,
  {Welker} C.,   {Volonteri} M.,  2016, \mn@doi [\mnras]
  {10.1093/mnras/stw2265}, \href
  {https://ui.adsabs.harvard.edu/abs/2016MNRAS.463.3948D} {463, 3948}

\bibitem[\protect\citeauthoryear{{El-Badry} et~al.,}{{El-Badry}
  et~al.}{2018}]{elbadry18}
{El-Badry} K.,  et~al., 2018, \mn@doi [\mnras] {10.1093/mnras/sty1864}, \href
  {https://ui.adsabs.harvard.edu/abs/2018MNRAS.480..652E} {480, 652}

\bibitem[\protect\citeauthoryear{{Elmegreen}, {Elmegreen}, {Ravindranath}  \&
  {Coe}}{{Elmegreen} et~al.}{2007}]{elmegreen07}
{Elmegreen} D.~M.,  {Elmegreen} B.~G.,  {Ravindranath} S.,   {Coe} D.~A.,
  2007, \mn@doi [\apj] {10.1086/511667}, \href
  {https://ui.adsabs.harvard.edu/abs/2007ApJ...658..763E} {658, 763}

\bibitem[\protect\citeauthoryear{{Elmegreen}, {Elmegreen}, {Marcus},
  {Shahinyan}, {Yau}  \& {Petersen}}{{Elmegreen} et~al.}{2009}]{elmegreen09}
{Elmegreen} D.~M.,  {Elmegreen} B.~G.,  {Marcus} M.~T.,  {Shahinyan} K.,  {Yau}
  A.,   {Petersen} M.,  2009, \mn@doi [\apj] {10.1088/0004-637X/701/1/306},
  \href {https://ui.adsabs.harvard.edu/abs/2009ApJ...701..306E} {701, 306}

\bibitem[\protect\citeauthoryear{{Fanidakis}, {Baugh}, {Benson}, {Bower},
  {Cole}, {Done}  \& {Frenk}}{{Fanidakis} et~al.}{2011}]{fanidakis2011}
{Fanidakis} N.,  {Baugh} C.~M.,  {Benson} A.~J.,  {Bower} R.~G.,  {Cole} S.,
  {Done} C.,   {Frenk} C.~S.,  2011, \mn@doi [\mnras]
  {10.1111/j.1365-2966.2010.17427.x}, \href
  {https://ui.adsabs.harvard.edu/abs/2011MNRAS.410...53F} {410, 53}

\bibitem[\protect\citeauthoryear{{Feldmeier-Krause} et~al.,}{{Feldmeier-Krause}
  et~al.}{2020}]{fk20}
{Feldmeier-Krause} A.,  et~al., 2020, \mn@doi [\mnras] {10.1093/mnras/staa703},
  \href {https://ui.adsabs.harvard.edu/abs/2020MNRAS.494..396F} {494, 396}

\bibitem[\protect\citeauthoryear{{Ferrarese} et~al.,}{{Ferrarese}
  et~al.}{2006}]{ferrarese06}
{Ferrarese} L.,  et~al., 2006, \mn@doi [\apjl] {10.1086/505388}, \href
  {https://ui.adsabs.harvard.edu/abs/2006ApJ...644L..21F} {644, L21}

\bibitem[\protect\citeauthoryear{{Fisher} \& {Drory}}{{Fisher} \&
  {Drory}}{2010}]{fisherdrory}
{Fisher} D.~B.,  {Drory} N.,  2010, \mn@doi [\apj]
  {10.1088/0004-637X/716/2/942}, \href
  {https://ui.adsabs.harvard.edu/abs/2010ApJ...716..942F} {716, 942}

\bibitem[\protect\citeauthoryear{{Frankel}, {Sanders}, {Ting}  \&
  {Rix}}{{Frankel} et~al.}{2020}]{frankel20}
{Frankel} N.,  {Sanders} J.,  {Ting} Y.-S.,   {Rix} H.-W.,  2020, \mn@doi
  [\apj] {10.3847/1538-4357/ab910c}, \href
  {https://ui.adsabs.harvard.edu/abs/2020ApJ...896...15F} {896, 15}

\bibitem[\protect\citeauthoryear{{Frankel} et~al.,}{{Frankel}
  et~al.}{2022}]{frankel22}
{Frankel} N.,  et~al., 2022, arXiv e-prints, \href
  {https://ui.adsabs.harvard.edu/abs/2022arXiv220108406F} {p. arXiv:2201.08406}

\bibitem[\protect\citeauthoryear{{Gadotti}}{{Gadotti}}{2009}]{gadotti09a}
{Gadotti} D.~A.,  2009, \mn@doi [\mnras] {10.1111/j.1365-2966.2008.14257.x},
  \href {https://ui.adsabs.harvard.edu/abs/2009MNRAS.393.1531G} {393, 1531}

\bibitem[\protect\citeauthoryear{{Gadotti} \& {Kauffmann}}{{Gadotti} \&
  {Kauffmann}}{2009}]{gadotti09b}
{Gadotti} D.~A.,  {Kauffmann} G.,  2009, \mn@doi [\mnras]
  {10.1111/j.1365-2966.2009.15328.x}, \href
  {https://ui.adsabs.harvard.edu/abs/2009MNRAS.399..621G} {399, 621}

\bibitem[\protect\citeauthoryear{{Gadotti} et~al.,}{{Gadotti}
  et~al.}{2020}]{gadotti20}
{Gadotti} D.~A.,  et~al., 2020, \mn@doi [\aap] {10.1051/0004-6361/202038448},
  \href {https://ui.adsabs.harvard.edu/abs/2020A&A...643A..14G} {643, A14}

\bibitem[\protect\citeauthoryear{{Gallazzi}, {Charlot}, {Brinchmann}, {White}
  \& {Tremonti}}{{Gallazzi} et~al.}{2005}]{gallazzi05}
{Gallazzi} A.,  {Charlot} S.,  {Brinchmann} J.,  {White} S. D.~M.,   {Tremonti}
  C.~A.,  2005, \mn@doi [\mnras] {10.1111/j.1365-2966.2005.09321.x}, \href
  {https://ui.adsabs.harvard.edu/abs/2005MNRAS.362...41G} {362, 41}

\bibitem[\protect\citeauthoryear{{Gallazzi}, {Pasquali}, {Zibetti}  \&
  {Barbera}}{{Gallazzi} et~al.}{2021}]{gallazzi20}
{Gallazzi} A.~R.,  {Pasquali} A.,  {Zibetti} S.,   {Barbera} F.~L.,  2021,
  \mn@doi [\mnras] {10.1093/mnras/stab265}, \href
  {https://ui.adsabs.harvard.edu/abs/2021MNRAS.502.4457G} {502, 4457}

\bibitem[\protect\citeauthoryear{{Gao}, {Loeb}, {Peebles}, {White}  \&
  {Jenkins}}{{Gao} et~al.}{2004}]{gao04}
{Gao} L.,  {Loeb} A.,  {Peebles} P.~J.~E.,  {White} S. D.~M.,   {Jenkins} A.,
  2004, \mn@doi [\apj] {10.1086/423444}, \href
  {https://ui.adsabs.harvard.edu/abs/2004ApJ...614...17G} {614, 17}

\bibitem[\protect\citeauthoryear{{Gargiulo} et~al.,}{{Gargiulo}
  et~al.}{2021}]{gargiulo21}
{Gargiulo} I.~D.,  et~al., 2021, arXiv e-prints, \href
  {https://ui.adsabs.harvard.edu/abs/2021arXiv211113712G} {p. arXiv:2111.13712}

\bibitem[\protect\citeauthoryear{{Genel} et~al.,}{{Genel}
  et~al.}{2012}]{genel12}
{Genel} S.,  et~al., 2012, \mn@doi [\apj] {10.1088/0004-637X/745/1/11}, \href
  {https://ui.adsabs.harvard.edu/abs/2012ApJ...745...11G} {745, 11}

\bibitem[\protect\citeauthoryear{{Genel} et~al.,}{{Genel}
  et~al.}{2014}]{genel14}
{Genel} S.,  et~al., 2014, \mn@doi [\mnras] {10.1093/mnras/stu1654}, \href
  {https://ui.adsabs.harvard.edu/abs/2014MNRAS.445..175G} {445, 175}

\bibitem[\protect\citeauthoryear{{Genel} et~al.,}{{Genel}
  et~al.}{2018}]{genel18}
{Genel} S.,  et~al., 2018, \mn@doi [\mnras] {10.1093/mnras/stx3078}, \href
  {https://ui.adsabs.harvard.edu/abs/2018MNRAS.474.3976G} {474, 3976}

\bibitem[\protect\citeauthoryear{{Georgiev} \& {B{\"o}ker}}{{Georgiev} \&
  {B{\"o}ker}}{2014}]{georgiev14}
{Georgiev} I.~Y.,  {B{\"o}ker} T.,  2014, \mn@doi [\mnras]
  {10.1093/mnras/stu797}, \href
  {https://ui.adsabs.harvard.edu/abs/2014MNRAS.441.3570G} {441, 3570}

\bibitem[\protect\citeauthoryear{{Georgiev}, {B{\"o}ker}, {Leigh},
  {L{\"u}tzgendorf}  \& {Neumayer}}{{Georgiev} et~al.}{2016}]{georgiev16}
{Georgiev} I.~Y.,  {B{\"o}ker} T.,  {Leigh} N.,  {L{\"u}tzgendorf} N.,
  {Neumayer} N.,  2016, \mn@doi [\mnras] {10.1093/mnras/stw093}, \href
  {https://ui.adsabs.harvard.edu/abs/2016MNRAS.457.2122G} {457, 2122}

\bibitem[\protect\citeauthoryear{{Grand} et~al.,}{{Grand}
  et~al.}{2017}]{grand17}
{Grand} R. J.~J.,  et~al., 2017, \mn@doi [\mnras] {10.1093/mnras/stx071}, \href
  {https://ui.adsabs.harvard.edu/abs/2017MNRAS.467..179G} {467, 179}

\bibitem[\protect\citeauthoryear{{Grand} et~al.,}{{Grand}
  et~al.}{2021}]{grand21}
{Grand} R. J.~J.,  et~al., 2021, \mn@doi [\mnras] {10.1093/mnras/stab2492},
  \href {https://ui.adsabs.harvard.edu/abs/2021MNRAS.507.4953G} {507, 4953}

\bibitem[\protect\citeauthoryear{{Guedes}, {Callegari}, {Madau}  \&
  {Mayer}}{{Guedes} et~al.}{2011}]{guedes11}
{Guedes} J.,  {Callegari} S.,  {Madau} P.,   {Mayer} L.,  2011, \mn@doi [\apj]
  {10.1088/0004-637X/742/2/76}, \href
  {https://ui.adsabs.harvard.edu/abs/2011ApJ...742...76G} {742, 76}

\bibitem[\protect\citeauthoryear{{Guedes}, {Mayer}, {Carollo}  \&
  {Madau}}{{Guedes} et~al.}{2013}]{guedes13}
{Guedes} J.,  {Mayer} L.,  {Carollo} M.,   {Madau} P.,  2013, \mn@doi [\apj]
  {10.1088/0004-637X/772/1/36}, \href
  {https://ui.adsabs.harvard.edu/abs/2013ApJ...772...36G} {772, 36}

\bibitem[\protect\citeauthoryear{{Guillard}, {Emsellem}  \&
  {Renaud}}{{Guillard} et~al.}{2016}]{guillard16}
{Guillard} N.,  {Emsellem} E.,   {Renaud} F.,  2016, \mn@doi [\mnras]
  {10.1093/mnras/stw1570}, \href
  {https://ui.adsabs.harvard.edu/abs/2016MNRAS.461.3620G} {461, 3620}

\bibitem[\protect\citeauthoryear{{Guo} et~al.,}{{Guo} et~al.}{2015}]{guo15}
{Guo} Y.,  et~al., 2015, \mn@doi [\apj] {10.1088/0004-637X/800/1/39}, \href
  {https://ui.adsabs.harvard.edu/abs/2015ApJ...800...39G} {800, 39}

\bibitem[\protect\citeauthoryear{{H{\"a}ring} \& {Rix}}{{H{\"a}ring} \&
  {Rix}}{2004}]{neumayer04}
{H{\"a}ring} N.,  {Rix} H.-W.,  2004, \mn@doi [\apjl] {10.1086/383567}, \href
  {https://ui.adsabs.harvard.edu/abs/2004ApJ...604L..89H} {604, L89}

\bibitem[\protect\citeauthoryear{{Hartmann}, {Debattista}, {Seth}, {Cappellari}
   \& {Quinn}}{{Hartmann} et~al.}{2011}]{hartmann11}
{Hartmann} M.,  {Debattista} V.~P.,  {Seth} A.,  {Cappellari} M.,   {Quinn}
  T.~R.,  2011, \mn@doi [\mnras] {10.1111/j.1365-2966.2011.19659.x}, \href
  {https://ui.adsabs.harvard.edu/abs/2011MNRAS.418.2697H} {418, 2697}

\bibitem[\protect\citeauthoryear{{Hirschmann}, {Dolag}, {Saro}, {Bachmann},
  {Borgani}  \& {Burkert}}{{Hirschmann} et~al.}{2014}]{hirschmann14}
{Hirschmann} M.,  {Dolag} K.,  {Saro} A.,  {Bachmann} L.,  {Borgani} S.,
  {Burkert} A.,  2014, \mn@doi [\mnras] {10.1093/mnras/stu1023}, \href
  {https://ui.adsabs.harvard.edu/abs/2014MNRAS.442.2304H} {442, 2304}

\bibitem[\protect\citeauthoryear{{Hopkins}}{{Hopkins}}{2013}]{hopkins13a}
{Hopkins} P.~F.,  2013, \mn@doi [\mnras] {10.1093/mnras/sts704}, \href
  {https://ui.adsabs.harvard.edu/abs/2013MNRAS.430.1653H} {430, 1653}

\bibitem[\protect\citeauthoryear{{Hopkins} et~al.,}{{Hopkins}
  et~al.}{2009}]{hopkins09}
{Hopkins} P.~F.,  et~al., 2009, \mn@doi [\mnras]
  {10.1111/j.1365-2966.2009.14983.x}, \href
  {https://ui.adsabs.harvard.edu/abs/2009MNRAS.397..802H} {397, 802}

\bibitem[\protect\citeauthoryear{{Hopkins} et~al.,}{{Hopkins}
  et~al.}{2010}]{hopkins10}
{Hopkins} P.~F.,  et~al., 2010, \mn@doi [\apj] {10.1088/0004-637X/715/1/202},
  \href {https://ui.adsabs.harvard.edu/abs/2010ApJ...715..202H} {715, 202}

\bibitem[\protect\citeauthoryear{{Hopkins}, {Kere{\v{s}}}, {Murray}, {Quataert}
   \& {Hernquist}}{{Hopkins} et~al.}{2012}]{hopkins12}
{Hopkins} P.~F.,  {Kere{\v{s}}} D.,  {Murray} N.,  {Quataert} E.,   {Hernquist}
  L.,  2012, \mn@doi [\mnras] {10.1111/j.1365-2966.2012.21981.x}, \href
  {https://ui.adsabs.harvard.edu/abs/2012MNRAS.427..968H} {427, 968}

\bibitem[\protect\citeauthoryear{{Hopkins}, {Cox}, {Hernquist}, {Narayanan},
  {Hayward}  \& {Murray}}{{Hopkins} et~al.}{2013a}]{hopkins13b}
{Hopkins} P.~F.,  {Cox} T.~J.,  {Hernquist} L.,  {Narayanan} D.,  {Hayward}
  C.~C.,   {Murray} N.,  2013a, \mn@doi [\mnras] {10.1093/mnras/stt017}, \href
  {https://ui.adsabs.harvard.edu/abs/2013MNRAS.430.1901H} {430, 1901}

\bibitem[\protect\citeauthoryear{{Hopkins}, {Narayanan}  \& {Murray}}{{Hopkins}
  et~al.}{2013b}]{hopkins13c}
{Hopkins} P.~F.,  {Narayanan} D.,   {Murray} N.,  2013b, \mn@doi [\mnras]
  {10.1093/mnras/stt723}, \href
  {https://ui.adsabs.harvard.edu/abs/2013MNRAS.432.2647H} {432, 2647}

\bibitem[\protect\citeauthoryear{{Hopkins} et~al.,}{{Hopkins}
  et~al.}{2018}]{hopkins18}
{Hopkins} P.~F.,  et~al., 2018, \mn@doi [\mnras] {10.1093/mnras/sty1690}, \href
  {https://ui.adsabs.harvard.edu/abs/2018MNRAS.480..800H} {480, 800}

\bibitem[\protect\citeauthoryear{{Huang}, {Leauthaud}, {Greene}, {Bundy},
  {Lin}, {Tanaka}, {Miyazaki}  \& {Komiyama}}{{Huang} et~al.}{2018}]{huang18}
{Huang} S.,  {Leauthaud} A.,  {Greene} J.~E.,  {Bundy} K.,  {Lin} Y.-T.,
  {Tanaka} M.,  {Miyazaki} S.,   {Komiyama} Y.,  2018, \mn@doi [\mnras]
  {10.1093/mnras/stx3200}, \href
  {https://ui.adsabs.harvard.edu/abs/2018MNRAS.475.3348H} {475, 3348}

\bibitem[\protect\citeauthoryear{{Joshi}, {Pillepich}, {Nelson}, {Marinacci},
  {Springel}, {Rodriguez-Gomez}, {Vogelsberger}  \& {Hernquist}}{{Joshi}
  et~al.}{2020}]{joshi20}
{Joshi} G.~D.,  {Pillepich} A.,  {Nelson} D.,  {Marinacci} F.,  {Springel} V.,
  {Rodriguez-Gomez} V.,  {Vogelsberger} M.,   {Hernquist} L.,  2020, \mn@doi
  [\mnras] {10.1093/mnras/staa1668}, \href
  {https://ui.adsabs.harvard.edu/abs/2020MNRAS.496.2673J} {496, 2673}

\bibitem[\protect\citeauthoryear{{Kormendy} \& {Ho}}{{Kormendy} \&
  {Ho}}{2013}]{reviewbh}
{Kormendy} J.,  {Ho} L.~C.,  2013, \mn@doi [\araa]
  {10.1146/annurev-astro-082708-101811}, \href
  {https://ui.adsabs.harvard.edu/abs/2013ARA&A..51..511K} {51, 511}

\bibitem[\protect\citeauthoryear{{Kormendy} \& {Kennicutt}}{{Kormendy} \&
  {Kennicutt}}{2004}]{reviewbulge}
{Kormendy} J.,  {Kennicutt} Robert~C. J.,  2004, \mn@doi [\araa]
  {10.1146/annurev.astro.42.053102.134024}, \href
  {https://ui.adsabs.harvard.edu/abs/2004ARA&A..42..603K} {42, 603}

\bibitem[\protect\citeauthoryear{{Kraljic}, {Bournaud}  \& {Martig}}{{Kraljic}
  et~al.}{2012}]{kraljic12}
{Kraljic} K.,  {Bournaud} F.,   {Martig} M.,  2012, \mn@doi [\apj]
  {10.1088/0004-637X/757/1/60}, \href
  {https://ui.adsabs.harvard.edu/abs/2012ApJ...757...60K} {757, 60}

\bibitem[\protect\citeauthoryear{{Lapiner}, {Dekel}  \& {Dubois}}{{Lapiner}
  et~al.}{2021}]{lapiner21}
{Lapiner} S.,  {Dekel} A.,   {Dubois} Y.,  2021, \mn@doi [\mnras]
  {10.1093/mnras/stab1205}, \href
  {https://ui.adsabs.harvard.edu/abs/2021MNRAS.505..172L} {505, 172}

\bibitem[\protect\citeauthoryear{{L{\"a}sker}, {Greene}, {Seth}, {van de Ven},
  {Braatz}, {Henkel}  \& {Lo}}{{L{\"a}sker} et~al.}{2016}]{lasker16}
{L{\"a}sker} R.,  {Greene} J.~E.,  {Seth} A.,  {van de Ven} G.,  {Braatz}
  J.~A.,  {Henkel} C.,   {Lo} K.~Y.,  2016, \mn@doi [\apj]
  {10.3847/0004-637X/825/1/3}, \href
  {https://ui.adsabs.harvard.edu/abs/2016ApJ...825....3L} {825, 3}

\bibitem[\protect\citeauthoryear{{Laurikainen}, {Peletier}  \&
  {Gadotti}}{{Laurikainen} et~al.}{2016}]{bulges}
{Laurikainen} E.,  {Peletier} R.,   {Gadotti} D.,  eds, 2016, {Galactic Bulges}
   Astrophysics and Space Science Library Vol. 418,
  \mn@doi{10.1007/978-3-319-19378-6.
}

\bibitem[\protect\citeauthoryear{{Leaman} \& {van de Ven}}{{Leaman} \& {van de
  Ven}}{2021}]{leaman21}
{Leaman} R.,  {van de Ven} G.,  2021, \mn@doi [\mnras]
  {10.1093/mnras/stab1966}, \href
  {https://ui.adsabs.harvard.edu/abs/2021MNRAS.tmp.2125L} {}

\bibitem[\protect\citeauthoryear{{Maciejewski}}{{Maciejewski}}{2004}]{maciejewski04}
{Maciejewski} W.,  2004, \mn@doi [\mnras] {10.1111/j.1365-2966.2004.08254.x},
  \href {https://ui.adsabs.harvard.edu/abs/2004MNRAS.354..892M} {354, 892}

\bibitem[\protect\citeauthoryear{{Malbon}, {Baugh}, {Frenk}  \&
  {Lacey}}{{Malbon} et~al.}{2007}]{malbon07}
{Malbon} R.~K.,  {Baugh} C.~M.,  {Frenk} C.~S.,   {Lacey} C.~G.,  2007, \mn@doi
  [\mnras] {10.1111/j.1365-2966.2007.12317.x}, \href
  {https://ui.adsabs.harvard.edu/abs/2007MNRAS.382.1394M} {382, 1394}

\bibitem[\protect\citeauthoryear{{Mandelker}, {Dekel}, {Ceverino}, {Tweed},
  {Moody}  \& {Primack}}{{Mandelker} et~al.}{2014}]{mandelker14}
{Mandelker} N.,  {Dekel} A.,  {Ceverino} D.,  {Tweed} D.,  {Moody} C.~E.,
  {Primack} J.,  2014, \mn@doi [\mnras] {10.1093/mnras/stu1340}, \href
  {https://ui.adsabs.harvard.edu/abs/2014MNRAS.443.3675M} {443, 3675}

\bibitem[\protect\citeauthoryear{{Mandelker}, {Dekel}, {Ceverino}, {DeGraf},
  {Guo}  \& {Primack}}{{Mandelker} et~al.}{2017}]{mandelker17}
{Mandelker} N.,  {Dekel} A.,  {Ceverino} D.,  {DeGraf} C.,  {Guo} Y.,
  {Primack} J.,  2017, \mn@doi [\mnras] {10.1093/mnras/stw2358}, \href
  {https://ui.adsabs.harvard.edu/abs/2017MNRAS.464..635M} {464, 635}

\bibitem[\protect\citeauthoryear{{Marinacci}, {Pakmor}  \&
  {Springel}}{{Marinacci} et~al.}{2014}]{marinacci14}
{Marinacci} F.,  {Pakmor} R.,   {Springel} V.,  2014, \mn@doi [\mnras]
  {10.1093/mnras/stt2003}, \href
  {https://ui.adsabs.harvard.edu/abs/2014MNRAS.437.1750M} {437, 1750}

\bibitem[\protect\citeauthoryear{{Marinacci} et~al.,}{{Marinacci}
  et~al.}{2018}]{marinacci18}
{Marinacci} F.,  et~al., 2018, \mn@doi [\mnras] {10.1093/mnras/sty2206}, \href
  {https://ui.adsabs.harvard.edu/abs/2018MNRAS.480.5113M} {480, 5113}

\bibitem[\protect\citeauthoryear{{Mart{\'\i}n-Navarro}, {Vazdekis},
  {Falc{\'o}n-Barroso}, {La Barbera}, {Y{\i}ld{\i}r{\i}m}  \& {van de
  Ven}}{{Mart{\'\i}n-Navarro} et~al.}{2018a}]{nacho18b}
{Mart{\'\i}n-Navarro} I.,  {Vazdekis} A.,  {Falc{\'o}n-Barroso} J.,  {La
  Barbera} F.,  {Y{\i}ld{\i}r{\i}m} A.,   {van de Ven} G.,  2018a, \mn@doi
  [\mnras] {10.1093/mnras/stx3346}, \href
  {https://ui.adsabs.harvard.edu/abs/2018MNRAS.475.3700M} {475, 3700}

\bibitem[\protect\citeauthoryear{{Mart{\'\i}n-Navarro}, {Brodie}, {Romanowsky},
  {Ruiz-Lara}  \& {van de Ven}}{{Mart{\'\i}n-Navarro} et~al.}{2018b}]{nacho18}
{Mart{\'\i}n-Navarro} I.,  {Brodie} J.~P.,  {Romanowsky} A.~J.,  {Ruiz-Lara}
  T.,   {van de Ven} G.,  2018b, \mn@doi [\nat] {10.1038/nature24999}, \href
  {https://ui.adsabs.harvard.edu/abs/2018Natur.553..307M} {553, 307}

\bibitem[\protect\citeauthoryear{{Mart{\'\i}n-Navarro}
  et~al.,}{{Mart{\'\i}n-Navarro} et~al.}{2019}]{nacho19}
{Mart{\'\i}n-Navarro} I.,  et~al., 2019, \mn@doi [\aap]
  {10.1051/0004-6361/201935360}, \href
  {https://ui.adsabs.harvard.edu/abs/2019A&A...626A.124M} {626, A124}

\bibitem[\protect\citeauthoryear{{Mart{\'\i}n-Navarro}
  et~al.,}{{Mart{\'\i}n-Navarro} et~al.}{2021}]{nacho21}
{Mart{\'\i}n-Navarro} I.,  et~al., 2021, \mn@doi [\aap]
  {10.1051/0004-6361/202141348}, \href
  {https://ui.adsabs.harvard.edu/abs/2021A&A...654A..59M} {654, A59}

\bibitem[\protect\citeauthoryear{{Mayer}, {Tamburello}, {Lupi}, {Keller},
  {Wadsley}  \& {Madau}}{{Mayer} et~al.}{2016}]{mayer16}
{Mayer} L.,  {Tamburello} V.,  {Lupi} A.,  {Keller} B.,  {Wadsley} J.,
  {Madau} P.,  2016, \mn@doi [\apjl] {10.3847/2041-8205/830/1/L13}, \href
  {https://ui.adsabs.harvard.edu/abs/2016ApJ...830L..13M} {830, L13}

\bibitem[\protect\citeauthoryear{{Merritt}, {van Dokkum}, {Abraham}  \&
  {Zhang}}{{Merritt} et~al.}{2016}]{merritt16}
{Merritt} A.,  {van Dokkum} P.,  {Abraham} R.,   {Zhang} J.,  2016, \mn@doi
  [\apj] {10.3847/0004-637X/830/2/62}, \href
  {https://ui.adsabs.harvard.edu/abs/2016ApJ...830...62M} {830, 62}

\bibitem[\protect\citeauthoryear{{Merritt}, {Pillepich}, {van Dokkum},
  {Nelson}, {Hernquist}, {Marinacci}  \& {Vogelsberger}}{{Merritt}
  et~al.}{2020}]{merritt20}
{Merritt} A.,  {Pillepich} A.,  {van Dokkum} P.,  {Nelson} D.,  {Hernquist} L.,
   {Marinacci} F.,   {Vogelsberger} M.,  2020, \mn@doi [\mnras]
  {10.1093/mnras/staa1164}, \href
  {https://ui.adsabs.harvard.edu/abs/2020MNRAS.495.4570M} {495, 4570}

\bibitem[\protect\citeauthoryear{{Mihos} \& {Hernquist}}{{Mihos} \&
  {Hernquist}}{1996}]{mihos96}
{Mihos} J.~C.,  {Hernquist} L.,  1996, \mn@doi [\apj] {10.1086/177353}, \href
  {https://ui.adsabs.harvard.edu/abs/1996ApJ...464..641M} {464, 641}

\bibitem[\protect\citeauthoryear{{Minchev} \& {Famaey}}{{Minchev} \&
  {Famaey}}{2010}]{minchev10}
{Minchev} I.,  {Famaey} B.,  2010, \mn@doi [\apj]
  {10.1088/0004-637X/722/1/112}, \href
  {https://ui.adsabs.harvard.edu/abs/2010ApJ...722..112M} {722, 112}

\bibitem[\protect\citeauthoryear{{Monachesi}, {Bell}, {Radburn-Smith},
  {Bailin}, {de Jong}, {Holwerda}, {Streich}  \& {Silverstein}}{{Monachesi}
  et~al.}{2016}]{monachesi16}
{Monachesi} A.,  {Bell} E.~F.,  {Radburn-Smith} D.~J.,  {Bailin} J.,  {de Jong}
  R.~S.,  {Holwerda} B.,  {Streich} D.,   {Silverstein} G.,  2016, \mn@doi
  [\mnras] {10.1093/mnras/stv2987}, \href
  {https://ui.adsabs.harvard.edu/abs/2016MNRAS.457.1419M} {457, 1419}

\bibitem[\protect\citeauthoryear{{Naiman} et~al.,}{{Naiman}
  et~al.}{2018}]{naiman18}
{Naiman} J.~P.,  et~al., 2018, \mn@doi [\mnras] {10.1093/mnras/sty618}, \href
  {https://ui.adsabs.harvard.edu/abs/2018MNRAS.477.1206N} {477, 1206}

\bibitem[\protect\citeauthoryear{{Nelson} et~al.,}{{Nelson}
  et~al.}{2012}]{nelson12}
{Nelson} E.~J.,  et~al., 2012, \mn@doi [\apjl] {10.1088/2041-8205/747/2/L28},
  \href {https://ui.adsabs.harvard.edu/abs/2012ApJ...747L..28N} {747, L28}

\bibitem[\protect\citeauthoryear{{Nelson} et~al.,}{{Nelson}
  et~al.}{2018}]{nelson18}
{Nelson} D.,  et~al., 2018, \mn@doi [\mnras] {10.1093/mnras/stx3040}, \href
  {https://ui.adsabs.harvard.edu/abs/2018MNRAS.475..624N} {475, 624}

\bibitem[\protect\citeauthoryear{{Nelson} et~al.,}{{Nelson}
  et~al.}{2019a}]{publictng}
{Nelson} D.,  et~al., 2019a, \mn@doi [Computational Astrophysics and Cosmology]
  {10.1186/s40668-019-0028-x}, \href
  {https://ui.adsabs.harvard.edu/abs/2019ComAC...6....2N} {6, 2}

\bibitem[\protect\citeauthoryear{{Nelson} et~al.,}{{Nelson}
  et~al.}{2019b}]{nelson19}
{Nelson} D.,  et~al., 2019b, \mn@doi [\mnras] {10.1093/mnras/stz2306}, \href
  {https://ui.adsabs.harvard.edu/abs/2019MNRAS.490.3234N} {490, 3234}

\bibitem[\protect\citeauthoryear{{Nelson} et~al.,}{{Nelson}
  et~al.}{2021}]{nelson21}
{Nelson} E.~J.,  et~al., 2021, \mn@doi [\mnras] {10.1093/mnras/stab2131}, \href
  {https://ui.adsabs.harvard.edu/abs/2021MNRAS.508..219N} {508, 219}

\bibitem[\protect\citeauthoryear{{Ness} et~al.,}{{Ness} et~al.}{2013}]{ness13}
{Ness} M.,  et~al., 2013, \mn@doi [\mnras] {10.1093/mnras/stt533}, \href
  {https://ui.adsabs.harvard.edu/abs/2013MNRAS.432.2092N} {432, 2092}

\bibitem[\protect\citeauthoryear{{Neumayer}, {Seth}  \& {B{\"o}ker}}{{Neumayer}
  et~al.}{2020}]{reviewnsc}
{Neumayer} N.,  {Seth} A.,   {B{\"o}ker} T.,  2020, \mn@doi [\aapr]
  {10.1007/s00159-020-00125-0}, \href
  {https://ui.adsabs.harvard.edu/abs/2020A&ARv..28....4N} {28, 4}

\bibitem[\protect\citeauthoryear{{Nogueras-Lara} et~al.,}{{Nogueras-Lara}
  et~al.}{2020}]{paco20}
{Nogueras-Lara} F.,  et~al., 2020, \mn@doi [Nature Astronomy]
  {10.1038/s41550-019-0967-9}, \href
  {https://ui.adsabs.harvard.edu/abs/2020NatAs...4..377N} {4, 377}

\bibitem[\protect\citeauthoryear{{Nogueras-Lara}, {Sch{\"o}del}  \&
  {Neumayer}}{{Nogueras-Lara} et~al.}{2021}]{paco21}
{Nogueras-Lara} F.,  {Sch{\"o}del} R.,   {Neumayer} N.,  2021, \mn@doi [\apj]
  {10.3847/1538-4357/ac185e}, \href
  {https://ui.adsabs.harvard.edu/abs/2021ApJ...920...97N} {920, 97}

\bibitem[\protect\citeauthoryear{{Okamoto}}{{Okamoto}}{2013}]{okamoto13}
{Okamoto} T.,  2013, \mn@doi [\mnras] {10.1093/mnras/sts067}, \href
  {https://ui.adsabs.harvard.edu/abs/2013MNRAS.428..718O} {428, 718}

\bibitem[\protect\citeauthoryear{{Oklop{\v{c}}i{\'c}}, {Hopkins}, {Feldmann},
  {Kere{\v{s}}}, {Faucher-Gigu{\`e}re}  \& {Murray}}{{Oklop{\v{c}}i{\'c}}
  et~al.}{2017}]{oklop17}
{Oklop{\v{c}}i{\'c}} A.,  {Hopkins} P.~F.,  {Feldmann} R.,  {Kere{\v{s}}} D.,
  {Faucher-Gigu{\`e}re} C.-A.,   {Murray} N.,  2017, \mn@doi [\mnras]
  {10.1093/mnras/stw2754}, \href
  {https://ui.adsabs.harvard.edu/abs/2017MNRAS.465..952O} {465, 952}

\bibitem[\protect\citeauthoryear{{Pakmor} \& {Springel}}{{Pakmor} \&
  {Springel}}{2013}]{pakmor13}
{Pakmor} R.,  {Springel} V.,  2013, \mn@doi [\mnras] {10.1093/mnras/stt428},
  \href {https://ui.adsabs.harvard.edu/abs/2013MNRAS.432..176P} {432, 176}

\bibitem[\protect\citeauthoryear{{Pakmor}, {Bauer}  \& {Springel}}{{Pakmor}
  et~al.}{2011}]{pakmor11}
{Pakmor} R.,  {Bauer} A.,   {Springel} V.,  2011, \mn@doi [\mnras]
  {10.1111/j.1365-2966.2011.19591.x}, \href
  {https://ui.adsabs.harvard.edu/abs/2011MNRAS.418.1392P} {418, 1392}

\bibitem[\protect\citeauthoryear{{Pakmor}, {Springel}, {Bauer}, {Mocz},
  {Munoz}, {Ohlmann}, {Schaal}  \& {Zhu}}{{Pakmor} et~al.}{2016}]{pakmor16}
{Pakmor} R.,  {Springel} V.,  {Bauer} A.,  {Mocz} P.,  {Munoz} D.~J.,
  {Ohlmann} S.~T.,  {Schaal} K.,   {Zhu} C.,  2016, \mn@doi [\mnras]
  {10.1093/mnras/stv2380}, \href
  {https://ui.adsabs.harvard.edu/abs/2016MNRAS.455.1134P} {455, 1134}

\bibitem[\protect\citeauthoryear{{Perets} \& {Mastrobuono-Battisti}}{{Perets}
  \& {Mastrobuono-Battisti}}{2014}]{perets14}
{Perets} H.~B.,  {Mastrobuono-Battisti} A.,  2014, \mn@doi [\apjl]
  {10.1088/2041-8205/784/2/L44}, \href
  {https://ui.adsabs.harvard.edu/abs/2014ApJ...784L..44P} {784, L44}

\bibitem[\protect\citeauthoryear{{Peschken} \& {{\L}okas}}{{Peschken} \&
  {{\L}okas}}{2019}]{peschken19}
{Peschken} N.,  {{\L}okas} E.~L.,  2019, \mn@doi [\mnras]
  {10.1093/mnras/sty3277}, \href
  {https://ui.adsabs.harvard.edu/abs/2019MNRAS.483.2721P} {483, 2721}

\bibitem[\protect\citeauthoryear{{Pillepich} et~al.,}{{Pillepich}
  et~al.}{2018a}]{pillepich18a}
{Pillepich} A.,  et~al., 2018a, \mn@doi [\mnras] {10.1093/mnras/stx2656}, \href
  {https://ui.adsabs.harvard.edu/abs/2018MNRAS.473.4077P} {473, 4077}

\bibitem[\protect\citeauthoryear{{Pillepich} et~al.,}{{Pillepich}
  et~al.}{2018b}]{pillepich18b}
{Pillepich} A.,  et~al., 2018b, \mn@doi [\mnras] {10.1093/mnras/stx3112}, \href
  {https://ui.adsabs.harvard.edu/abs/2018MNRAS.475..648P} {475, 648}

\bibitem[\protect\citeauthoryear{{Pillepich} et~al.,}{{Pillepich}
  et~al.}{2019}]{pillepich19}
{Pillepich} A.,  et~al., 2019, \mn@doi [\mnras] {10.1093/mnras/stz2338}, \href
  {https://ui.adsabs.harvard.edu/abs/2019MNRAS.490.3196P} {490, 3196}

\bibitem[\protect\citeauthoryear{{Planck Collaboration} et~al.,}{{Planck
  Collaboration} et~al.}{2016}]{planck15}
{Planck Collaboration} et~al., 2016, \mn@doi [\aap]
  {10.1051/0004-6361/201525830}, \href
  {https://ui.adsabs.harvard.edu/abs/2016A&A...594A..13P} {594, A13}

\bibitem[\protect\citeauthoryear{{Powell}, {Bournaud}, {Chapon}  \&
  {Teyssier}}{{Powell} et~al.}{2013}]{powell13}
{Powell} L.~C.,  {Bournaud} F.,  {Chapon} D.,   {Teyssier} R.,  2013, \mn@doi
  [\mnras] {10.1093/mnras/stt1036}, \href
  {https://ui.adsabs.harvard.edu/abs/2013MNRAS.434.1028P} {434, 1028}

\bibitem[\protect\citeauthoryear{{Pulsoni} et~al.,}{{Pulsoni}
  et~al.}{2018}]{pulsoni18}
{Pulsoni} C.,  et~al., 2018, \mn@doi [\aap] {10.1051/0004-6361/201732473},
  \href {https://ui.adsabs.harvard.edu/abs/2018A&A...618A..94P} {618, A94}

\bibitem[\protect\citeauthoryear{{Pulsoni}, {Gerhard}, {Arnaboldi},
  {Pillepich}, {Rodriguez-Gomez}, {Nelson}, {Hernquist}  \&
  {Springel}}{{Pulsoni} et~al.}{2021}]{pulsoni21}
{Pulsoni} C.,  {Gerhard} O.,  {Arnaboldi} M.,  {Pillepich} A.,
  {Rodriguez-Gomez} V.,  {Nelson} D.,  {Hernquist} L.,   {Springel} V.,  2021,
  \mn@doi [\aap] {10.1051/0004-6361/202039166}, \href
  {https://ui.adsabs.harvard.edu/abs/2021A&A...647A..95P} {647, A95}

\bibitem[\protect\citeauthoryear{{Reines} \& {Volonteri}}{{Reines} \&
  {Volonteri}}{2015}]{reines15}
{Reines} A.~E.,  {Volonteri} M.,  2015, \mn@doi [\apj]
  {10.1088/0004-637X/813/2/82}, \href
  {https://ui.adsabs.harvard.edu/abs/2015ApJ...813...82R} {813, 82}

\bibitem[\protect\citeauthoryear{{Remus} \& {Forbes}}{{Remus} \&
  {Forbes}}{2021}]{remus21}
{Remus} R.-S.,  {Forbes} D.~A.,  2021, arXiv e-prints, \href
  {https://ui.adsabs.harvard.edu/abs/2021arXiv210112216R} {p. arXiv:2101.12216}

\bibitem[\protect\citeauthoryear{{Renaud}, {Boily}, {Naab}  \&
  {Theis}}{{Renaud} et~al.}{2009}]{renaud09}
{Renaud} F.,  {Boily} C.~M.,  {Naab} T.,   {Theis} C.,  2009, \mn@doi [\apj]
  {10.1088/0004-637X/706/1/67}, \href
  {https://ui.adsabs.harvard.edu/abs/2009ApJ...706...67R} {706, 67}

\bibitem[\protect\citeauthoryear{{Rodriguez-Gomez} et~al.,}{{Rodriguez-Gomez}
  et~al.}{2015}]{rodriguesgomez15}
{Rodriguez-Gomez} V.,  et~al., 2015, \mn@doi [\mnras] {10.1093/mnras/stv264},
  \href {http://adsabs.harvard.edu/abs/2015MNRAS.449...49R} {449, 49}

\bibitem[\protect\citeauthoryear{{Rodriguez-Gomez} et~al.,}{{Rodriguez-Gomez}
  et~al.}{2016}]{rodriguesgomez16}
{Rodriguez-Gomez} V.,  et~al., 2016, \mn@doi [\mnras] {10.1093/mnras/stw456},
  \href {http://adsabs.harvard.edu/abs/2016MNRAS.458.2371R} {458, 2371}

\bibitem[\protect\citeauthoryear{{Rodriguez-Gomez} et~al.,}{{Rodriguez-Gomez}
  et~al.}{2017}]{rodriguesgomez17}
{Rodriguez-Gomez} V.,  et~al., 2017, \mn@doi [\mnras] {10.1093/mnras/stx305},
  \href {http://adsabs.harvard.edu/abs/2017MNRAS.467.3083R} {467, 3083}

\bibitem[\protect\citeauthoryear{{Romano-D{\'\i}az}, {Shlosman}, {Heller}  \&
  {Hoffman}}{{Romano-D{\'\i}az} et~al.}{2008}]{romano08}
{Romano-D{\'\i}az} E.,  {Shlosman} I.,  {Heller} C.,   {Hoffman} Y.,  2008,
  \mn@doi [\apjl] {10.1086/593168}, \href
  {https://ui.adsabs.harvard.edu/abs/2008ApJ...687L..13R} {687, L13}

\bibitem[\protect\citeauthoryear{{Rosas-Guevara} et~al.,}{{Rosas-Guevara}
  et~al.}{2020}]{rosas-guevara20}
{Rosas-Guevara} Y.,  et~al., 2020, \mn@doi [\mnras] {10.1093/mnras/stz3180},
  \href {https://ui.adsabs.harvard.edu/abs/2020MNRAS.491.2547R} {491, 2547}

\bibitem[\protect\citeauthoryear{{Rosas-Guevara} et~al.,}{{Rosas-Guevara}
  et~al.}{2021}]{rosasguevara21}
{Rosas-Guevara} Y.,  et~al., 2021, arXiv e-prints, \href
  {https://ui.adsabs.harvard.edu/abs/2021arXiv211004537R} {p. arXiv:2110.04537}

\bibitem[\protect\citeauthoryear{{Sahu}, {Graham}  \& {Davis}}{{Sahu}
  et~al.}{2019}]{sahu19}
{Sahu} N.,  {Graham} A.~W.,   {Davis} B.~L.,  2019, \mn@doi [\apj]
  {10.3847/1538-4357/ab0f32}, \href
  {https://ui.adsabs.harvard.edu/abs/2019ApJ...876..155S} {876, 155}

\bibitem[\protect\citeauthoryear{{S{\'a}nchez-Janssen}
  et~al.,}{{S{\'a}nchez-Janssen} et~al.}{2019}]{sanchez19}
{S{\'a}nchez-Janssen} R.,  et~al., 2019, \mn@doi [\apj]
  {10.3847/1538-4357/aaf4fd}, \href
  {https://ui.adsabs.harvard.edu/abs/2019ApJ...878...18S} {878, 18}

\bibitem[\protect\citeauthoryear{{Sani}, {Marconi}, {Hunt}  \&
  {Risaliti}}{{Sani} et~al.}{2011}]{sani11}
{Sani} E.,  {Marconi} A.,  {Hunt} L.~K.,   {Risaliti} G.,  2011, \mn@doi
  [\mnras] {10.1111/j.1365-2966.2011.18229.x}, \href
  {https://ui.adsabs.harvard.edu/abs/2011MNRAS.413.1479S} {413, 1479}

\bibitem[\protect\citeauthoryear{{Schaye} et~al.,}{{Schaye}
  et~al.}{2015}]{schaye15}
{Schaye} J.,  et~al., 2015, \mn@doi [\mnras] {10.1093/mnras/stu2058}, \href
  {https://ui.adsabs.harvard.edu/abs/2015MNRAS.446..521S} {446, 521}

\bibitem[\protect\citeauthoryear{{Schultheis} et~al.,}{{Schultheis}
  et~al.}{2021}]{schultheis21}
{Schultheis} M.,  et~al., 2021, \mn@doi [\aap] {10.1051/0004-6361/202140499},
  \href {https://ui.adsabs.harvard.edu/abs/2021A&A...650A.191S} {650, A191}

\bibitem[\protect\citeauthoryear{{Schulze}, {Remus}, {Dolag}, {Bellstedt},
  {Burkert}  \& {Forbes}}{{Schulze} et~al.}{2020}]{schulze20}
{Schulze} F.,  {Remus} R.-S.,  {Dolag} K.,  {Bellstedt} S.,  {Burkert} A.,
  {Forbes} D.~A.,  2020, \mn@doi [\mnras] {10.1093/mnras/staa511}, \href
  {https://ui.adsabs.harvard.edu/abs/2020MNRAS.493.3778S} {493, 3778}

\bibitem[\protect\citeauthoryear{{Schweizer}, {Seitzer}, {Whitmore}, {Kelson}
  \& {Villanueva}}{{Schweizer} et~al.}{2018}]{schweizer18}
{Schweizer} F.,  {Seitzer} P.,  {Whitmore} B.~C.,  {Kelson} D.~D.,
  {Villanueva} E.~V.,  2018, \mn@doi [\apj] {10.3847/1538-4357/aaa424}, \href
  {https://ui.adsabs.harvard.edu/abs/2018ApJ...853...54S} {853, 54}

\bibitem[\protect\citeauthoryear{{Scott} \& {Graham}}{{Scott} \&
  {Graham}}{2013}]{scott13}
{Scott} N.,  {Graham} A.~W.,  2013, \mn@doi [\apj]
  {10.1088/0004-637X/763/2/76}, \href
  {https://ui.adsabs.harvard.edu/abs/2013ApJ...763...76S} {763, 76}

\bibitem[\protect\citeauthoryear{{Sellwood} \& {Binney}}{{Sellwood} \&
  {Binney}}{2002}]{sellwood02}
{Sellwood} J.~A.,  {Binney} J.~J.,  2002, \mn@doi [\mnras]
  {10.1046/j.1365-8711.2002.05806.x}, \href
  {https://ui.adsabs.harvard.edu/abs/2002MNRAS.336..785S} {336, 785}

\bibitem[\protect\citeauthoryear{{Seo}, {Kim}, {Kwak}, {Hsieh}, {Han}  \&
  {Hopkins}}{{Seo} et~al.}{2019}]{seo19}
{Seo} W.-Y.,  {Kim} W.-T.,  {Kwak} S.,  {Hsieh} P.-Y.,  {Han} C.,   {Hopkins}
  P.~F.,  2019, \mn@doi [\apj] {10.3847/1538-4357/aafc5f}, \href
  {https://ui.adsabs.harvard.edu/abs/2019ApJ...872....5S} {872, 5}

\bibitem[\protect\citeauthoryear{{S{\'e}rsic}}{{S{\'e}rsic}}{1968}]{sersic}
{S{\'e}rsic} J.~L.,  1968, {Atlas de Galaxias Australes}

\bibitem[\protect\citeauthoryear{{Sheth} et~al.,}{{Sheth} et~al.}{2008}]{bars1}
{Sheth} K.,  et~al., 2008, \mn@doi [\apj] {10.1086/524980}, \href
  {https://ui.adsabs.harvard.edu/abs/2008ApJ...675.1141S} {675, 1141}

\bibitem[\protect\citeauthoryear{{Simmons} et~al.,}{{Simmons}
  et~al.}{2014}]{bars2}
{Simmons} B.~D.,  et~al., 2014, \mn@doi [\mnras] {10.1093/mnras/stu1817}, \href
  {https://ui.adsabs.harvard.edu/abs/2014MNRAS.445.3466S} {445, 3466}

\bibitem[\protect\citeauthoryear{{Smith}}{{Smith}}{2021}]{smith21a}
{Smith} M.~C.,  2021, \mn@doi [\mnras] {10.1093/mnras/stab291}, \href
  {https://ui.adsabs.harvard.edu/abs/2021MNRAS.502.5417S} {502, 5417}

\bibitem[\protect\citeauthoryear{{Smith}, {Sijacki}  \& {Shen}}{{Smith}
  et~al.}{2018}]{smith18}
{Smith} M.~C.,  {Sijacki} D.,   {Shen} S.,  2018, \mn@doi [\mnras]
  {10.1093/mnras/sty994}, \href
  {https://ui.adsabs.harvard.edu/abs/2018MNRAS.478..302S} {478, 302}

\bibitem[\protect\citeauthoryear{{Smith}, {Bryan}, {Somerville}, {Hu},
  {Teyssier}, {Burkhart}  \& {Hernquist}}{{Smith} et~al.}{2021}]{smith21b}
{Smith} M.~C.,  {Bryan} G.~L.,  {Somerville} R.~S.,  {Hu} C.-Y.,  {Teyssier}
  R.,  {Burkhart} B.,   {Hernquist} L.,  2021, \mn@doi [\mnras]
  {10.1093/mnras/stab1896}, \href
  {https://ui.adsabs.harvard.edu/abs/2021MNRAS.506.3882S} {506, 3882}

\bibitem[\protect\citeauthoryear{{Somerville} \& {Dav{\'e}}}{{Somerville} \&
  {Dav{\'e}}}{2015}]{reviewsim}
{Somerville} R.~S.,  {Dav{\'e}} R.,  2015, \mn@doi [\araa]
  {10.1146/annurev-astro-082812-140951}, \href
  {https://ui.adsabs.harvard.edu/abs/2015ARA&A..53...51S} {53, 51}

\bibitem[\protect\citeauthoryear{{Sormani}, {Tress}, {Glover}, {Klessen},
  {Battersby}, {Clark}, {Hatchfield}  \& {Smith}}{{Sormani}
  et~al.}{2020}]{sormani20}
{Sormani} M.~C.,  {Tress} R.~G.,  {Glover} S. C.~O.,  {Klessen} R.~S.,
  {Battersby} C.~D.,  {Clark} P.~C.,  {Hatchfield} H.~P.,   {Smith} R.~J.,
  2020, \mn@doi [\mnras] {10.1093/mnras/staa1999}, \href
  {https://ui.adsabs.harvard.edu/abs/2020MNRAS.497.5024S} {497, 5024}

\bibitem[\protect\citeauthoryear{{Sotillo-Ramos} et~al.,}{{Sotillo-Ramos}
  et~al.}{2022}]{diego22}
{Sotillo-Ramos} D.,  et~al., 2022, \mn@doi [\mnras] {10.1093/mnras/stac2586},
  \href {https://ui.adsabs.harvard.edu/abs/2022MNRAS.516.5404S} {516, 5404}

\bibitem[\protect\citeauthoryear{{Spavone} et~al.,}{{Spavone}
  et~al.}{2017}]{spavone17}
{Spavone} M.,  et~al., 2017, \mn@doi [\aap] {10.1051/0004-6361/201629111},
  \href {https://ui.adsabs.harvard.edu/abs/2017A&A...603A..38S} {603, A38}

\bibitem[\protect\citeauthoryear{{Spavone} et~al.,}{{Spavone}
  et~al.}{2020}]{spavone20}
{Spavone} M.,  et~al., 2020, \mn@doi [\aap] {10.1051/0004-6361/202038015},
  \href {https://ui.adsabs.harvard.edu/abs/2020A&A...639A..14S} {639, A14}

\bibitem[\protect\citeauthoryear{{Springel}}{{Springel}}{2010}]{springel10}
{Springel} V.,  2010, \mn@doi [\mnras] {10.1111/j.1365-2966.2009.15715.x},
  \href {https://ui.adsabs.harvard.edu/abs/2010MNRAS.401..791S} {401, 791}

\bibitem[\protect\citeauthoryear{{Springel}, {White}, {Tormen}  \&
  {Kauffmann}}{{Springel} et~al.}{2001}]{subfind1}
{Springel} V.,  {White} S.~D.~M.,  {Tormen} G.,   {Kauffmann} G.,  2001,
  \mn@doi [\mnras] {10.1046/j.1365-8711.2001.04912.x}, \href
  {http://adsabs.harvard.edu/abs/2001MNRAS.328..726S} {328, 726}

\bibitem[\protect\citeauthoryear{{Springel}, {Di Matteo}  \&
  {Hernquist}}{{Springel} et~al.}{2005}]{springel05}
{Springel} V.,  {Di Matteo} T.,   {Hernquist} L.,  2005, \mn@doi [\mnras]
  {10.1111/j.1365-2966.2005.09238.x}, \href
  {https://ui.adsabs.harvard.edu/abs/2005MNRAS.361..776S} {361, 776}

\bibitem[\protect\citeauthoryear{{Springel} et~al.,}{{Springel}
  et~al.}{2018}]{springel18}
{Springel} V.,  et~al., 2018, \mn@doi [\mnras] {10.1093/mnras/stx3304}, \href
  {https://ui.adsabs.harvard.edu/abs/2018MNRAS.475..676S} {475, 676}

\bibitem[\protect\citeauthoryear{{Tacchella} et~al.,}{{Tacchella}
  et~al.}{2019}]{tacchella19}
{Tacchella} S.,  et~al., 2019, \mn@doi [\mnras] {10.1093/mnras/stz1657}, \href
  {https://ui.adsabs.harvard.edu/abs/2019MNRAS.487.5416T} {487, 5416}

\bibitem[\protect\citeauthoryear{{Tal} \& {van Dokkum}}{{Tal} \& {van
  Dokkum}}{2011}]{tal11}
{Tal} T.,  {van Dokkum} P.~G.,  2011, \mn@doi [\apj]
  {10.1088/0004-637X/731/2/89}, \href
  {https://ui.adsabs.harvard.edu/abs/2011ApJ...731...89T} {731, 89}

\bibitem[\protect\citeauthoryear{{Teklu}, {Remus}, {Dolag}, {Beck}, {Burkert},
  {Schmidt}, {Schulze}  \& {Steinborn}}{{Teklu} et~al.}{2015}]{teklu15}
{Teklu} A.~F.,  {Remus} R.-S.,  {Dolag} K.,  {Beck} A.~M.,  {Burkert} A.,
  {Schmidt} A.~S.,  {Schulze} F.,   {Steinborn} L.~K.,  2015, \mn@doi [\apj]
  {10.1088/0004-637X/812/1/29}, \href
  {https://ui.adsabs.harvard.edu/abs/2015ApJ...812...29T} {812, 29}

\bibitem[\protect\citeauthoryear{{Toomre}}{{Toomre}}{1964}]{toomre64}
{Toomre} A.,  1964, \mn@doi [\apj] {10.1086/147861}, \href
  {https://ui.adsabs.harvard.edu/abs/1964ApJ...139.1217T} {139, 1217}

\bibitem[\protect\citeauthoryear{{Torrey} et~al.,}{{Torrey}
  et~al.}{2019}]{torrey17}
{Torrey} P.,  et~al., 2019, \mn@doi [\mnras] {10.1093/mnras/stz243}, \href
  {https://ui.adsabs.harvard.edu/abs/2019MNRAS.484.5587T} {484, 5587}

\bibitem[\protect\citeauthoryear{{Tress}, {Sormani}, {Glover}, {Klessen},
  {Battersby}, {Clark}, {Hatchfield}  \& {Smith}}{{Tress}
  et~al.}{2020}]{tress20}
{Tress} R.~G.,  {Sormani} M.~C.,  {Glover} S. C.~O.,  {Klessen} R.~S.,
  {Battersby} C.~D.,  {Clark} P.~C.,  {Hatchfield} H.~P.,   {Smith} R.~J.,
  2020, \mn@doi [\mnras] {10.1093/mnras/staa3120}, \href
  {https://ui.adsabs.harvard.edu/abs/2020MNRAS.499.4455T} {499, 4455}

\bibitem[\protect\citeauthoryear{{Vogelsberger} et~al.,}{{Vogelsberger}
  et~al.}{2014a}]{vogelsberger14a}
{Vogelsberger} M.,  et~al., 2014a, \mn@doi [\mnras] {10.1093/mnras/stu1536},
  \href {https://ui.adsabs.harvard.edu/abs/2014MNRAS.444.1518V} {444, 1518}

\bibitem[\protect\citeauthoryear{{Vogelsberger} et~al.,}{{Vogelsberger}
  et~al.}{2014b}]{vogelsberger14b}
{Vogelsberger} M.,  et~al., 2014b, \mn@doi [\nat] {10.1038/nature13316}, \href
  {https://ui.adsabs.harvard.edu/abs/2014Natur.509..177V} {509, 177}

\bibitem[\protect\citeauthoryear{{Vogelsberger}, {Marinacci}, {Torrey}  \&
  {Puchwein}}{{Vogelsberger} et~al.}{2020}]{vogelsberger20}
{Vogelsberger} M.,  {Marinacci} F.,  {Torrey} P.,   {Puchwein} E.,  2020,
  \mn@doi [Nature Reviews Physics] {10.1038/s42254-019-0127-2}, \href
  {https://ui.adsabs.harvard.edu/abs/2020NatRP...2...42V} {2, 42}

\bibitem[\protect\citeauthoryear{{Voggel} et~al.,}{{Voggel}
  et~al.}{2021}]{voggel21}
{Voggel} K.~T.,  et~al., 2021, arXiv e-prints, \href
  {https://ui.adsabs.harvard.edu/abs/2021arXiv211114854V} {p. arXiv:2111.14854}

\bibitem[\protect\citeauthoryear{{Wang}, {Dutton}, {Stinson}, {Macci{\`o}},
  {Penzo}, {Kang}, {Keller}  \& {Wadsley}}{{Wang} et~al.}{2015}]{wang15}
{Wang} L.,  {Dutton} A.~A.,  {Stinson} G.~S.,  {Macci{\`o}} A.~V.,  {Penzo} C.,
   {Kang} X.,  {Keller} B.~W.,   {Wadsley} J.,  2015, \mn@doi [\mnras]
  {10.1093/mnras/stv1937}, \href
  {https://ui.adsabs.harvard.edu/abs/2015MNRAS.454...83W} {454, 83}

\bibitem[\protect\citeauthoryear{{Weinberger} et~al.,}{{Weinberger}
  et~al.}{2017}]{weinberger17}
{Weinberger} R.,  et~al., 2017, \mn@doi [\mnras] {10.1093/mnras/stw2944}, \href
  {https://ui.adsabs.harvard.edu/abs/2017MNRAS.465.3291W} {465, 3291}

\bibitem[\protect\citeauthoryear{{Wetzel}, {Hopkins}, {Kim},
  {Faucher-Gigu{\`e}re}, {Kere{\v{s}}}  \& {Quataert}}{{Wetzel}
  et~al.}{2016}]{wetzel16}
{Wetzel} A.~R.,  {Hopkins} P.~F.,  {Kim} J.-h.,  {Faucher-Gigu{\`e}re} C.-A.,
  {Kere{\v{s}}} D.,   {Quataert} E.,  2016, \mn@doi [\apjl]
  {10.3847/2041-8205/827/2/L23}, \href
  {https://ui.adsabs.harvard.edu/abs/2016ApJ...827L..23W} {827, L23}

\bibitem[\protect\citeauthoryear{{Zhu} et~al.,}{{Zhu} et~al.}{2020}]{zhu20}
{Zhu} L.,  et~al., 2020, \mn@doi [\mnras] {10.1093/mnras/staa1584}, \href
  {https://ui.adsabs.harvard.edu/abs/2020MNRAS.496.1579Z} {496, 1579}

\bibitem[\protect\citeauthoryear{{Zhu} et~al.,}{{Zhu} et~al.}{2021}]{zhu21}
{Zhu} L.,  et~al., 2021, arXiv e-prints, \href
  {https://ui.adsabs.harvard.edu/abs/2021arXiv211013172Z} {p. arXiv:2110.13172}

\bibitem[\protect\citeauthoryear{{Zibetti}, {White}  \& {Brinkmann}}{{Zibetti}
  et~al.}{2004}]{zibetti04}
{Zibetti} S.,  {White} S. D.~M.,   {Brinkmann} J.,  2004, \mn@doi [\mnras]
  {10.1111/j.1365-2966.2004.07235.x}, \href
  {https://ui.adsabs.harvard.edu/abs/2004MNRAS.347..556Z} {347, 556}

\bibitem[\protect\citeauthoryear{{Zinger} et~al.,}{{Zinger}
  et~al.}{2020}]{zinger20}
{Zinger} E.,  et~al., 2020, \mn@doi [\mnras] {10.1093/mnras/staa2607}, \href
  {https://ui.adsabs.harvard.edu/abs/2020MNRAS.499..768Z} {499, 768}

\bibitem[\protect\citeauthoryear{{Zoccali} et~al.,}{{Zoccali}
  et~al.}{2017}]{zoccali17}
{Zoccali} M.,  et~al., 2017, \mn@doi [\aap] {10.1051/0004-6361/201629805},
  \href {https://ui.adsabs.harvard.edu/abs/2017A&A...599A..12Z} {599, A12}

\bibitem[\protect\citeauthoryear{{van den Bosch} \& {Ogiya}}{{van den Bosch} \&
  {Ogiya}}{2018}]{vandenbosch18}
{van den Bosch} F.~C.,  {Ogiya} G.,  2018, \mn@doi [\mnras]
  {10.1093/mnras/sty084}, \href
  {https://ui.adsabs.harvard.edu/abs/2018MNRAS.475.4066V} {475, 4066}

\makeatother
\end{thebibliography}
\input{tng.bbl}



\appendix

\section{Properties of TNG50 galaxies}\label{sec: properties_long}

Below we describe in detail the properties of TNG50 galaxies as well as their stellar particles that were mentioned in Section \ref{sec: properties_short}. They are either directly available from the corresponding halo/subhalo, stellar particle or supplementary data catalogues on the TNG website\footnote{\url{https://www.tng-project.org/data/docs/specifications/}}. The classification of `bar-like' signatures for galaxies is available upon reasonable request from the corresponding author.

\subsection{Bulk Properties}\label{sec: galprops}

We here describe how different galaxy bulk properties are defined and measured in order to define different galaxy populations that are analyzed in Section \ref{sec: demo}. All properties refer to $\mathrm{z}=0$.

\begin{itemize}
    \item \textit{Mass}: Generally all galaxy masses are reported to be the total mass of particles of a given type (or all types in the case of dynamical mass) bound to a specific subhalo as identified by the \textsc{Subfind} algorithm.
    \item \textit{Environment}: We crudely define the environment of a galaxy by distinguishing between centrals and satellites. A central galaxy is the most massive subhalo in its corresponding FoF halo, all other galaxies within the same FoF halo are satellite galaxies.
    \item \textit{Star formation activity}: Whether a galaxy is actively forming stars or not is classified according to \citet{pillepich19} \citep[see also][]{donnari19,donnari21a,donnari21b}, who determined the logarithmic distance from the star forming main sequence for each galaxy ($\Delta\log_{10}\mathrm{SFR}$). For this, the instantaneous star formation rate (SFR) of the gas cells as well as galaxy stellar masses were calculated within twice the stellar half mass radius. Star forming galaxies have $\Delta\log_{10}\mathrm{SFR}\geq-0.5$ and quenched galaxies have $\Delta\log_{10}\mathrm{SFR}\leq-1$, whereas galaxies in the green valley are in between those two values. Unless otherwise stated we will omit the distinction of green valley galaxies and also classify them as quenched.
    \item \textit{Morphology}: We quantify disk or bulge dominated galaxies based on the kinematic classification by \citet{genel18}. For each stellar particle the circularity parameter $\epsilon$ is calculated, which gives the ratio of the particle's specific angular momentum in z-direction and its theoretical maximum angular momentum at that specific binding energy \citep{abadi03,marinacci14}. Then the mass fraction of all stellar particles with $\epsilon>0.7$ and within ten stellar half mass radii is computed. If that fractional mass is above 0.4 we classify that galaxy as disky \citep[see][]{joshi20}, otherwise the galaxy is bulge dominated.
    \item \textit{Bar-like signatures}: We also provide a quick estimate of whether a galaxy has a bar-like structure in its center. For this, we calculate $A_2$, the ratio between the second and zeroth term of the amplitude of the Fourier expansion, from the face-on stellar surface density of each galaxy as a function of the 2D radius in $\sim 0.04\,\mathrm{dex}$ steps, where each bin is ensured to have at least 100 stellar particles. We then identify peaks in the $A_2$-radius plane with a prominence of at least 0.05. After that, the value of $A_2$ for the largest peak within a radius of 10\,kpc is recorded. We impose this radius cut to mitigate the effect of other $A_2$ features that may be present at larger radii \citep[see also][]{frankel22}. This is done for all snapshots between $\mathrm{z}=0$ (\texttt{SnapNum} 99) and $\mathrm{z}=4.2$ (\texttt{SnapNum} 20). Similarly to \citet{rosas-guevara20}\footnote{We find that our computed $A_2$ values are slightly lower than those measured by methods of \citet{rosas-guevara20}, hence we adopt their `weak bar' threshold of 0.2, whereas their `strong bars' have values of $A_2\geq0.3$.}, we then define a bar-like structure, when the maximum $A_2$ value (at a given time step) is above 0.2. Bar-like structures at $\mathrm{z}=0$ are defined solely based on their instantaneous $A_2$ value at that snapshot. While this method leads to accurate identification of symmetrically elongated `bar-like' features, we do not check if this is actually a `proper' bar in the astrophysical sense. Nevertheless, our classification leads to a bar fraction of around 40\% (50\%) for disk (all) galaxies above $10^{10}\,\mathrm{M}_{\odot}$ in stellar mass, which is consistent with observations \citep[see e.g.][]{bars1,bars2,bars3}.
    \item \textit{AGN feedback}: We quantify the severity of feedback from supermassive black holes by determining whether a specific galaxy lies below or above the median scaling relation of TNG50 galaxies \citep[e.g. similarly to][for observations]{nacho18}. Typical AGN feedback defining properties could be the black hole (BH) mass and the cumulative energy injection in the thermal and/or kinetic feedback modes \citep[see e.g.][for background]{weinberger17,zinger20}. Such scaling relations are always computed with respect to the total stellar mass of galaxies as well as for the total TNG50 galaxy sample above $5\times10^{8}\,\mathrm{M}_{\odot}$ (see Section \ref{sec: sample_choice}). Properties of black hole particles per galaxy are computed as the sum of all black holes particles associated to a given galaxy via \textsc{Subfind}.
    \item \textit{Physical size}: Similarly, we define extended or compact galaxies depending on whether they are below or above the median stellar mass-size relation of TNG50 galaxies. The size is the 3D stellar half mass radius.
\end{itemize}

\subsection{Stellar Particle Properties}\label{sec: starprops}

Stellar particle properties at $\mathrm{z}=0$ that are analyzed in Section \ref{sec: pop} are briefly described here:

\begin{itemize}
    \item \textit{Age}: We define the age of a stellar particle as the lookback time in Gyr. This quantity is calculated from the field \texttt{GFM\_StellarFormationTime}, which provides the \emph{exact} time of birth of a star in scale factors.
    \item \textit{Metallicity}: We convert the mass fraction in metals $Z$ as provided by the field \texttt{GFM\_Metallicity} to $\log_{10} Z/Z_{\odot}$ with $Z_{\odot}=0.02$. This follows conventions adopted in observations, e.g. \citet{gallazzi20}.
    \item \textit{[Mg/Fe]}: We calculate the magnesium-to-iron abundance from the mass fraction in magnesium $Z_{\mathrm{Mg}}$ and iron $Z_{\mathrm{Fe}}$ provided by the simulation (\texttt{GFM\_Metals}) via $[\mathrm{Mg/Fe}]=\log_{10}(Z_{\mathrm{Mg}}/Z_{\mathrm{Mg},\, \odot})-\log_{10}(Z_{\mathrm{Fe}}/Z_{\mathrm{Fe},\, \odot})$. The adopted solar values are $Z_{\mathrm{Mg},\, \odot}=0.00064298$ and $Z_{\mathrm{Fe},\, \odot}=0.001218$ respectively from \citet{asplund}.
    \item \textit{Circularity $\epsilon$}: We calculate the instantaneous circularity of each stellar particle following \citet{genel18}. For this we first compute the specific angular momentum of each particle in z-direction by aligning the z-axis of the simulation box with the total angular momentum of stellar particles within twice the stellar half mass radius of a given galaxy. The theoretical maximum angular momentum each stellar particle can have at its recorded specific binding energy (i.e. $\frac{1}{2}|\mathbf{v}|^2+\Phi$) is calculated by sliding a maximum filter across the particle list of specific angular momenta sorted by their total specific binding energy with a window size of one hundred. Stars with circularities around zero are on random motion dominated orbits, whereas values close to one indicate more circular orbits. Negative circularities depict counter rotating orbits.
    
\end{itemize}

\section{Validation for analysis of TNG50 galaxy centers}\label{appendix: validate}

\begin{figure*}
    \centering
    \includegraphics[width=0.45\textwidth]{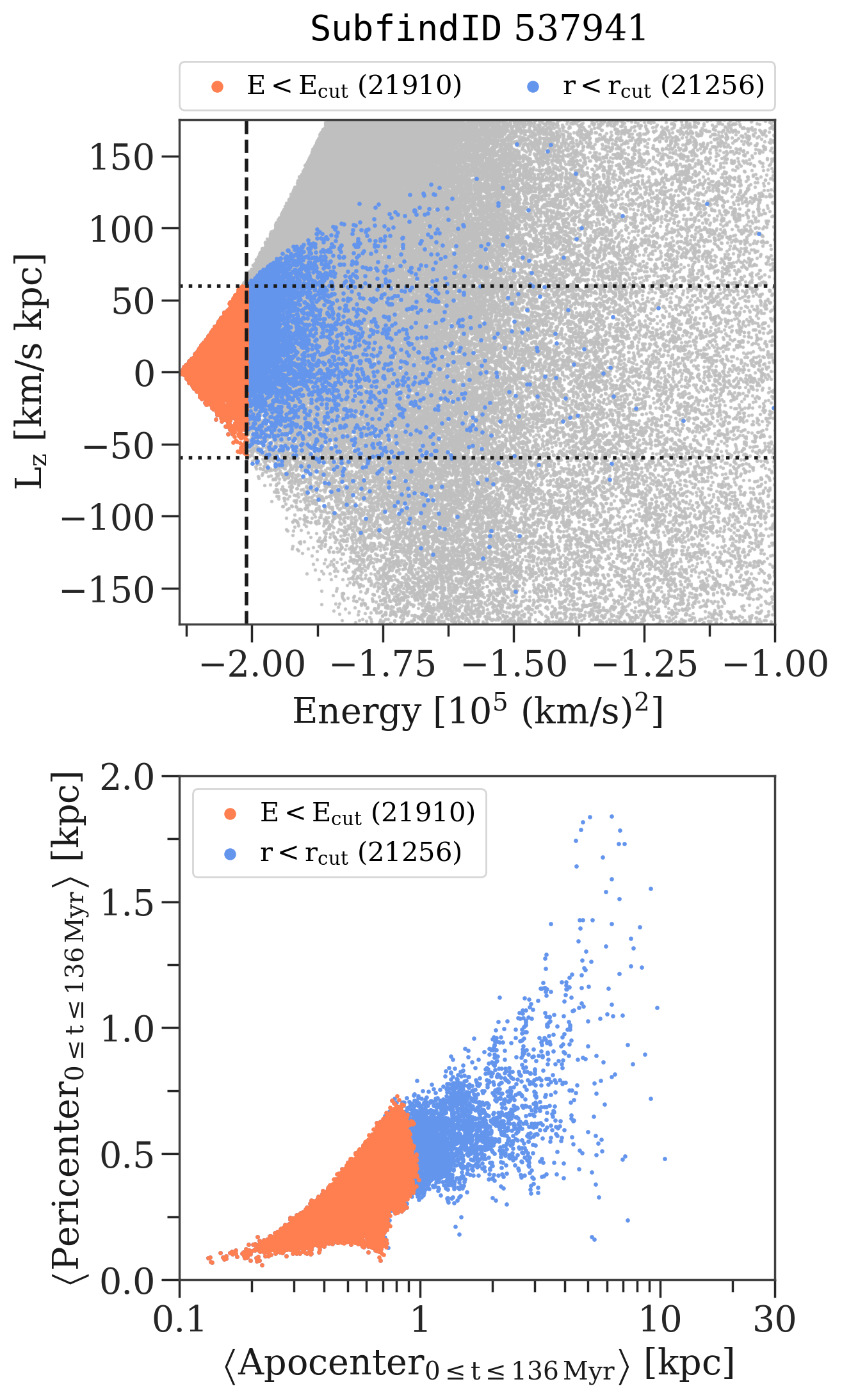}
    \includegraphics[width=0.45\textwidth]{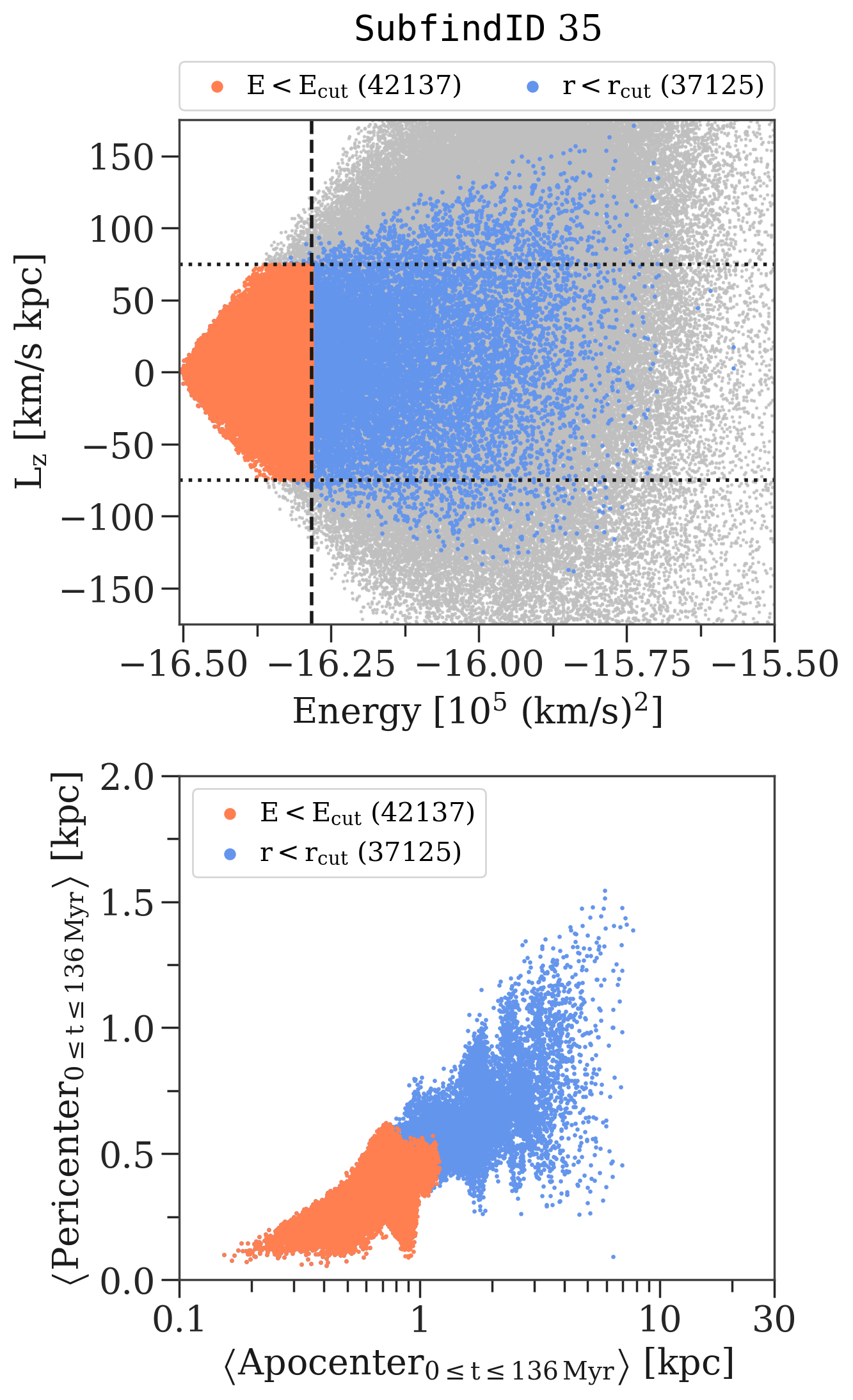}
    \caption{\textbf{Selection of stellar particles belonging to a galaxy's center}, \textit{left column} for \texttt{SubfindID} 537941 and \textit{right column} for \texttt{SubfindID} 35. Blue points show stellar particles that have radii smaller than $r_{\mathrm{cut}}=500\,\mathrm{pc}$ and orange points where selected based on their energies being smaller than $E_{\mathrm{cut}}$ according to Equation \protect\ref{eq: ecut}, also with $r_{\mathrm{cut}}=500\,\mathrm{pc}$. \textit{Top row}: Angular momentum of selected particles as a function of energy. The dashed lines emphasizes $E_{\mathrm{cut}}$, whereas the two dotted lines show $\mathrm{L}_{\mathrm{z}}=\pm r_{\mathrm{cut}}v_{\mathrm{circ}}(r_{\mathrm{cut}})$. Grey points show all other stellar particles belonging to the respective galaxy. \textit{Bottom row}: Pericenter of stellar particles versus their apocenter time averaged from subbox outputs between 0 (full box snapshot 99) and 136\,Myr (full box snapshot 98). Stellar particles purely selected on their radius have a wider distribution in their energies and are hence able to move much further outside our selected spherical volume of 500\,pc, which is also reflected by their larger apocenters.}
    \label{fig: ecutvsradcut}
\end{figure*}

We briefly validate and justify here the analysis choices we made in Sections \ref{sec: center} and \ref{sec: origindef} of the main text using two TNG50 galaxies as examples where applicable. Both of these galaxies are also contained within the subboxes, i.e. smaller regions of the full simulation box, which offer 3600 snapshots resulting in a time sampling of $2-3 \, \mathrm{Myr}$. At $\mathrm{z}=0$, \texttt{SubfindID} 537941 is a Milky Way like galaxy, found in `Subbox0', and \texttt{SubfindID} 35, found in `Subbox2', is a compact $\sim10^{10} \, \mathrm{M}_{\odot}$ in stellar mass galaxy, that quenched approximately 9\,Gyr ago and is now found in the most massive halo ($\sim10^{14} \, \mathrm{M}_{\odot}$) in TNG50. Both of these subhalos stay inside the subbox across their life time making it possible to compare their full histories in the higher cadence outputs to that of the hundred full box snapshot outputs. \par

\begin{figure}
    \centering
    \includegraphics[width=\columnwidth]{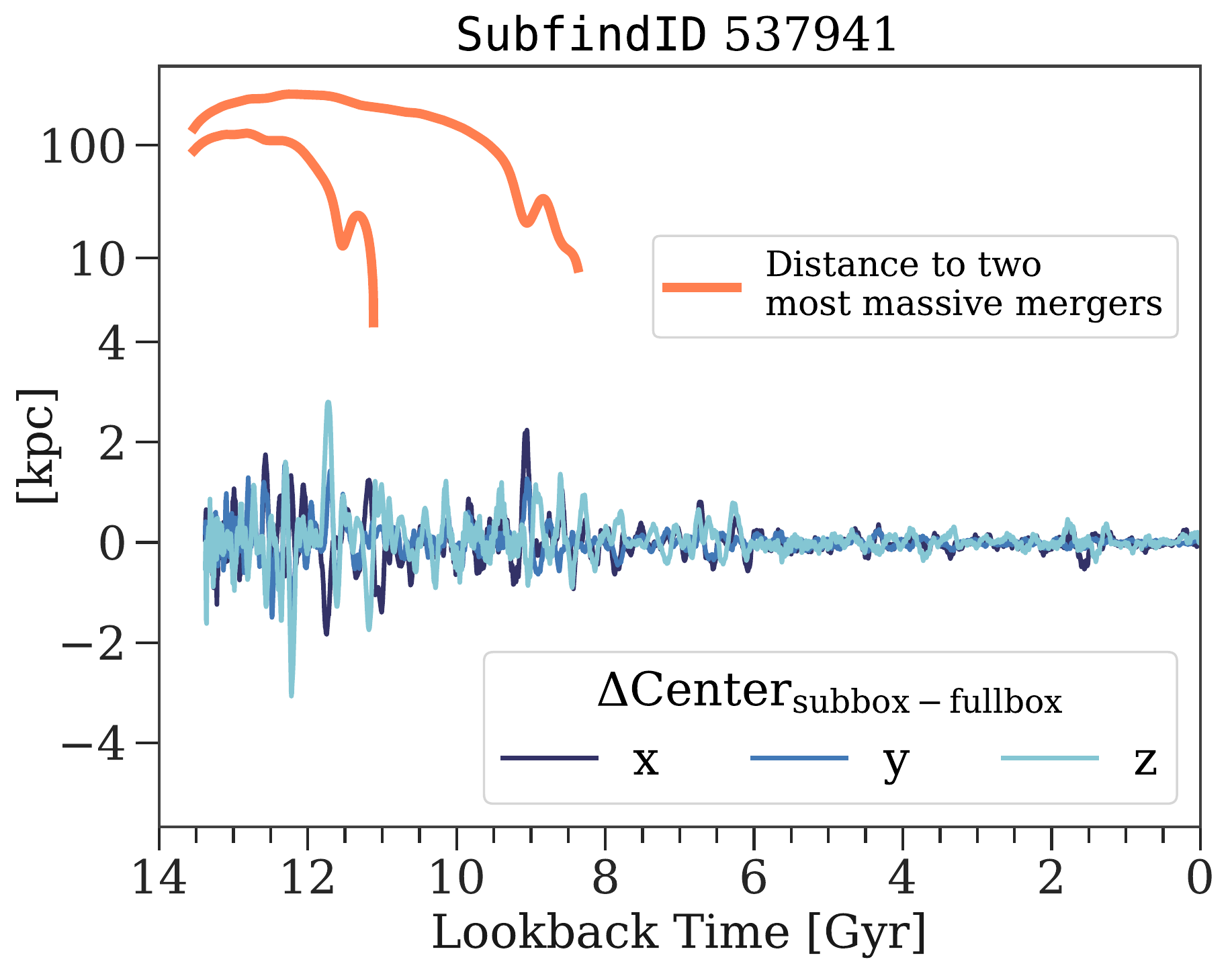}\\
    \vspace*{12pt}
    \includegraphics[width=\columnwidth]{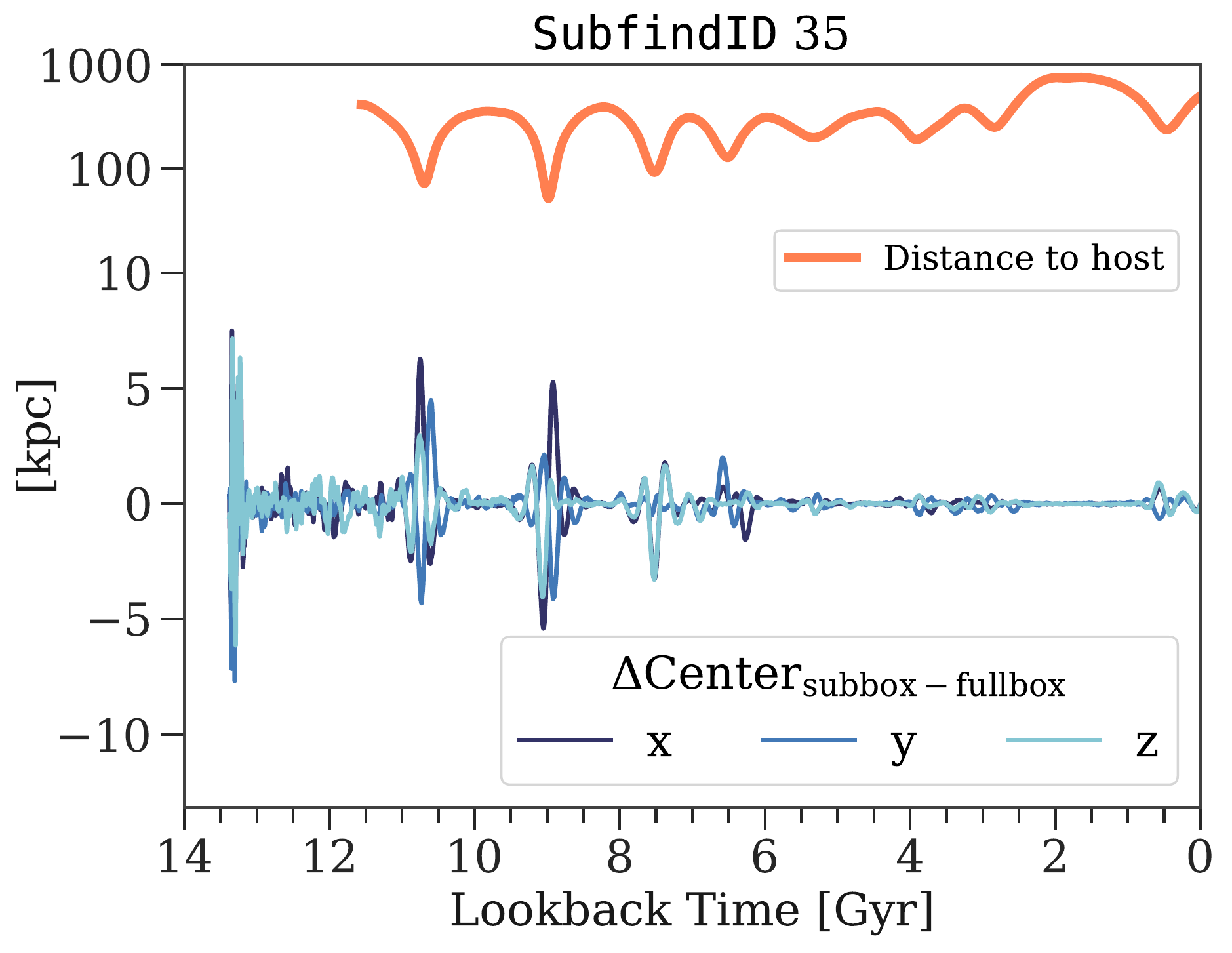}
    \caption{\textbf{Differences in galaxy centering between full box and subbox snapshots.} Comparison between the interpolated (cubic spline) center position from the 100 full box snapshots (using \texttt{SubhaloPos} from the subhalo catalogue) and the center position (most bound stellar particle) from the higher cadence subbox outputs for two example galaxies (\textit{blue shaded lines}). The \textit{top panel} shows \texttt{SubfindID} 537941, a Milky Way like galaxy and the \textit{bottom panel} shows \texttt{SubfindID} 35, a $\sim10^{10} \, \mathrm{M}_{\odot}$, quenched galaxy that is now a satellite of the most massive halo in TNG50. We see that the interpolated center of the galaxies starts to deviate significantly from the true center position when there are pericenter passages from satellites that merge with galaxy or when the galaxy itself is a satellite and approaches its host (\textit{thick orange lines}).}
    \label{fig: subbox_centering}
\end{figure}

\subsection{Selection of central stars}\label{appendix: validate1}

We show in Figure \ref{fig: ecutvsradcut} the energy and angular momentum distribution of stars in the galaxy's centers at $\mathrm{z}=0$ by selecting particles based on their current radius as well as on their energies according to Equation \ref{eq: ecut}. The selection was made with $r_{\mathrm{cut}}=500\,\mathrm{pc}$. We also show their peri- and apocenters which we calculated by recording their radii from all 16 subbox outputs between the last full box snapshot, i.e. $\mathrm{z}=0$ (\texttt{SnapNum} 99), and the one prior to that, i.e. $\approx 136\,\mathrm{Myr}$ (\texttt{SnapNum} 98) before. We then found all minima and maxima and took the average respectively. \par
Some particles selected by their instantaneous radius at $\mathrm{z}=0$ have large energies and are hence able to move away from the center to much larger radii, i.e. they are not actually spending the majority of their orbital time within our selected spherical aperture. This is also reflected by their larger (up to 10\,kpc) apocenters, whereas particles selected by their energy have time-averaged peri- and apocenters not larger than 1\,kpc. Even though that is larger than our selected $r_{\mathrm{cut}}$ value, probably due to our simplifying assumptions in calculating $E_{\mathrm{cut}}$, we argue that this selection gives a much cleaner selection of stars actually belonging to the center without interloping particles on much more eccentric orbits. Both selection criteria yield a comparable number of stars, with around 21000 stars for \texttt{SubfindID} 537941 (1.6\% of total amount of stars) and $\sim\,40000$ stars for \texttt{SubfindID} 35 ($\sim\,15\%$ of total amount of stars). \par

\subsection{Definition of birth radii of stellar particles}\label{appendix: validate3}

In Figure \ref{fig: subbox_centering} we show the difference between the center position of the two galaxies once taken from the high cadence subbox outputs (the most bound stellar particle)\footnote{As there is in fact no subhalo information from \textsc{Subfind} available for the subboxes, we start off with the interpolated center values from the full box snapshots and then recalculate the center in a 5\,kpc box around that, recenter and recalculate the center again in a 5\,kpc box around that. We verified that this gives the correct center by inspecting the galaxies by eye. We note however that this approach will not yield correct results in a general black box fashion.} and once from the \texttt{SubhaloPos} argument from the full box snapshots, which we interpolated onto the finer time sampling of the subbox outputs using a cubic spline. It is evident that deviations of around $2-5\,\mathrm{kpc}$ appear, specifically at times of pericenter passages of either satellites merging with the host (\texttt{SubfindID} 537941) or of the galaxy itself around its host (\texttt{SubfindID} 35). \par
\begin{figure*}
    \centering
    \includegraphics[width=\textwidth]{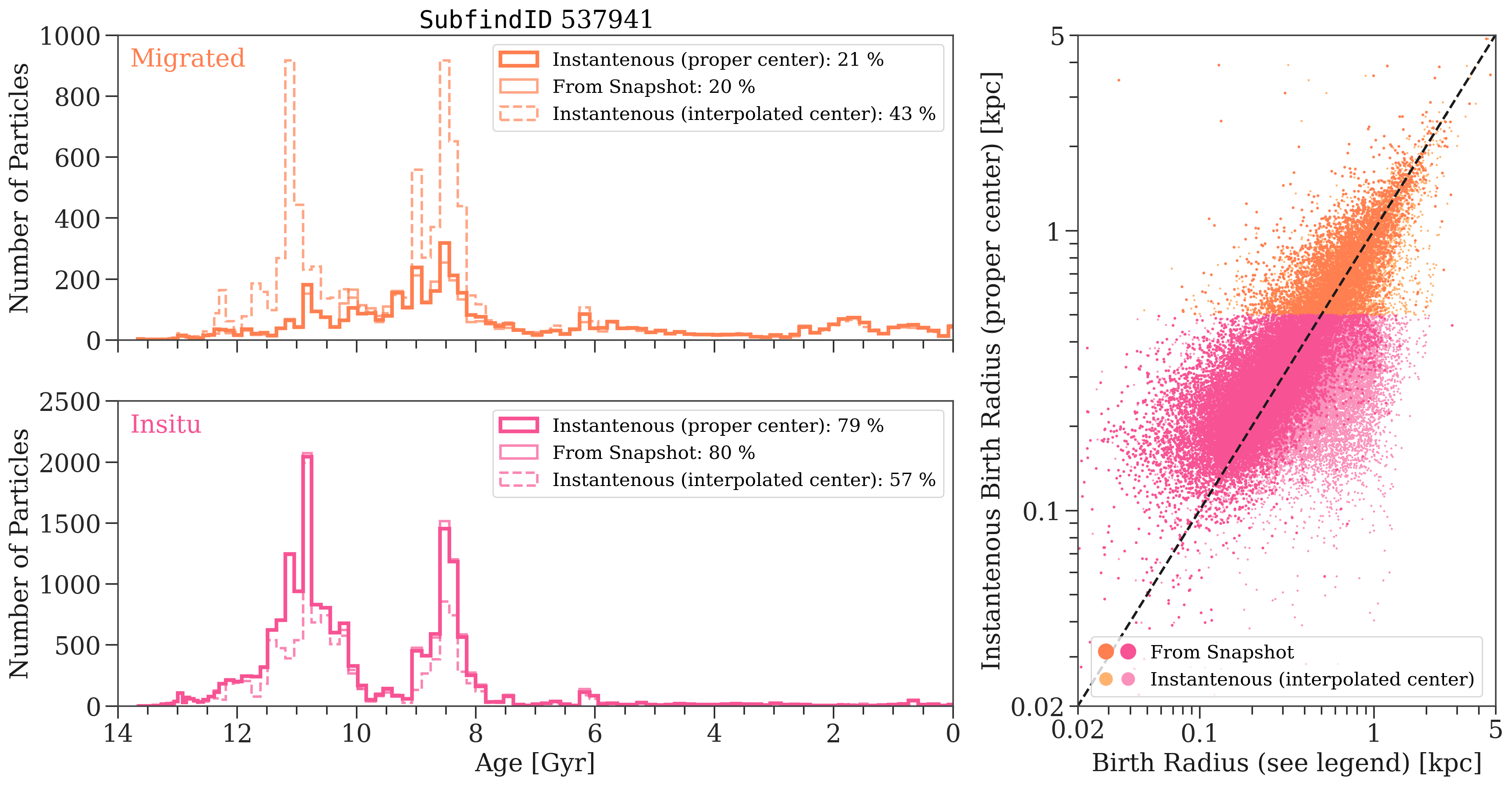} 
    \vspace*{12pt}
    \includegraphics[width=\textwidth]{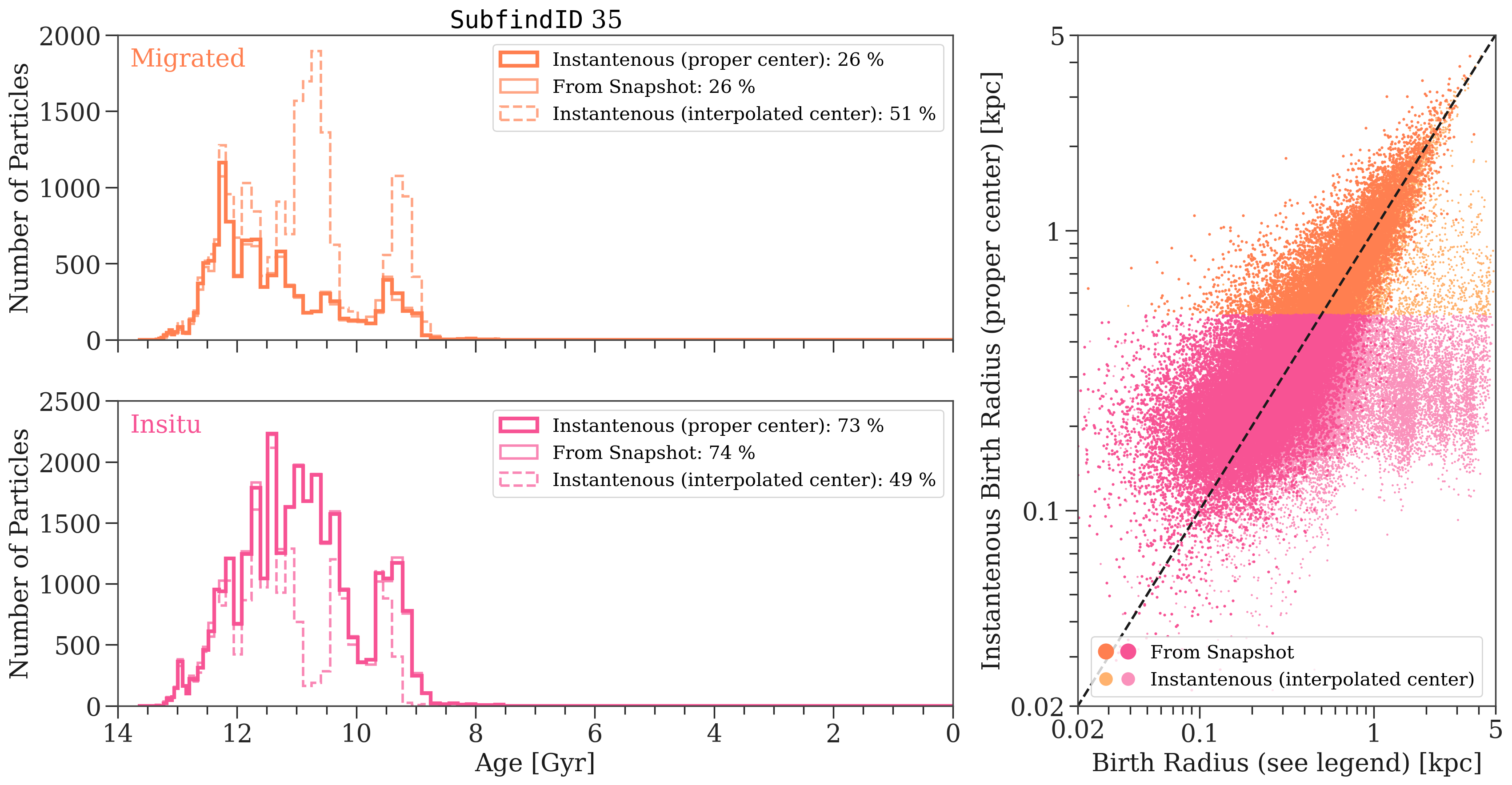}
    \caption{\textbf{Effect of erroneous galaxy centering on the classification of migrated stellar particles.} \textit{Left panels}: Histograms of particles being classified as migrated (\textit{orange}) and in-situ (\textit{pink}) using three different approaches: the instantaneous position at birth with the proper centering from the subbox outputs (\textit{thick solid line}), the same but with the interpolated center (\textit{thin dashed line}) as well as the position of particles taken from the full box snapshots they first appeared in (\textit{thin solid line}). \textit{Right panels}: Comparing the values of the instantaneous birth radii with the centering from the subbox outputs with the ones with the interpolated center (\textit{fainter, smaller points}) as well as the radii from the full box snapshots in which the particles first appeared in (\textit{bolder, bigger points}). The one-to-one relation is also shown (\textit{black dashed lines}). The \textit{top panel} shows \texttt{SubfindID} 537941, a Milky Way like galaxy and the \textit{bottom panel} shows \texttt{SubfindID} 35, a $\sim10^{10} \, \mathrm{M}_{\odot}$, quenched galaxy that fell into the most massive halo of TNG50. Using the instantaneous birth position with the interpolated center leads to a wrong classification of migrated and in-situ stars, whereas using the positions from the full box snapshots at birth gives comparable results to the instantaneous birth positions with the proper centering in the actual number of particles in the two categories as well as the values of their birth radii.}
    \label{fig: subbox_migrated}
\end{figure*}

These deviations of a galaxy's center position are enough to severely miss-classify migrated and in-situ stars, when their instantaneous birth positions, as provided by the simulations output (\texttt{BirthPos}), are used in conjunction with the interpolated subhalo center. For example, in Figure \ref{fig: subbox_migrated} we show histograms of migrated and in-situ stars selected according to their instantaneous birth position using the proper center from the subbox and the interpolated one from the full box snapshots. For both galaxies the number of migrated stars doubles when the interpolated center is used compared to the correct center. \par
Ideally, we would like to apply our analysis to as many galaxies as possible, but as only a handful of galaxies reside inside the subboxes during their whole life time, we need an alternative measure for classifying them into migrated and in-situ stars that can be applied to the full box of TNG50. We therefore take the particle's position at the full box snapshot it first appeared in as its ``birth'' position and compare it to the instantaneous one with the proper centering, also shown in Figure \ref{fig: subbox_migrated}. We see that the shapes of their histograms as well as their percentages match. Of course, their classification is not identical, as particles move between their exact formation time and the time of when the snapshot was taken, however this measure seems to be more accurate than using the instantaneous birth positions with the interpolated center. \par
Finally, we compare the instantaneous birth radii using the correct center of the migrated and in-situ stars (classified with the instantaneous positions using the correct center) with birth radii calculated by the other two methods. The birth radii determined from the full box snapshots scatter around the one-to-one relation, whereas the ones with the interpolated center do not. Hence, we conclude that applying the migrated and in-situ classification based on their birth snapshot position to the whole TNG50 box seems to provide us with similar knowledge about their origin as if we would have used their instantaneous birth position. \par

\section{The total ex-situ stellar mass fraction of TNG50}\label{appendix: validate2}

\begin{figure}
    \centering
    \includegraphics[width=\columnwidth]{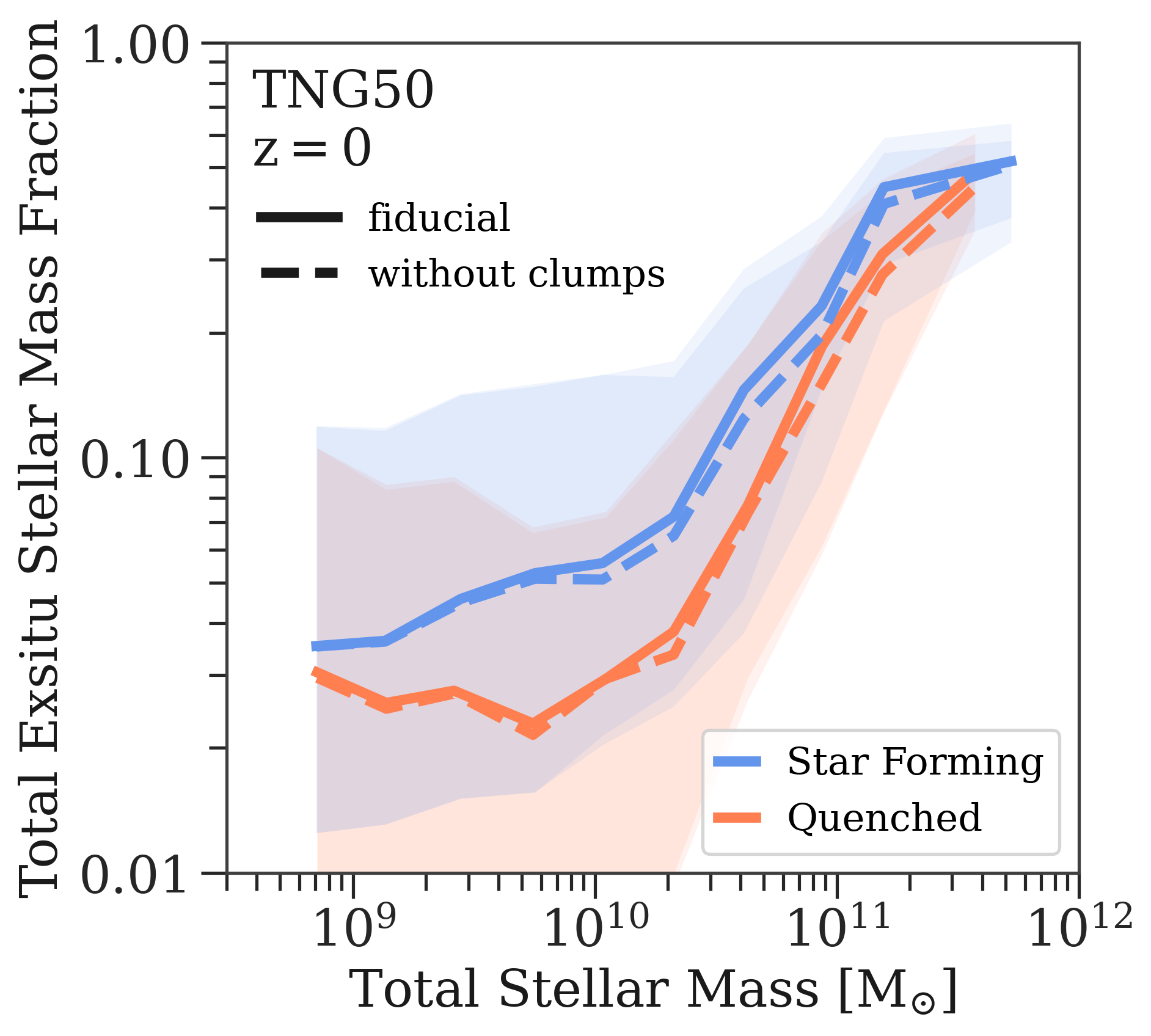}
    \caption{\textbf{Total ex-situ stellar mass fraction versus total stellar mass in TNG50 at} $\mathbf{z=0.}$ The solid lines show the fiducial ex-situ fraction as classified by methods in \protect\citet{rodriguesgomez16}, whereas the dashed lines show total ex-situ fractions excluding spurious ($\mathtt{SubhaloFlag}=0$) galaxies or ``clumps'' (see Section \protect\ref{sec: migrated}). The faint bands depict the 16th and 84th percentiles. There is no significant difference between the two classifications regarding the total ex-situ stellar mass fraction. We also divide by star-forming (\textit{blue} lines) and quenched galaxies (\textit{red} lines) at $\mathrm{z}=0$. The star forming galaxies have higher ex-situ fractions on average at fixed stellar mass than quenched galaxies.}
    \label{fig: tot_exsitu}
\end{figure}

\begin{figure}
    \centering
    \includegraphics[width=\columnwidth]{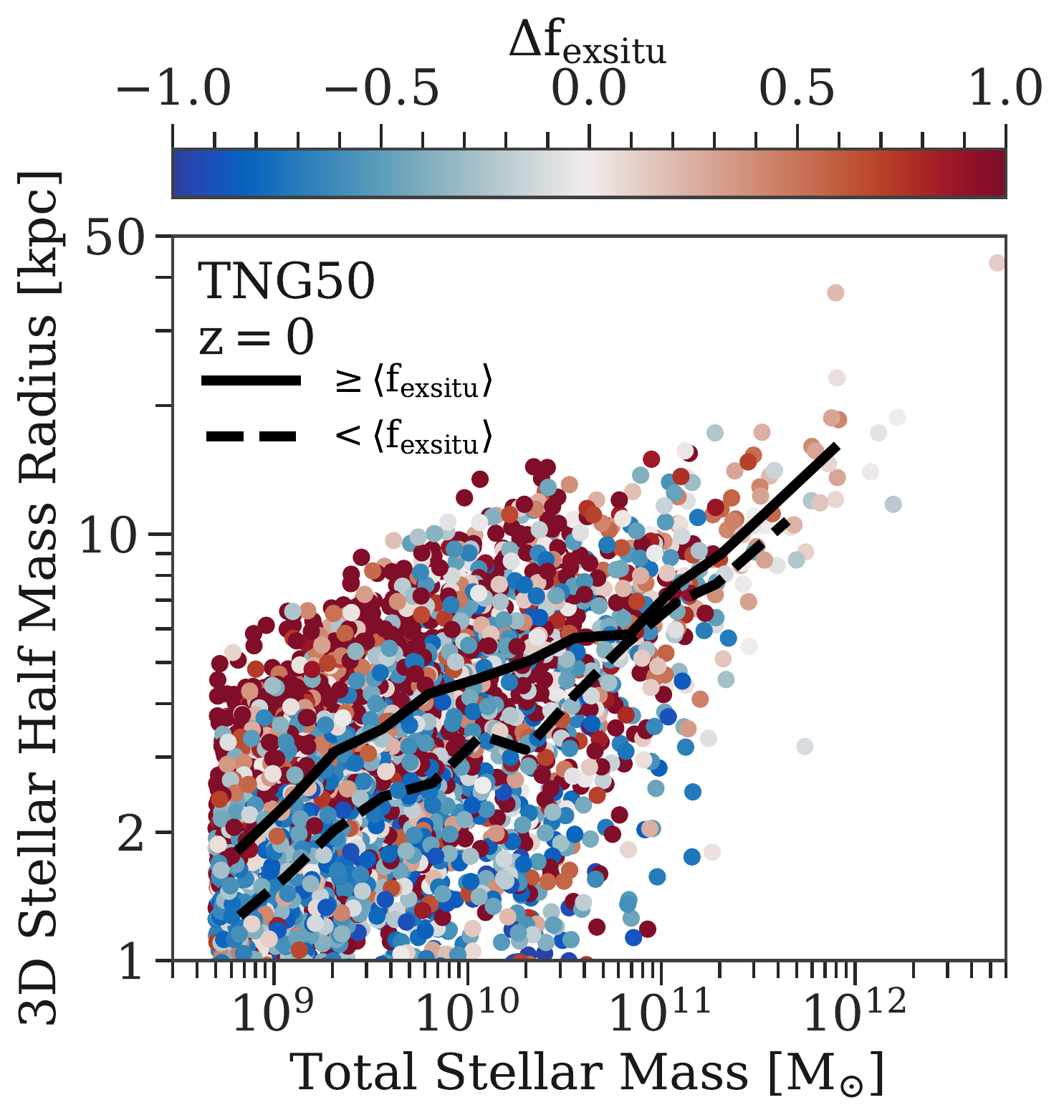}
    \caption{\textbf{Connection between the galaxy mass-size relation and total ex-situ fractions in TNG50 at} $\mathbf{z=0.}$ The 3D half mass radius against total stellar mass for 4344 galaxies in TNG50 ($\mathrm{M}_{\star,\,\mathrm{tot}}>5\times 10^{8}\,\mathrm{M}_{\odot}$). The points are color-coded according to the relative ex-situ fraction $\Delta\mathrm{f}_{\mathrm{exsitu}}$ indicating whether the total ex-situ fraction of a given galaxy above or below the average at fixed galaxy stellar mass. Median mass-size relations are shown separately for galaxies having above (\textit{solid black line}) and below (\textit{dashed black line}) average ex-situ fractions. Below $\sim5\times 10^{10}\,\mathrm{M}_{\odot}$, galaxies with high relative ex-situ fractions are on average more extended.}
    \label{fig: exsitu_rad}
\end{figure}

In Figure \ref{fig: tot_exsitu} we show the total ex-situ stellar mass fraction as a function of total stellar mass for TNG50 galaxies. They are further divided into star forming and quenched galaxies according to their distance to the star forming main sequence at $\mathrm{z}=0$.\par
Most importantly, we see that the total ex-situ stellar mass fraction does not change worryingly, when comparing the fiducial definition of ex-situ stars according to \citet{rodriguesgomez16} as well as our definition, where we exclude stars from accreted satellites classified as clumps (i.e. $\mathtt{SubhaloFlag}=0$). Even though the clumps do not play an important role for the definition of the ex-situ fraction, they are abundant in TNG50 (see Appendix \ref{appendix: validate4}). Only due to our definition, we could confirm the migration of these clumps contributing significantly to the build-up of the galaxy center for galaxies above $10^{11}\,\mathrm{M}_{\odot}$ in stellar mass. \par
Additionally, Figure \ref{fig: tot_exsitu} shows that star forming galaxies have a higher total ex-situ fraction on average than quenched galaxies across all stellar masses. Until galaxy stellar masses of $10^{10}\,\mathrm{M}_{\odot}$ the ex-situ fraction stays roughly constant with values around 4-5\% and 3\% for star forming and quenched galaxies respectively. There is however a large scatter associated with the ex-situ fraction for galaxies in this mass regime. Above $10^{10}\,\mathrm{M}_{\odot}$ the ex-situ fraction sharply increases with galaxy stellar mass reaching approximately 50\% for galaxy between $10^{11}-10^{12}\,\mathrm{M}_{\odot}$ for TNG50. The scatter decreases accordingly. We have checked this exact relation with the lower resolution run TNG50-2 as well as TNG100 and obtained similar results. \par
Figure \ref{fig: exsitu_rad} shows the mass-size relation for TNG50 galaxies colored according their relative ex-situ fraction, i.e. if they have high or low ex-situ fractions with respect to the average typical for their respective stellar mass \citep[following]{merritt20}. Galaxies with stellar masses $\lesssim5\times 10^{10}\,\mathrm{M}_{\odot}$ and above average ex-situ fractions are on average more extended. This median trend is not observed for the high-mass end, however compact galaxies at $10^{11}\,\mathrm{M}_{\odot}$ have almost exclusively below average ex-situ fractions in agreement with other studies (see \citealt[][for EAGLE]{davison20}; \citealt[][for TNG100]{merritt20}; \citealt[][for TNG50]{zhu21}).

\section{Resolution Convergence Tests}\label{appendix: validate5}

To test numerical convergence we perform the exact same analysis from the main text for three different resolution realizations of TNG50: TNG50-1 (flagship), TNG50-2 and TNG50-3. Throughout, the galaxy sample selection is the same as in Section \ref{sec: sample_choice} except we omit cutting galaxies with less then one hundred stellar particles in their centers. \par
For reference, the mass resolution of TNG50-2 and TNG50-3 is 8 and 64 times and the gravitational softening length is 2 and 4 times worse than TNG50-1 respectively. The latter translates to physical sizes of 288\,pc (TNG50-1), 576\,pc (TNG50-2) and 1.152\,kpc (TNG50-3) for collisionless particles at $\mathrm{z}=0$. \par

\begin{figure*}
    \centering
    \includegraphics[width=\textwidth]{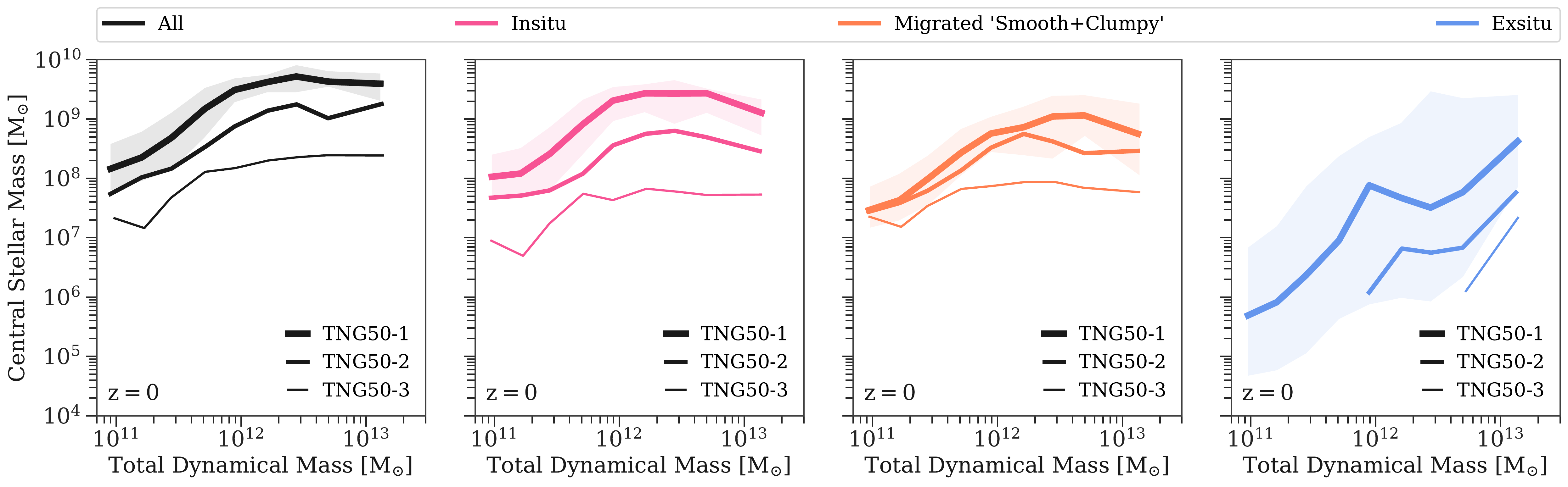}\\
    \vspace*{12pt}
    \includegraphics[width=0.75\textwidth]{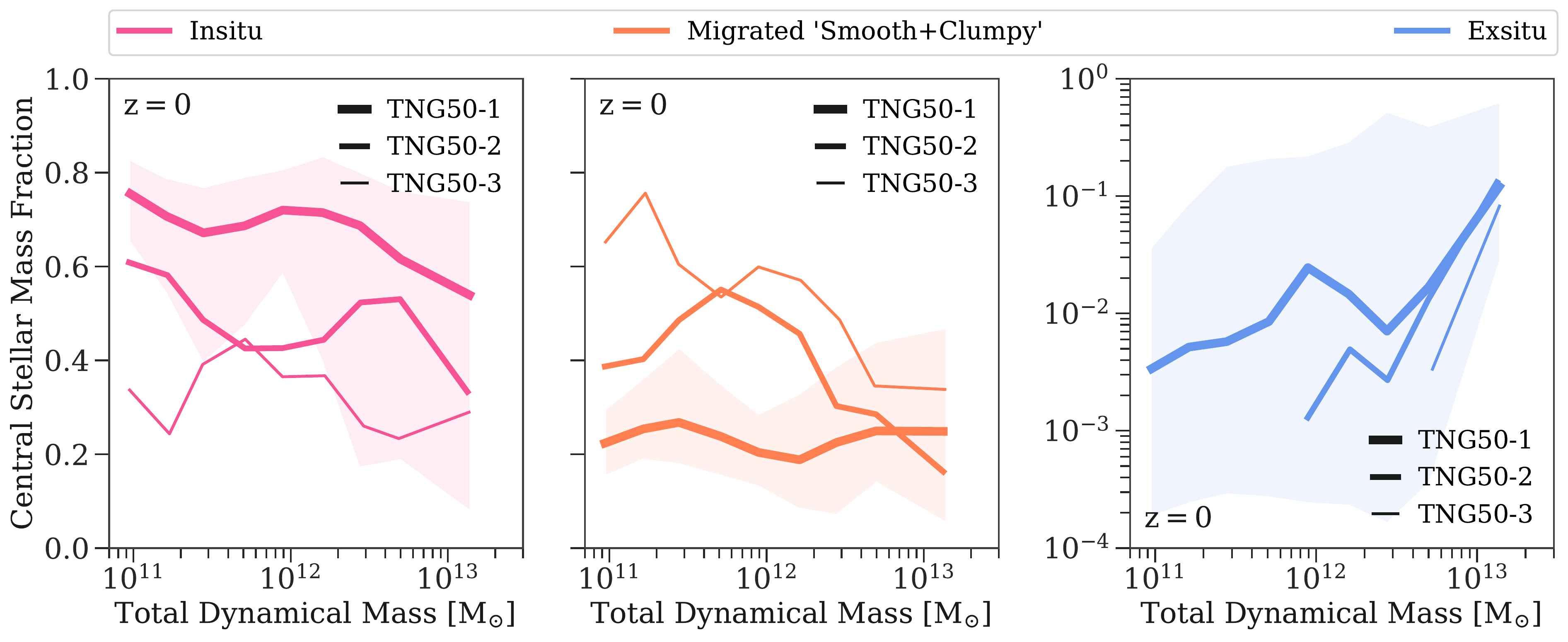}
    \caption{\textbf{Influence of the numerical resolution on the central (500\,pc) in-situ, migrated and ex-situ stars of TNG50 galaxies at} $\mathbf{z=0}.$ \textit{Top panel}: Median trends of the central stellar mass as a function of a galaxy's total dynamical mass for all \textit{(all)}, in-situ \textit{(pink)}, migrated \textit{(orange)} and ex-situ \textit{(blue)} stars and different numerical resolution realizations of the same cosmological volume. The thicker the line the better the numerical resolution. Shaded areas show the 16th and 84th percentiles for the highest resolution run. \textit{Bottom panel}: Instead of the absolute stellar mass we now show the central stellar mass fraction of the in-situ, migrated and ex-situ stars. The central ex-situ mass faction is converging, whereas the behaviour is more complex for the in-situ and migrated fraction.}
    \label{fig: convergence}
\end{figure*}

In the top row of Figure \ref{fig: convergence} we show the results of the central stellar mass of the in-situ, migrated and ex-situ stars as a function of total dynamical mass for all three resolutions. We see that the \emph{total} stellar mass in the central 500\,pc of TNG50 galaxies is converging at all halo masses, meaning that the distance between TNG50-1 and TNG50-2 ($\sim1\,\mathrm{dex}$) is around 50\% smaller than the distance between TNG50-2 and TNG50-3 ($\sim0.5\,\mathrm{dex}$). The same is true when looking at the central mass for just the in-situ stars, even though the converging of the lines becomes much less obvious. For the migrated stars it looks like the central mass is better converged for galaxies with total dynamical masses below $10^{12}\,\mathrm{M}_{\odot}$. However, when comparing the central stellar mass \emph{fraction} of in-situ and migrated stars (bottom row of Figure \ref{fig: convergence}), the convergence behaviour seems to be much more complex. For galaxies outside masses of $10^{12\pm0.5}\,\mathrm{M}_{\odot}$, the central stellar mass fractions seem to be converging. Overall though, the central migrated mass becomes on average larger than the in-situ mass with decreasing resolution for all galaxy masses. \par
This is due to the fact that the numerical resolution also influences which stars are classified as in-situ and migrated, which is a consequence of how (spatial and mass) resolution effects star formation and feedback in the TNG model. Better resolution allows for higher gas densities \emph{and} better spatial sampling of the gas cells, which produces galaxies with higher stellar mass \emph{and} more compact sizes. Thus, in an absolute sense more stellar mass resides overall in the center of TNG50 galaxies for higher resolution runs, however differentially more stellar mass comes from outside the fixed 500\,pc aperture for lower resolution runs (see also Section \ref{sec: moresims}). Interestingly, the complexity in the central in-situ and migrated fractions seen around $10^{12}\,\mathrm{M}_{\odot}$ coincides with the complexity in the convergence behaviour of the mass-size relation in TNG50 \citep[see][Figure B1]{pillepich19}. \par
For the \emph{absolute} stellar mass of the central ex-situ stars the start of convergence is not yet apparent, but the \emph{fractions} is clearly converging. Again, the ex-situ mass in the center increases with increasing resolution. Additionally, the minimum galaxy mass that exhibits any ex-situ stellar mass in the center is extended towards lower mass galaxies for increasing resolution. Again, this behaviour is a consequence of the complex interplay between mass and spatial resolution. Subhalos that were either not sufficiently resolved with enough stellar particles or even remained dark at lower resolution are able to form enough stars at higher resolution. As a consequence, galaxies that become accreted are not only more massive in stellar mass, but also more abundant, especially at the low mass end. Thus, the ex-situ stellar mass is higher in general and also contributes at the lower galaxy masses when resolution is increased. Numerical resolution has also been shown to impact the disruption of merging satellite galaxies, which could lead to overmerging, especially in the outskirts of galaxies \citep[see][]{vandenbosch18,merritt20}. Hence, a higher resolution will result in more accreted stellar material in the center of galaxies. Furthermore, the resolution of TNG50 will impact the ability to resolve the dynamics of its resolution elements (stars, clumps, gas cells, etc.), including dynamical friction, at scales much below the spatial resolution of the simulation, i.e. certainly at the tens of pc scales. \par
Even though TNG50 is a tremendous improvement in resolution for cosmological box simulations, we still need a higher resolution to fully assess the convergence behaviour of the absolute central ex-situ stellar mass \citep[see][for a study of the satellite galaxy population of a highly resolved Milky Way like galaxy in a cosmological context and comparison to lower resolution models]{grand21}.

\section{Clumps in TNG50}\label{appendix: validate4}

\begin{figure*}
    \centering
    \includegraphics[width=0.85\textwidth]{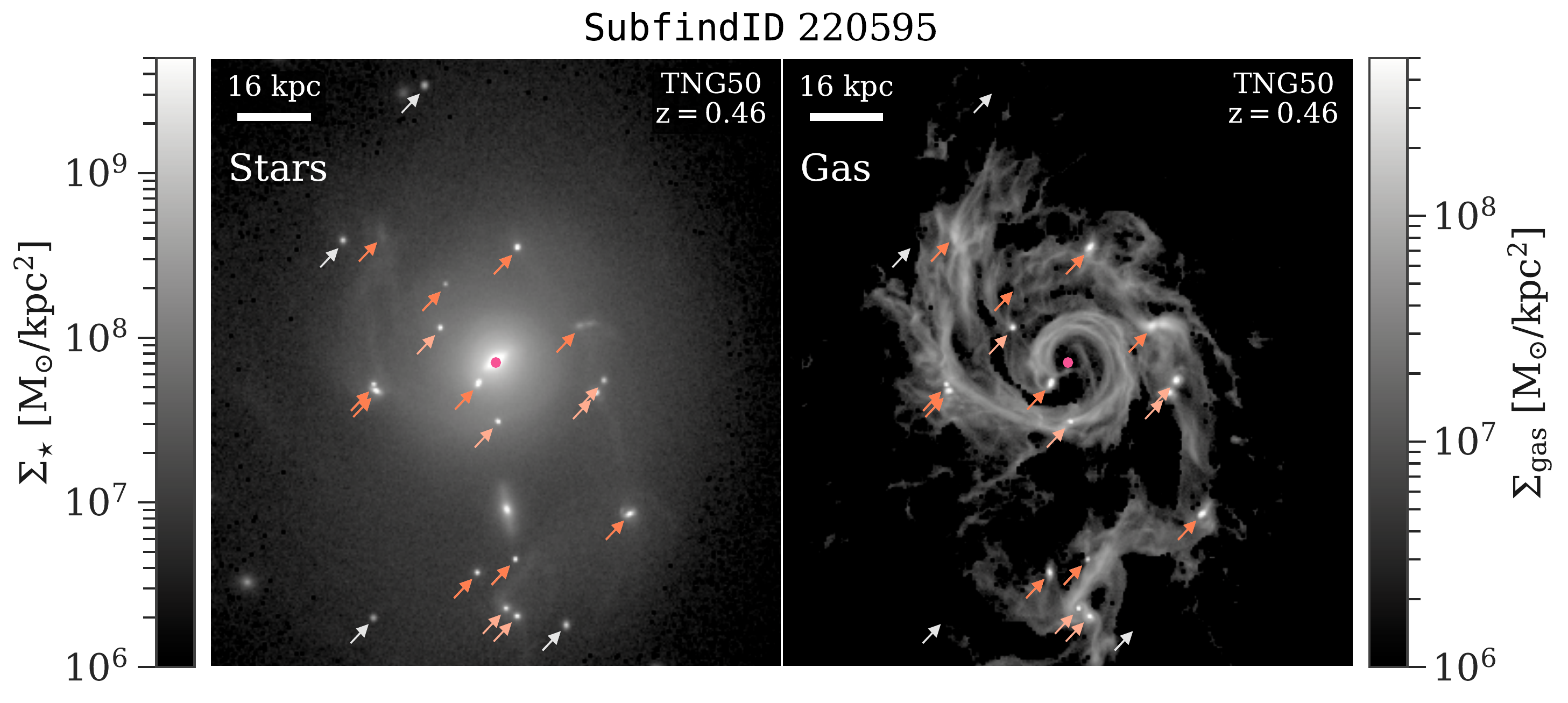}\\
    \vspace*{12pt}
    \includegraphics[width=0.85\textwidth]{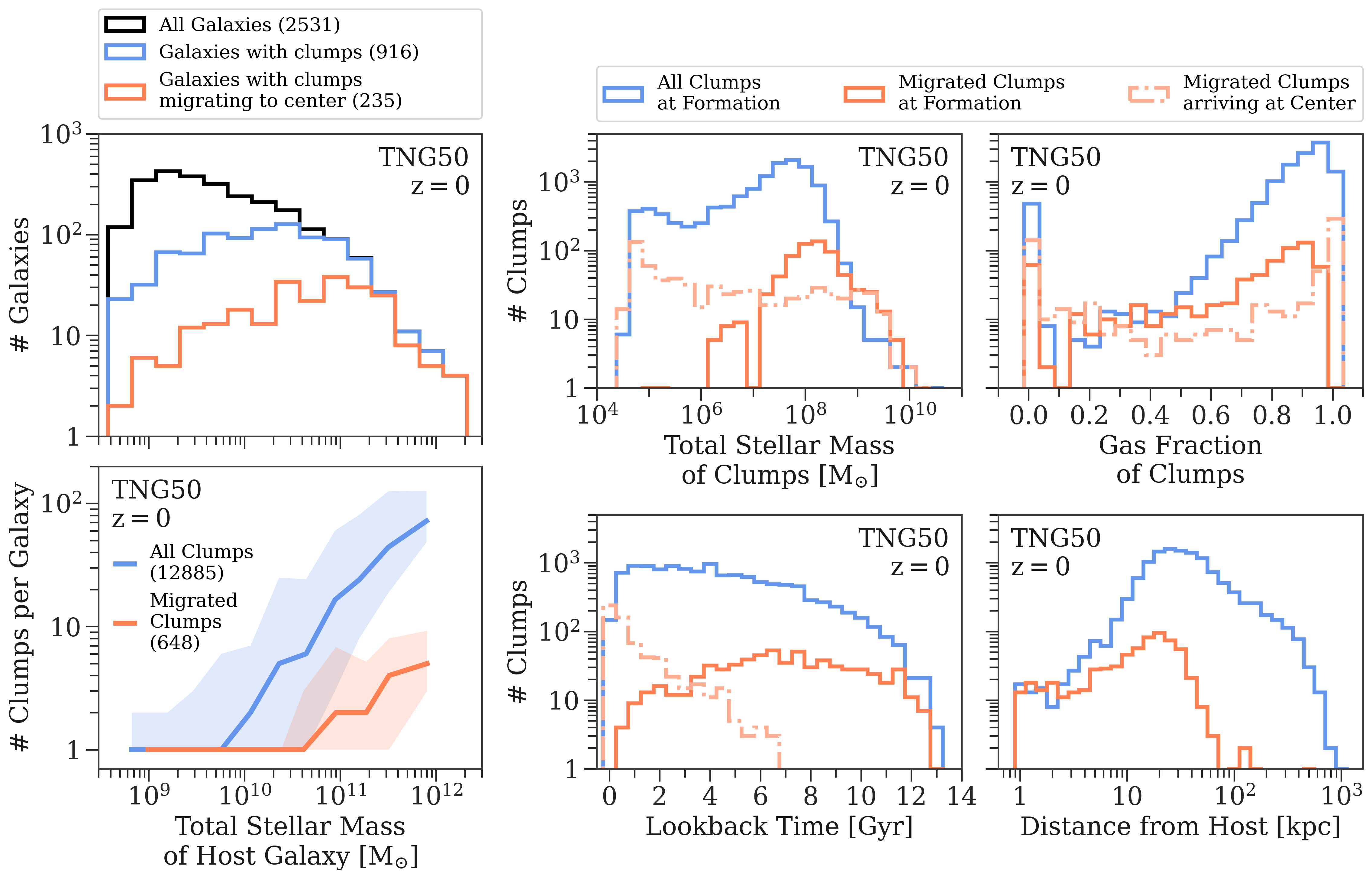}
    \caption{\textbf{Example and summary statistics of clumps in TNG50 galaxies.} \textit{Top panel}: Stellar (\textit{left}, full projection) and gas (\textit{right}, thin slice with $|z|\leq5\,\mathrm{kpc}$) surface mass density of a TNG50 galaxy (\texttt{SubfindID} 220595) in face-on projection exhibiting clumps at $\mathrm{z}=0.47$. Clumps migrating to the center (pink dot) are marked with orange arrows, whereas clumps not belonging to the galaxy are marked with white arrows. Clumps with lighter orange arrows will first merge with other clumps (darker orange arrows), before arriving at the center. \textit{Lower panel, left}: Distribution of host galaxy stellar masses (\textit{top}) displaying clumps in general (\textit{blue}) and clumps that migrate to the center (\textit{orange}) \textit{at any time}, compared to the total galaxy sample (\textit{black}). Numbers inside the brackets display the total amount. All galaxies in our sample above $\sim 5\times10^{10}\,\mathrm{M}_{\odot}$ exhibit clumps at some point in their life time. The median number of clumps that ever existed per galaxy as a function of host galaxy stellar mass are also shown (\textit{bottom}). The shaded area shows the 16th and 84th percentile. The number of clumps arriving at a galaxy's center is about a decade lower than the total amount of clumps formed. Note however that the true number of all clumps ever formed is higher due to clump-clump mergers. \textit{Lower panel, right}: Distribution of certain clump properties (\textit{blue solid line}: all clumps at the time of formation, \textit{orange solid line}: clumps that migrate to the center at the time of formation, \textit{light orange dashed-dotted line}: clumps that migrated to the center at the time of arrival at the center): \textit{from left to right}: total stellar mass, gas fraction (ratio of gas mass to total baryonic mass), formation time (\textit{solid lines})/time taken to arrive at the center (\textit{dashed line}) and distance from the host galaxy of the clumps.}
    \label{fig: clumps_sum}
\end{figure*}

Here, we show some examples and properties of clumps, i.e. subhalos detected by the \textsc{Subfind} algorithm, but which were formed not through the collapse of a dark matter halo, that are present in TNG50. Instead these clumps form in the gaseous disk of galaxies or fragments of it. Particularly, we could observe qualitatively that significant clump formation predominantly occurs during galaxy interactions at $\mathrm{z>1}$, such as mergers or fly-bys, but it can also take place when a galaxy evolves in isolation \citep[see also][for a similar finding]{dimatteo08}. \par
We present a visual example of such clumps at the top of Figure \ref{fig: clumps_sum} for a galaxy that lives in an environment with many galaxy interactions. Clumps embedded in the stellar disk of that galaxy clearly exhibit a gaseous counterpart, whereas clumps further out and not within the disk are only seen in the stellar surface mass density (at the time of inspection). The clumps visible in both the stars and gas are clearly distributed along the spiral (or tidal) arms of that galaxy. \emph{All} of them migrate to the center and deposit stars (and also gas) there (marked by orange arrows). In that process some of them (marked by lighter orange arrows) merge with other clumps, which we reconstructed with the merger trees. \par
The statistics on clumps in TNG50 shown on the bottom left hand side of Figure \ref{fig: clumps_sum} reveals that around 36\% of all TNG50 galaxies posses clumps at any point in their lifetime of which a fourth have clumps that migrate to the center. Above $5\times10^{10}\,\mathrm{M}_{\odot}$ in stellar mass all galaxies exhibit clumps at some point. The amount of clumps per galaxy starts to rise above one for galaxy stellar masses larger than $10^{10}\,\mathrm{M}_{\odot}$. On average a Milky Way mass galaxy has five clumps, whereas the most massive galaxies in TNG50 can have up to a hundred clumps. The number of clumps that actually migrate to a galaxy's center starts to rise at higher galaxy masses at around $5\times10^{10}\,\mathrm{M}_{\odot}$ reaching an average of about 5 migrated clumps per galaxy at the high mass end. We remind the reader that these and following numbers \emph{do not} account for clumps that merged with other clumps, thus the true number of all clumps ever formed is actually even higher. \par
We also show the distribution of four properties of the clumps in TNG50 on the bottom right hand side of Figure \ref{fig: clumps_sum}. Each property is shown for all clumps at the time of formation, for all clumps that migrated at the time of formation and for all clumps that migrated at the time they arrive at their galaxy's center. \par
We see that clumps are formed with a broad distribution of stellar masses with a peak at $10^{8}\,\mathrm{M}_{\odot}$\footnote{We caution the reader to not trust the few clumps that have stellar masses of around $10^{10}\,\mathrm{M}_{\odot}$, which likely originate from switching between the host galaxy and the actual clump during the halo/subhalo finding process.}. Only clumps that are formed with such stellar masses or higher are able to migrate to the center. The stellar mass that they then actually deposit at the center is a flattened out distribution all the way down to a few stellar particles. Thus clumps can suffer significant stellar mass loss while travelling to the galaxy center.\par
Even though there is a broader peak of clumps formed with high gas fractions ($0.6-1$), the gas fraction distribution of clumps that migrate to the center is significantly flatter. Thus, a high gas fraction is not necessarily an indication of whether a clump is able to migrate to the center or not.\par
Furthermore, the clumps are formed all throughout cosmic time in TNG50, with a slight increase towards younger lookback times. The number of clumps that migrated with formation times younger than 6\,Gyr ago drops compared to clumps formed at older ages. This is understandable as it takes a few Gyr for the clumps to migrate to the center. Most clumps need around 1\,Gyr to do so, but there is a long tail towards higher migration times up until 6\,Gyr. \par
The peak distance from the host galaxy, where clumps form, is around 20\,kpc and flattens out towards smaller distances. Whereas the number of migrated clumps drop sharply after that distance, the distance of clumps formed overall extends all the way until several hundred kpc. Hence, clumps in TNG50 can form in the halo of galaxies, possibly in gaseous tidal or ram-pressure-stripped tails and gas fragments during galaxy interactions. We empathize here that the clumps formed at such large distances are \emph{not} part of any satellite galaxy that then merged with the host (at least according to the \textsc{Subfind} algorithm and the merger trees). This is because we identify clumps as subhalos with $\mathtt{SubhaloFlag}=0$ that \emph{directly} merged onto the main progenitor branch of the host galaxy. Hence, clumps that are part of satellite galaxies that are then brought in by merging with the host galaxy, are also not accounted for in our statistics. \par
We provide further discussion and comparison to clumps in other simulations in Section \ref{sec: clumpdiscuss}.

\bsp	
\label{lastpage}
\end{document}